\RequirePackage[table]{xcolor}

\documentclass[12pt]{article}
\usepackage{amssymb}   
\usepackage{amsmath}
\usepackage{graphicx}
\usepackage{axodraw4j}
\usepackage[sort]{cite}

\usepackage{multirow}
\usepackage{grffile}

\def\beq{\begin{equation}}   
\def\eeq{\end{equation}}
\def\bea{\begin{eqnarray}}  
\def\eea{\end{eqnarray}} 
\def\nn{\nonumber}

\def\f21{{}_2F_{1}}

\RequirePackage[colorlinks=true
,urlcolor=blue
,anchorcolor=blue
,citecolor=blue
,filecolor=blue
,linkcolor=blue
,menucolor=blue
,pagecolor=blue
,linktocpage=true
,pdfproducer=medialab
,pdfa=true
]{hyperref}

\newcommand{\df}{\mathrm{d}}

\def\beq{\begin{equation}}
\def\eeq{\end{equation}}
\def\bsp#1\esp{\begin{split}#1\end{split}}

\newcommand\barparen[1]{\overset{\textbf{\fontsize{2pt}{2pt}\selectfont(--)}}{#1}}

\usepackage[section]{placeins}

\usepackage{color}
\newcounter{RSQ}

\newlength{\parwidth}
\newcommand{\cpcsub}[1]
{%
\setlength{\parwidth}{\textwidth}\addtolength{\parwidth}{-2.1em}%
\begin{center}
\begin{tabular}[t]{@{}p{\parwidth}@{}}
#1
\end{tabular}
\end{center}
}%

\catcode`\@=11
\font\manfnt=manfnt
\def\Watchout{\@ifnextchar [{\W@tchout}{\W@tchout[1]}}
\def\W@tchout[#1]{{\manfnt\@tempcnta#1\relax%
  \@whilenum\@tempcnta>\z@\do{%
    \char"7F\hskip 0.3em\advance\@tempcnta\m@ne}}}
\let\foo\W@tchout
\def\dubious{\@ifnextchar[{\@dubious}{\@dubious[1]}}

\def\@dubious[#1]{%
  \setbox\@tempboxa\hbox{\@W@tchout#1}
  \@tempdima\wd\@tempboxa
  \list{}{\leftmargin\@tempdima}\item[\hbox to 0pt{\hss\@W@tchout#1}]}
\def\@W@tchout#1{\W@tchout[#1]}
\catcode`\@=12

\usepackage{longtable}

\begin{document}

\begin{flushright}
{\small
CERN-TH-2022-109,
SLAC-PUB-17699,
BONN-TH-2022-22

}
\end{flushright}

\vskip1cm
\begin{center}
{\Large \bf \boldmath Inclusive Production Cross Sections at N$^3$LO}
\end{center}

  \vspace{0.5cm}
\begin{center}
{\sc Julien~Baglio,$^{a}$ \sc Claude~Duhr,$^{b}$ \sc Bernhard~Mistlberger,$^{c}$  and Robert~Szafron$^{d}$} 
\\[6mm]
{\it ${}^a$Theoretical Physics Department, CERN, Esplanade des Particules 1, 1217 Meyrin, Switzerland}\\[1mm]
{\it ${}^b$Bethe Center for Theoretical Physics, Universit\"at Bonn, D-53115, Germany}\\[1mm]
{\it ${}^c$SLAC National Accelerator Laboratory, Stanford University, Stanford, CA 94039, U.S.A.}\\[1mm]
{\it ${}^d$Department of Physics, Brookhaven National Laboratory, Upton, N.Y., 11973, U.S.A.}
\\[0.3cm]

\end{center}

\begin{abstract}
We present for the first time the inclusive cross section for associated Higgs boson production with a massive gauge boson at next-to-next-to-next-to-leading order in QCD.
Furthermore, we introduce {\tt n3loxs}, a public, numerical program for the evaluation of inclusive cross sections at the third order in the strong coupling constant.
Our tool allows to derive predictions for charged- and neutral-current Drell-Yan production, gluon- and bottom-quark-fusion Higgs boson production and Higgs boson associated production with a heavy gauge boson.
We discuss perturbative and parton distribution function (PDF) uncertainties of the aforementioned processes.
We perform a comparison of global PDF sets for a variety of process including associated Higgs boson production and observe $1\sigma$ deviations among predictions for several processes.
\end{abstract}
 \thispagestyle{empty} 
 
\newpage
\tableofcontents
\newpage
\section{Introduction}
Experimental high-energy physics is at the dawn of a new precision era. 
Run 3 of the Large Hadron Collider (LHC) at CERN and its subsequent upgrade to the High-Luminosity LHC will bring a plethora of further data. 
These efforts on the experimental side must be met with a solid push towards more accurate and precise theoretical predictions. 
So far, the Standard Model (SM) has been highly successful in describing data across many scales and processes. 
This success has been possible thanks (among other things) to reasonable control over QCD effects, which dominate at hadron colliders. 
Next-to-leading order (NLO) corrections are now routinely available for the SM backgrounds and predicted signals of new physics. 
The past few years have seen a surge of next-to-next-to-leading order (NNLO) computations, and nowadays, many differential observables are known at NNLO accuracy. 
Still, very little is known about the third-order corrections (N$^3$LO), with the first prediction for single Higgs production at the LHC becoming available in 2015~\cite{Anastasiou:2015vya}, and the very first differential observable for the same process having been computed only very recently~\cite{Chen:2021isd}. Since this very first differential calculation more results focusing on Drell-Yan processes have been released in the past few months, see refs.
~\cite{Chen:2021vtu,Camarda:2021ict,Neumann:2022lft,Chen:2022cgv,Chen:2022lwc}.
Inclusive processes for the production of a colorless final state are typically the simplest to compute, and thus they are the first step in pushing the frontier of QCD perturbative computations. 
Recent years have brought significant progress in evaluating the N$^3$LO corrections to inclusive color singlet production cross sections. 
This endeavor started with the evaluation of the gluon fusion Higgs production cross section~\cite{Anastasiou:2014vaa,Anastasiou:2015vya,Anastasiou:2016cez,Mistlberger:2018etf,Anastasiou:2013srw,Anastasiou:2015yha,Anastasiou:2013mca,Anastasiou:2014lda}. 
Shortly after the Vector Boson Fusion (VBF) cross section for Higgs boson production was obtained at N$^3$LO with very different methods~\cite{Dreyer:2016oyx}.
Today $W$-boson and photon mediated Drell-Yan cross sections are also known at the third order in the strong coupling constant $\alpha_S$~\cite{Duhr:2020seh,Duhr:2020sdp}, as well as the computation of the N$^3$LO QCD predictions for Higgs production in bottom-quark fusion in the five-flavor scheme~\cite{Duhr:2019kwi} (including the matching to the four-flavor scheme~\cite{Duhr:2020kzd}) and the calculation of the gluon fusion and VBF double Higgs boson production cross sections~\cite{Chen:2019fhs,Dreyer:2018qbw}.
Recently, the computation of the Drell-Yan cross section was completed at N$^3$LO with the calculation of the $Z$ boson production cross section~\cite{Duhr:2021vwj}. All these computations share many similarities. 
One of the main results of this paper is to introduce the new public computer code {\tt n3loxs}, which allows us to evaluate the N$^3$LO inclusive cross sections for a variety of hadron collider processes.\footnote{There already exist dedicated codes for evaluation of the N$^3$LO Higgs production cross section, such as {\tt iHixs 2}~\cite{Dulat:2018rbf}, {\tt ggHiggs}~\cite{Bonvini:2016frm} and {\tt SusHi}~\cite{Harlander:2016hcx}. Note also that for Higgs production in vector-boson fusion, which is not included in our tool, there also exists a public tool for the evaluation of the cross section called {\tt proVBFH}~\cite{Dreyer:2016oyx}} We then use our code and present a comprehensive phenomenological study of these processes, at $pp$ colliders, and, for the first time, also at $p\bar{p}$ colliders.

Inclusive cross sections play a variety of roles in high-energy phenomenology, ranging from precision tests of the SM to serving as standard candles for luminosity measurements.
Furthermore, they provide a unique testing ground to study the progression of predictions obtained with increasing order in the perturbative expansion of scattering cross sections.
Deriving reliable estimates of the uncertainty due to the truncation of the perturbative expansion plays a role of increasing importance as experimental results become more precise.
Today's standard paradigm is to estimate the size of the corrections neglected due to the truncation of the perturbative expansion by varying the unphysical renormalization and factorization scales. 
Only terms beyond the highest computed perturbative order are sensitive to the value of these scales, and the magnitude of the variation of the prediction is interpreted as an uncertainty.
As the perturbative order is increased, the size of scale variations is expected to diminish, leading to quantitatively better results.
Studying the progression of the perturbative series then serves as a test of this method to estimate uncertainties. 
Note that alternative approaches to estimating missing higher-order uncertainties were discussed for example in refs.~\cite{Cacciari:2011ze,Bonvini:2020xeo,Duhr:2021mfd}.

In refs.~\cite{Duhr:2020seh,Duhr:2020sdp,Duhr:2021vwj}, a significant cancellation of contributions to the total cross section from different partonic initial-state channels was observed. 
Individual partonic channels are weighted (or convoluted) with their respective parton distribution functions (PDFs).
A small change in the PDF of one parton with respect to others might consequently result in a significant change of the predicted cross section.
This observation raises a series of questions:
Are these cancellation relevant only for proton-proton collisions, or do they also appear in proton-antiproton initiated processes, such as those studied at the former Fermilab Tevatron collider? 
Do missing N$^3$LO PDFs play an essential role, and how much do our N$^3$LO predictions depend on the choice of the PDF set? 
We are going to discuss these questions on examples of different processes using the N$^3$LO perturbative results. 
This paper provides a comprehensive study of the impact of the choice of the PDF and the associated 68\% confidence level (CL) PDF uncertainties on a variety of processes at N$^3$LO.
We compare the latest releases of the major PDF sets for hadron colliders to a nominal PDF set released in 2015 as a statistical combination of three global sets.

We will also present for the first time results for associated $W^\pm H$ and $Z H$ production at N$^3$LO in QCD. 
These so-called Higgsstrahlung processes are characterized by large masses of the final-state particles, which cannot be neglected, as it is customarily done for leptons. 
We provide both the total cross section and results binned according to the invariant mass of the pair of a massive weak boson with a Higgs boson. 
We discuss the theoretical uncertainties due to scale variation and PDFs, providing also the correlated PDF+$\alpha_S$ uncertainty as well as the uncertainty related to the missing N$^3$LO PDF set.

This paper is organized as follows. In section~\ref{sec:theory} we introduce the main calculational framework and the needed formulae for inclusive color singlet production at hadron colliders. The phenomenological results for Higgs and Drell-Yan production are given in section~\ref{sec:pheno}. We present the analysis of the variation of the PDF sets and the corresponding PDF uncertainties at the LHC at 13 TeV, including also validation plots for scale variation. The new predictions at N$^3$LO in QCD for inclusive Higgsstrahlung production at hadron colliders, $p p / p \bar{p} \to  VH$ with $V=W^\pm, Z$, are given in section~\ref{sec:vh}.
We compare N$^3$LO predictions for different processes in section~\ref{sec:N3LOComp}.
 A conclusion is provided in section~\ref{sec:conclusion}. We also include three appendices. Appendix~\ref{app:code} explains in detail how to install and use our public code {\tt n3loxs}. Appendix~\ref{app:parameters} provides the reader and user of {\tt n3loxs} with the physical parameters used in our predictions. Appendices~\ref{app:vhadditional} and~\ref{app:binned} collect additional results for associated Higgs production at proton-proton colliders for various center-of-mass energies as well as binned invariant-mass distributions for Drell-Yan processes.



\section{Inclusive color singlet production in QCD}
\label{sec:theory}

The focus of this paper are QCD corrections to hadron collider
processes for the $s$-channel production of a colorless final-state
$B$ of invariant mass $Q$:
\begin{align}
N_1(P_1) +N_2(P_2) \to B(Q)+X\,,
\end{align}
where $N_i(P_i)$ is either a proton or antiproton with momentum $P_i$,
and $X$ represents arbitrary QCD final-state radiation. In the
following we will be inclusive in the QCD radiation $X$ and we are
interested in the production cross section differential in $Q^2$.

\begin{figure}[htb!]
  \begin{center}
    \includegraphics[scale=0.15]{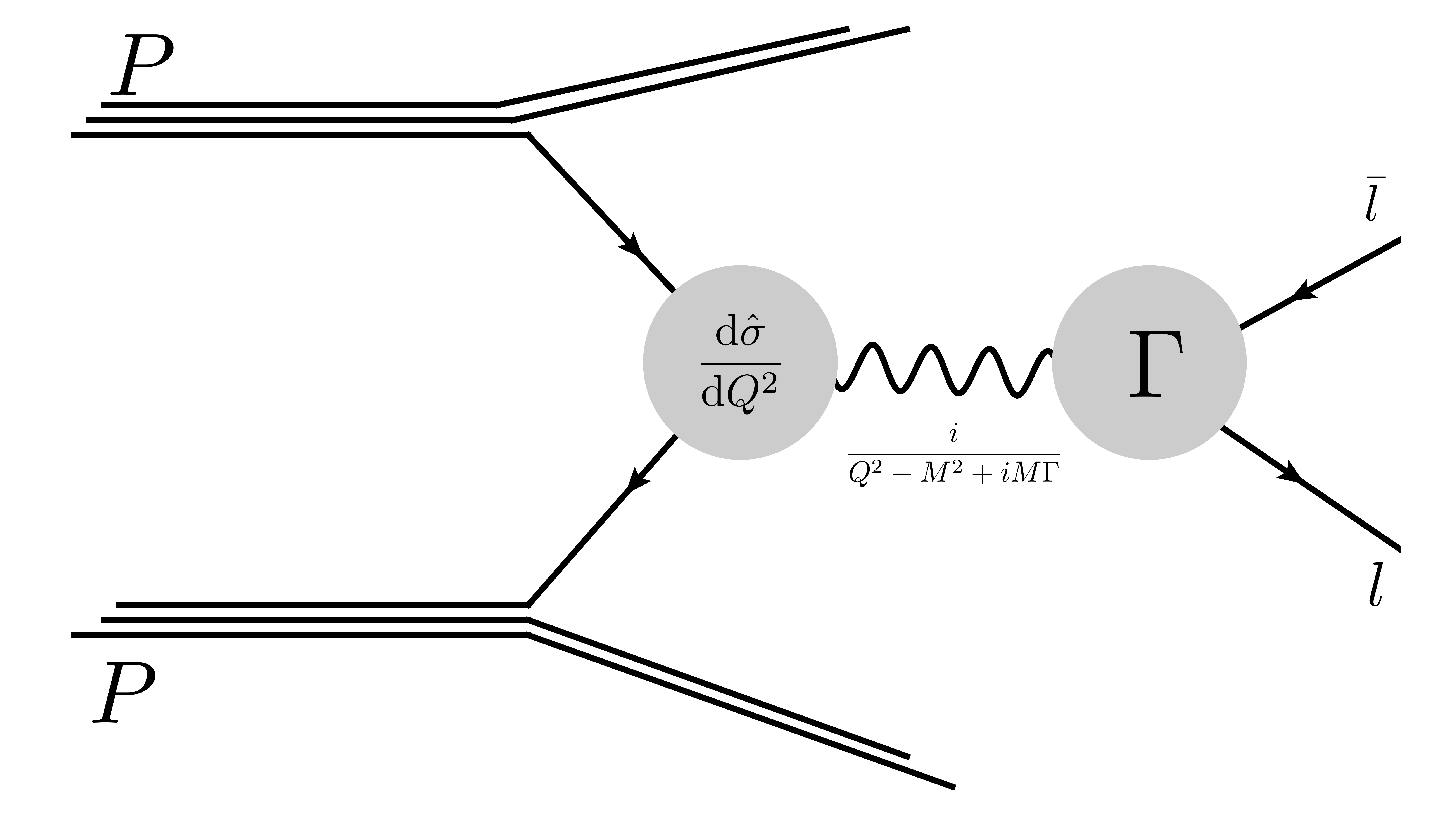}
    \caption{\label{fig:masterform}
      Schematic depiction of the Drell-Yan process and its
      factorization into the production probability of a virtual gauge
      boson and subsequent decay to final-state leptons.}
  \end{center}
\end{figure}

If the widths of all intermediate $s$-channel states are small, we can
think of this process in a factorized form involving three distinct
steps.
First, energetic partons collide and produce one (or more) off-shell
colorless $s$-channel resonances.
This part is described by short-distance partonic cross sections
convoluted with the parton distribution functions (PDFs). These resonances then propagate over a short distance. Since we assume
that the widths are much smaller than the mass, the propagation can be
described by a relativistic Breit-Wigner formula.
Finally, the intermediate particles decay into the color-singlet
final-state particles with a probability determined by the partial
decay width. The prototypical example of such a process is the
Drell-Yan production process \cite{Drell:1970wh}, which describes the
production of a lepton/antilepton pair in proton-proton collisions
via an off-shell photon or $Z$-boson (see
fig.~\ref{fig:masterform}). In the case where the final state $B$
consists of a pair $f_1f_2$ of particles with mass $m_{f_i}$, the
cross section differential in the invariant mass $Q^2$ can be cast in
the form:
\begin{align}
\label{eq:master}
&\Sigma_{N_1 N_2\to f_1 f_2+X}(Q^2) \equiv Q^2 \frac{\df \sigma_{N_1 N_2\to f_1 f_2+X}}{\df Q^2}=\nn\\
&= \sum_{V,V^\prime}Q^2 \frac{\df \sigma_{N_1\, N_2\rightarrow V/ V^\prime+X}}{\df Q^2}\times \text{BW}(Q^2,m_V,m_V^\prime) \times \Gamma_{V / V^\prime \to f_1 f_2}(Q^2 ,m_{f_1}^2, m_{f_2}^2)  \,.
\end{align}
The propagation of the virtual $s$-channel states $V$ and $V^\prime$ is described by the relativistic Breit-Wigner distribution,
\beq
\text{BW}(Q^2,m_V,m_V^\prime) =\frac{Q^3}{\pi} \Re\left[\frac{1}{(Q^2-m_{V}^2)+i m_V \Gamma_V }\frac{1}{(Q^2-m_{{V^\prime}}^2)-i m_{V^\prime} \Gamma_{V^\prime} }\right]\,,
\eeq
where $\Re$ denotes the real part and $m_{V^{(\prime)}}$ and
$\Gamma_{V^{(\prime)}}$ the on-shell mass and total decay width of the
state ${V^{(\prime)}}$ respectively. The decay is described by the
interference of the decay matrix elements integrated over the
phase-space of the final-state particles $f_1\,f_2$:
\beq
\Gamma_{V / V^\prime \to f_1f_2}(Q ,m_{f_1}, m_{f_2}) =\frac{1}{2QN_{\textrm{spin}}}\int \df\Phi_2\,
 \mathcal{M}_{V \to f_1f_2} \cdot \mathcal{M}_{V^\prime \to f_1f_2}^* \,,
\eeq
where $N_{\textrm{spin}}$ denotes the number of spin states of $V$,
i.e., $N_{\textrm{spin}} = 3$ if $V^{(\prime)}$ is a massive
(axial-)vector and $N_{\textrm{spin}} = 1$ if it is a scalar, and
$\df\Phi_n$ denotes the $n$-body phase-space measure in $d$
dimensions:
\beq
\df\Phi_n =(2\pi)^d\,\delta^d\left(Q-\sum_{f=1}^n p_f\right) \prod_{f=1}^n\frac{\df^dp_f}{(2\pi)^{d-1}}\,\delta_+(p_f^2-m_f^2)\,.
\eeq
The quantity $ \mathcal{M}_{V \to f_1f_2} \cdot \mathcal{M}_{V^\prime
  \to f_1f_2}^*$ denotes the interference of the matrix elements,
summed over all color and spin quantum numbers of all external
particles.
The above definition is chosen such that, if the formula is evaluated
on-shell, $Q=m_V$ and $V=V^\prime$, then it corresponds to the partial
width $\Gamma_{V\to f_1f_2}$ of the state $V$ in its rest-frame,
\beq
\Gamma_{V\to f_1f_2}=\Gamma_{V \to f_1f_2}(m_V ,m_{f_1}, m_{f_2})\,.
\eeq

The remaining factor in eq.~\eqref{eq:master} is the total inclusive
two-to-one cross section for the production of a color-singlet
(axial-) vector or (pseudo-) scalar state $V$ with virtuality $Q^2$
(or the interference in the case $V\neq V^{\prime}$). It is obtained
by convolution of the partonic coefficient functions with the PDFs:
\beq
Q^2 \frac{\df \sigma_{N_1 N_2\rightarrow V/ V^\prime+X}}{\df Q^2}= \tau \sum_{i,j} \int_0^1 \df x_1\, \df x_2\,\df z\,  \delta(\tau-z x_1x_2)\,f_{i/N_1}(x_1)\,f_{j/N_2}(x_2)\,  \eta_{ij\to V / V^\prime +X}(z)\,,
\eeq
with 
\beq
\tau = \frac{Q^2}{s} \textrm{~~~and~~~} s = (P_1+P_2)^2\,,
\eeq
and $f_{i/N_k}(x_k)$ denotes the PDF to find a parton species $i$ with
momentum fraction $x_k$ inside the hadron $N_k$. We see that the computation of inclusive cross sections
for $s$-channel mediated color-singlet production can be reduced to
the computation of the coefficient functions for inclusive two-to-one
processes. The coefficient functions for (axial-)vector production
from quark annihilation and (pseudo-)scalar production from quark
annihilation as well as gluon fusion were computed up to NNLO in the
strong coupling constant more than two decades
ago~\cite{Altarelli:1978id,Kubar-Andre:1978eri,Altarelli:1979ub,Matsuura:1987wt,Matsuura:1988nd,Matsuura:1988sm,Hamberg:1990np,Matsuura:1990ba,vanNeerven:1991gh,Dicus:1988cx,Dicus:1998hs,Balazs:1998sb,Campbell:2002zm,Maltoni:2003pn,Harlander:2003ai,Georgi:1977gs,Djouadi:1991tka,Dawson:1990zj,Spira:1995rr,Harlander:2002wh,Anastasiou:2002yz,Ravindran:2003um,Harlander:2002vv,
Marzani:2008az,Pak:2009dg,Harlander:2009my,Czakon:2021yub}. Very recently also
N$^3$LO corrections to inclusive color-singlet production of scalar,
vector and axial-vector states have become available.
In the next sections, we discuss phenomenological predictions for the following processes.
\begin{itemize}
\item Drell-Yan production: $N_1N_2 \to \gamma^*/Z\to
  \ell^+\ell^-$~\cite{Duhr:2020seh,Duhr:2021vwj},
\item Charged-Current Drell-Yan production: $N_1N_2 \to W^{\pm}\to
  \ell^{\pm}\barparen{\nu}_{\ell}$~\cite{Duhr:2020sdp},
\item On-shell Higgs production $N_1N_2 \to H$ from both gluon and
  bottom-quark fusion~\cite{Anastasiou:2015vya,Duhr:2019kwi},
  \item Higgsstrahlung $N_1N_2 \to HV$, with $V\in \{Z,W^{\pm}\}$. These results are new and are presented here for the first time.
\end{itemize}



\section{Phenomenological results for Higgs and Drell-Yan production}
\label{sec:pheno}

In this section we present phenomenological results for the SM processes discussed in the previous section. We note that this list of (inclusive) processes exhausts the set of SM processes that can be computed using inclusive results for two-to-one processes. All results are obtained using the code {\tt n3loxs}, which is publicly available as a repository at \href{https://github.com/jubaglio/n3loxs}{https://github.com/jubaglio/n3loxs}.
We refer to appendix~\ref{app:code} for a detailed description of how to use the code.
We work with $n_f=5$ massless quark flavors. 
For the gluon fusion process, we consider an effective coupling of the Higgs boson to gluons in the large $m_t$ limit (heavy top-quark limit, or HTL)~\cite{Wilczek1977,Spiridonov:1988md,Shifman1978,Inami1983}, with the possibility of a Born-improved prediction where the heavy-top quark predictions are rescaled by the exact Born matrix elements at one loop.\footnote{Note, that we can choose either the on-shell scheme or the $\overline{\text{MS}}$ scheme (marked in the plots in this case) for treating the top-quark mass in the Wilson coefficients~\cite{Schroder:2005hy,Kramer:1996iq,Chetyrkin:2005ia,Chetyrkin:1997un} of the HTL Lagrangian.}
In the case of the neutral-current Drell-Yan process, we include non-decoupling top-mass effects due to the axial anomaly~\cite{Ju:2021lah,Chen:2021rft} and for the rest assume that the top quark is infinitely heavy and its degrees of freedom are integrated out. 
Unless stated otherwise, we present results for a proton-proton collider with a center-of-mass (c.m.) energy $\sqrt{s}=13$ TeV, and we use the \verb+PDF4LHC15_nnlo_mc+ set~\cite{Butterworth:2015oua}. 
The cross sections depend on the factorization scale $\mu_F$ and the renormalization scale $\mu_R$. 
The central scale choices for $\mu_F$ and $\mu_R$ are the following:
\begin{itemize}
\item $\mu_R^0 = \mu_F^0 = Q$ for Drell-Yan processes, $Q$ being the off-shellness of the intermediate vector boson,
\item $\mu_R^0 = \mu_F^0 = \frac12 m_H$ for Higgs production in gluon fusion,
\item $\mu_R^0 = m_H$, $\mu_F^0 = \frac14 (m_H + 2 m_b)$ for Higgs production in bottom-quark fusion.
\end{itemize}
The strong coupling is evolved from $\alpha_S(m_Z)=0.118$ using the five-loop QCD beta function~\cite{VanRitbergen:1997va,Czakon:2004bu,Baikov:2016tgj,Herzog:2017ohr} in the $\overline{\textrm{MS}}$-scheme using $N_f=5$ active massless quark flavors. For all other numerical input parameters (e.g., electroweak coupling constant, quark and boson masses, etc.) we refer to appendix~\ref{app:parameters}. Some of the results discussed in this section have already appeared earlier in the literature~\cite{Anastasiou:2015vya,Anastasiou:2016cez,Duhr:2019kwi,Duhr:2020sdp,Duhr:2020seh,Duhr:2021vwj}. We reproduce them here as an illustration and validation of the results one can obtain with {\tt n3loxs}. We will also perform an extensive study of the various sets of parton distribution functions (PDFs), including the latest sets that have been released and for which no results can be previously found in the literature. 

\subsection{Scale dependence and scale uncertainties}

\begin{figure}[!h]
\begin{center}
\includegraphics[width=0.45 \textwidth]{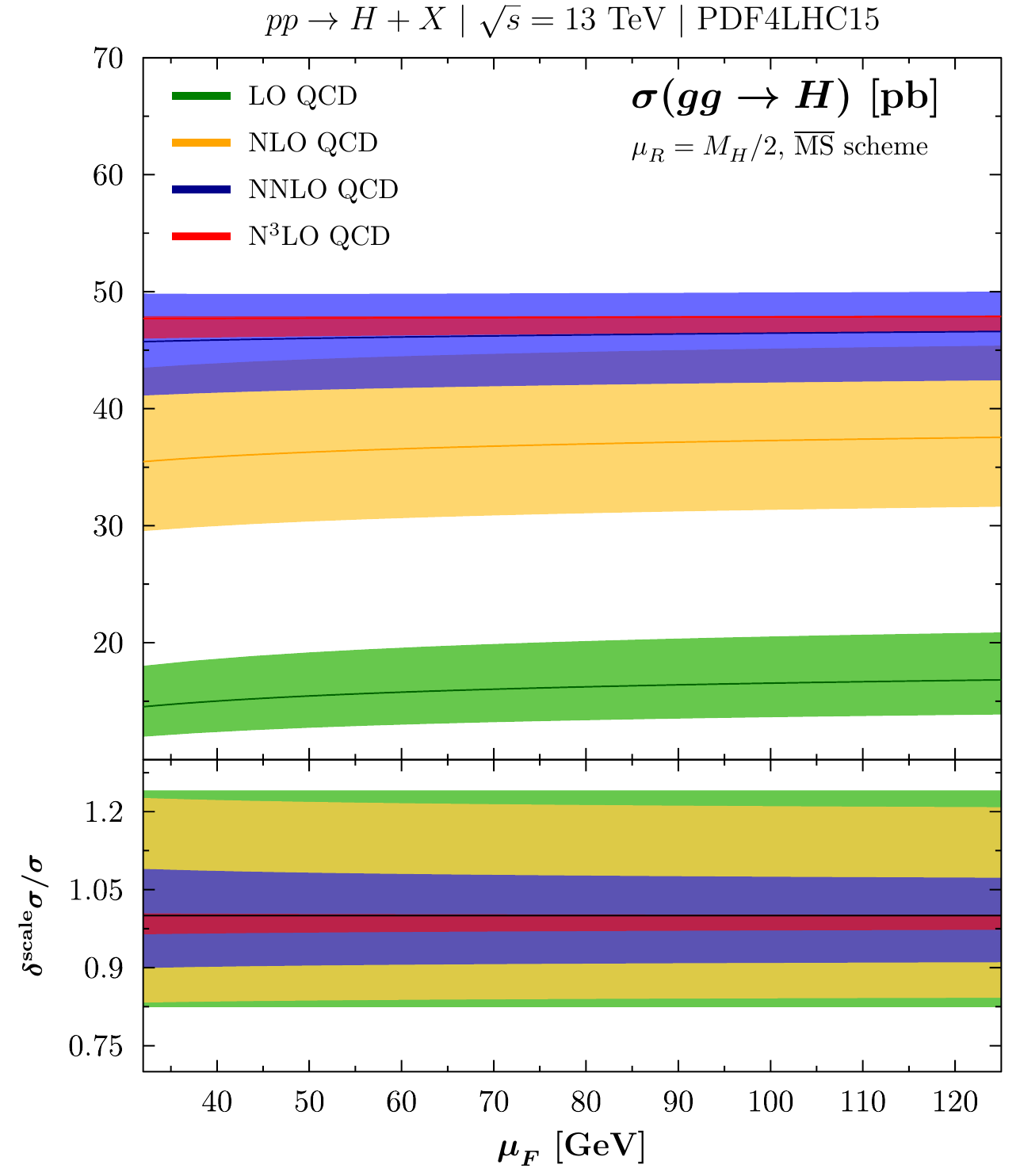} 
\includegraphics[width=0.45 \textwidth]{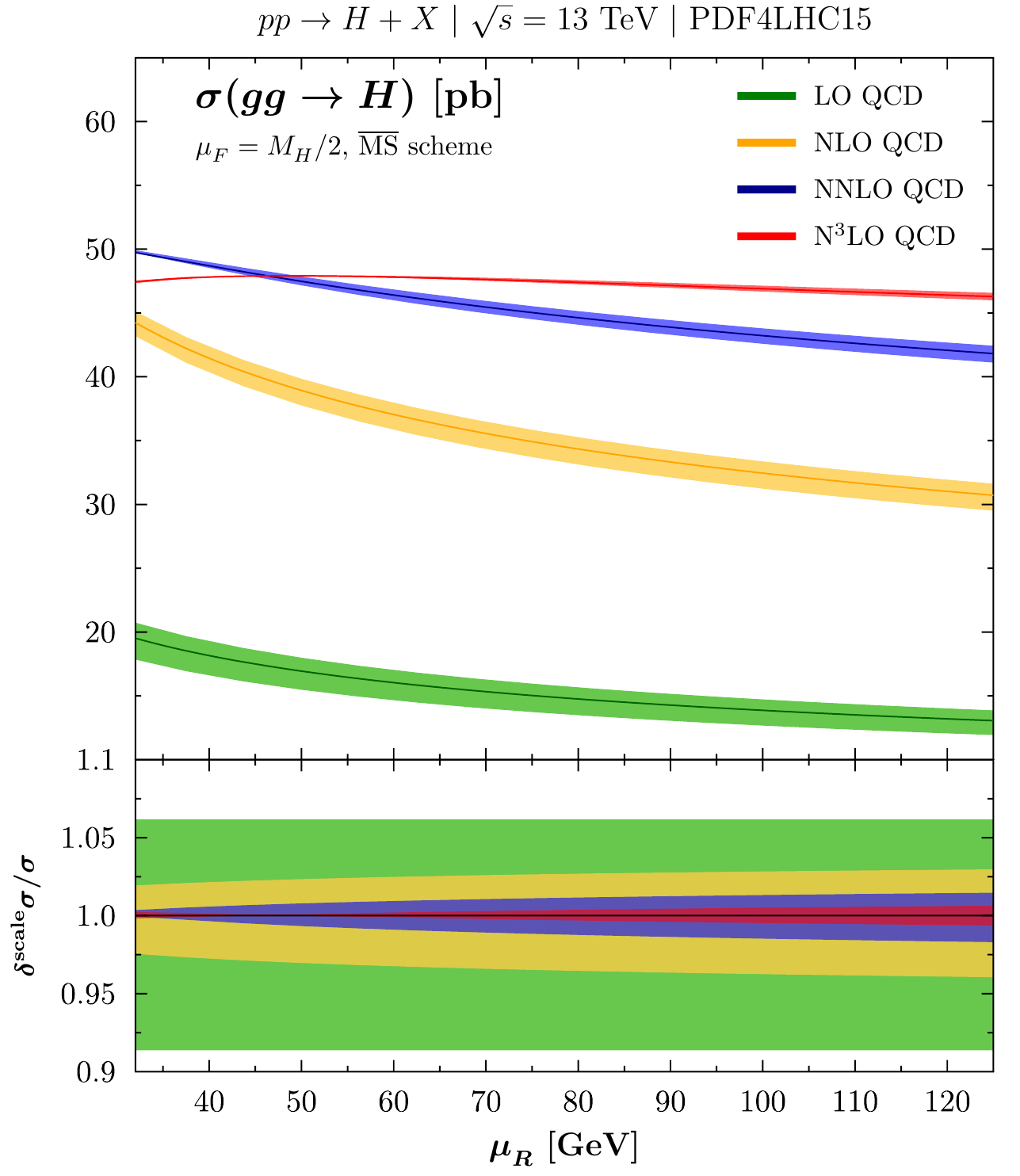}\\
\vspace*{4mm}
\includegraphics[width=0.45 \textwidth]{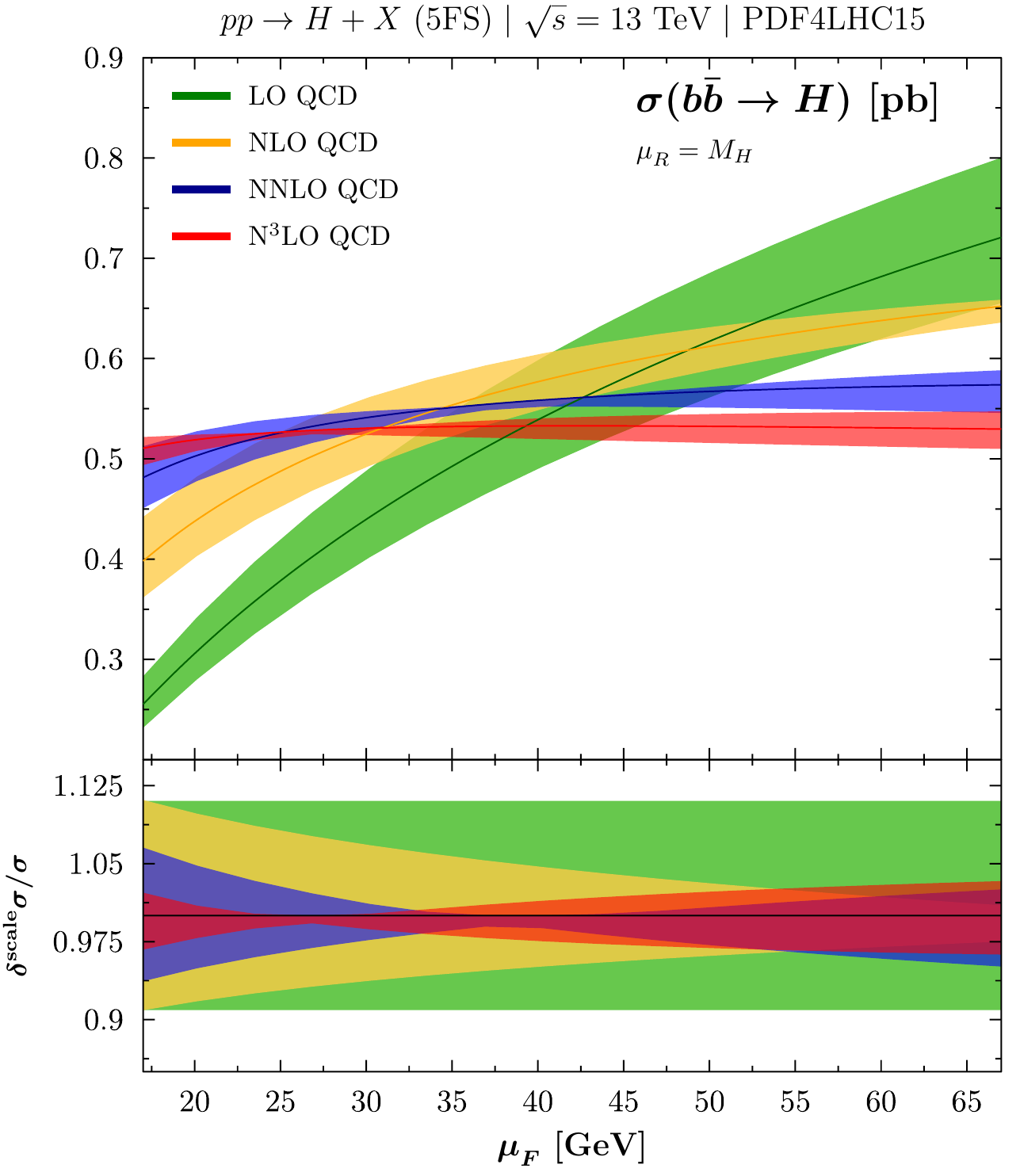}
\includegraphics[width=0.45 \textwidth]{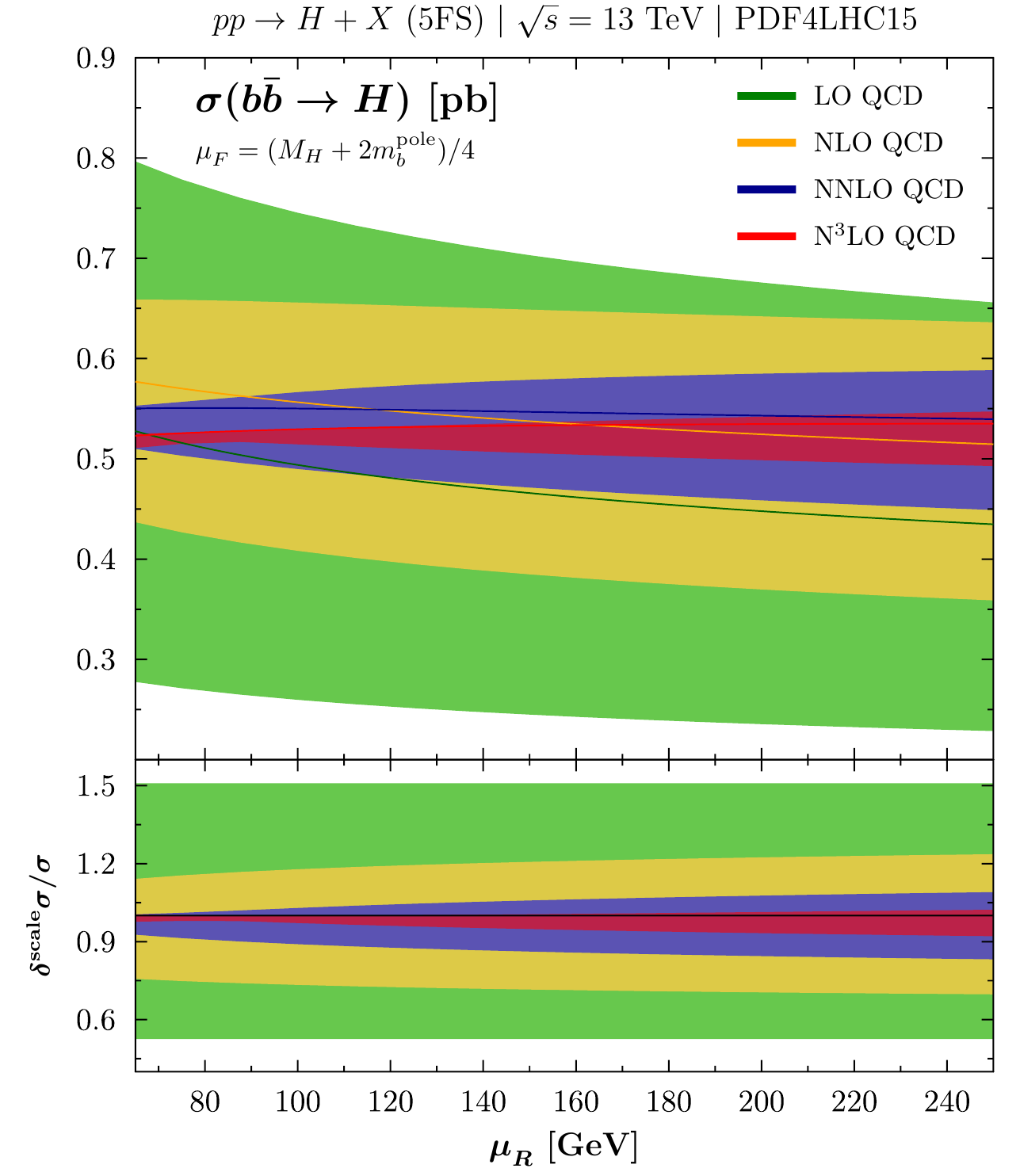}
\caption{\label{fig:scaleHiggs_ggH_bbH} Scale dependence for Higgs production in gluon fusion (upper row) and $b\bar{b}$ annihilation (lower row) at the 13 TeV LHC. The sub-panels display the ratio of the predictions to the central prediction for $\mu_R=\mu_R^0$, $\mu_F=\mu_F^0$. The left panels display the dependence over the factorization scale $\mu_F$ while the renormalization scale $\mu_R$ is fixed to $\mu_R^0$, the right panels display the dependence over $\mu_R$ while $\mu_F$ is fixed to $\mu_F^0$.
Bands represent the variation of the respective other scale by a factor of two up and down w.r.t. its central value.}
\end{center}
\end{figure}

In fig.~\ref{fig:scaleHiggs_ggH_bbH} we show the dependence of the cross section for Higgs production in bottom or gluon fusion on one of the two perturbative scales $\mu_F$ or $\mu_R$, with the other held fixed to the corresponding central value. The bands are obtained by varying the fixed scale up and down by a factor of two around its central value. We note that the scale dependence for Higgs production in bottom or gluon fusion at N$^3$LO had already been considered in refs.~\cite{Anastasiou:2015vya,Anastasiou:2016cez,Duhr:2019kwi}, and we reproduce those results perfectly. This serves as a validation of the code {\tt n3loxs}.

\begin{figure}[!h]
\begin{center}
\includegraphics[width=0.48 \textwidth]{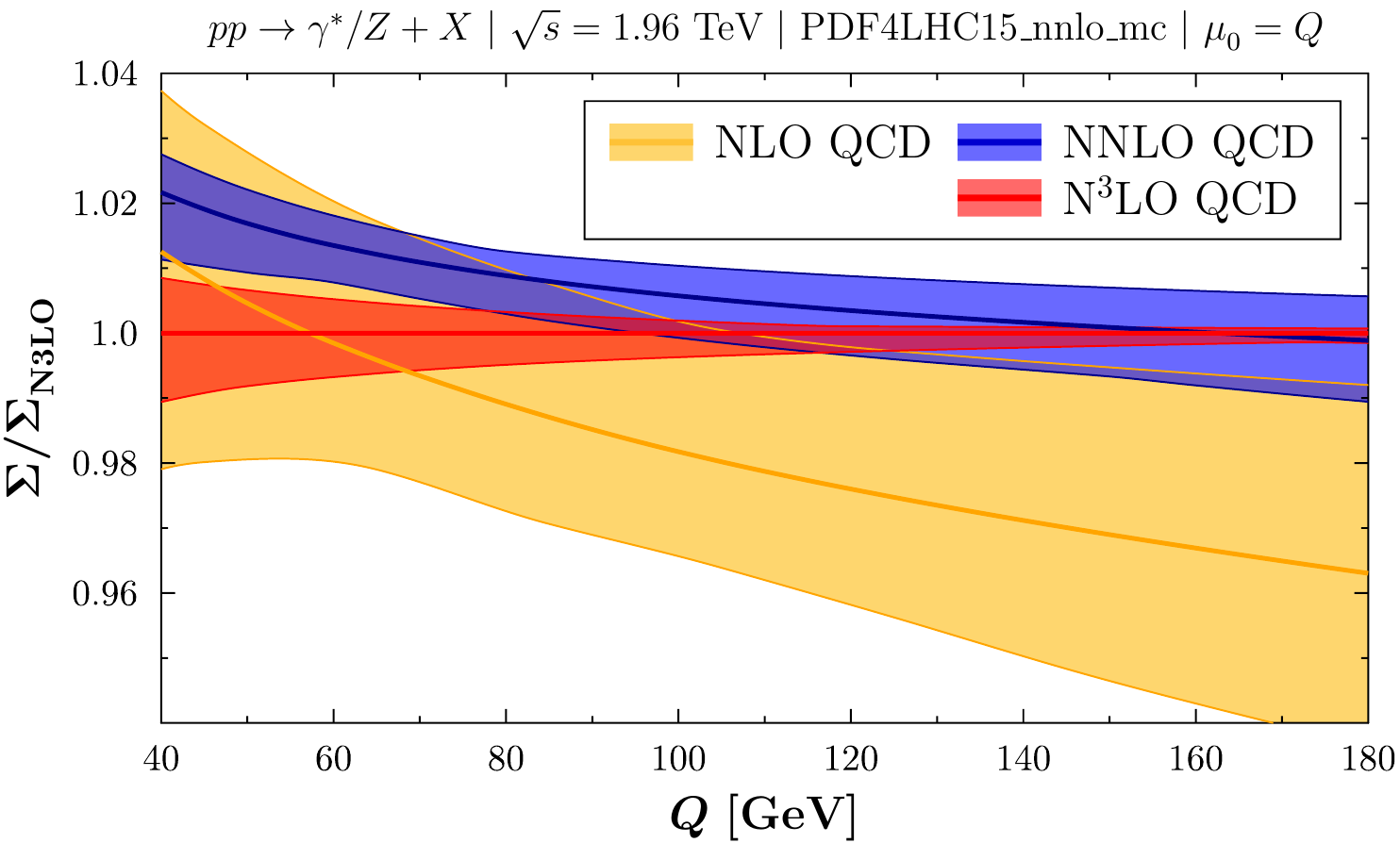} 
\includegraphics[width=0.48 \textwidth]{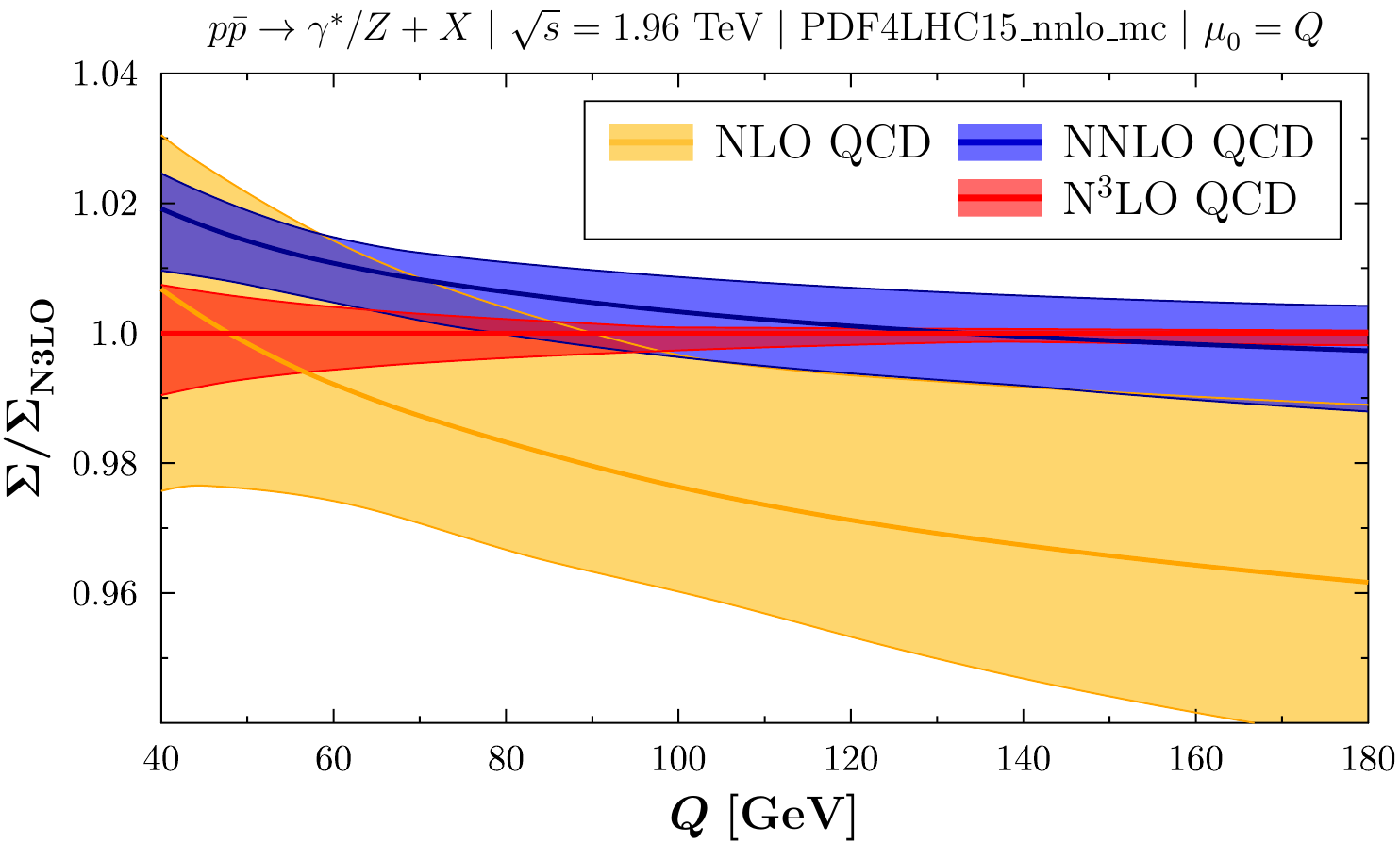} \\
\includegraphics[width=0.48 \textwidth]{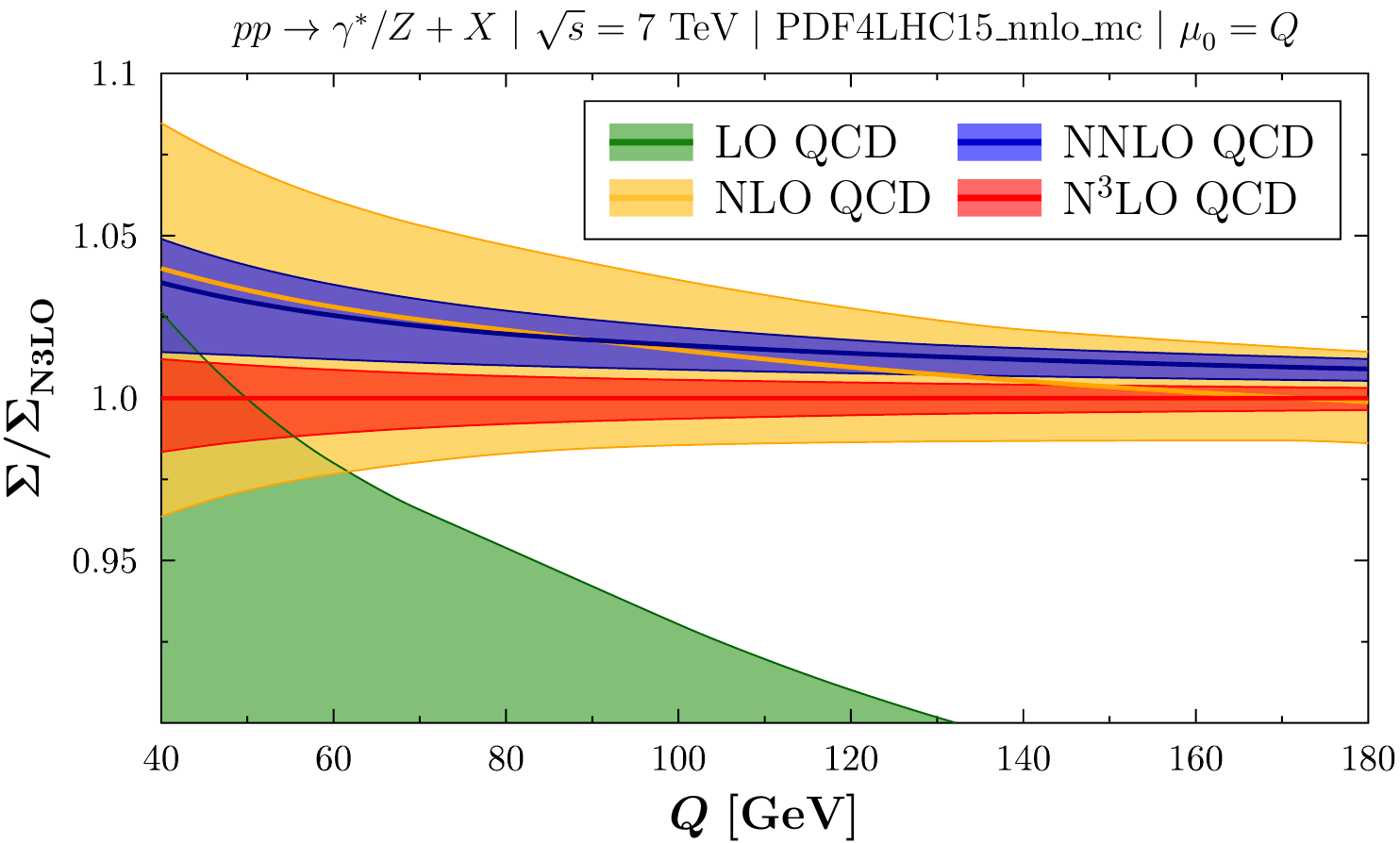} 
\includegraphics[width=0.48 \textwidth]{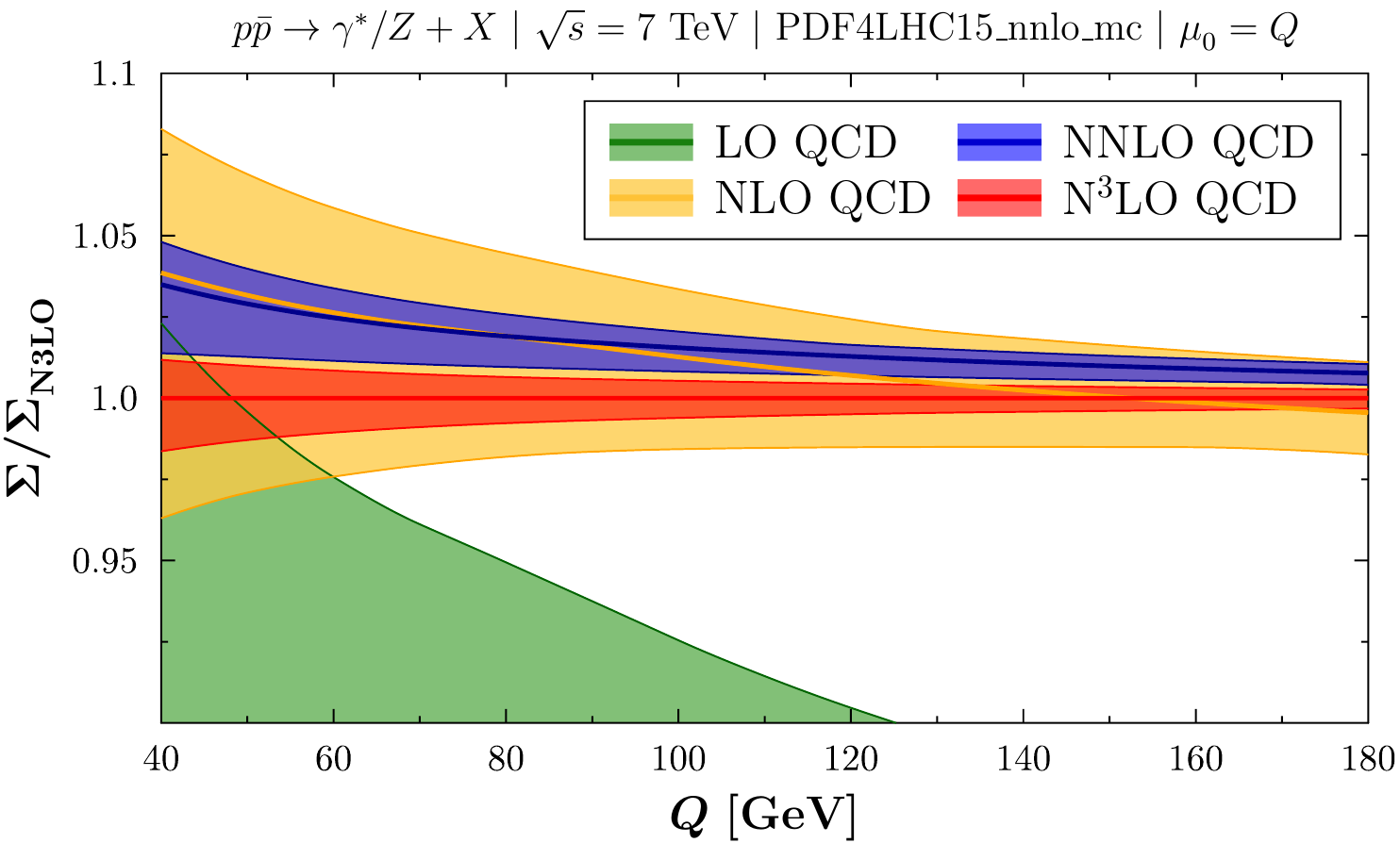} \\
\includegraphics[width=0.48 \textwidth]{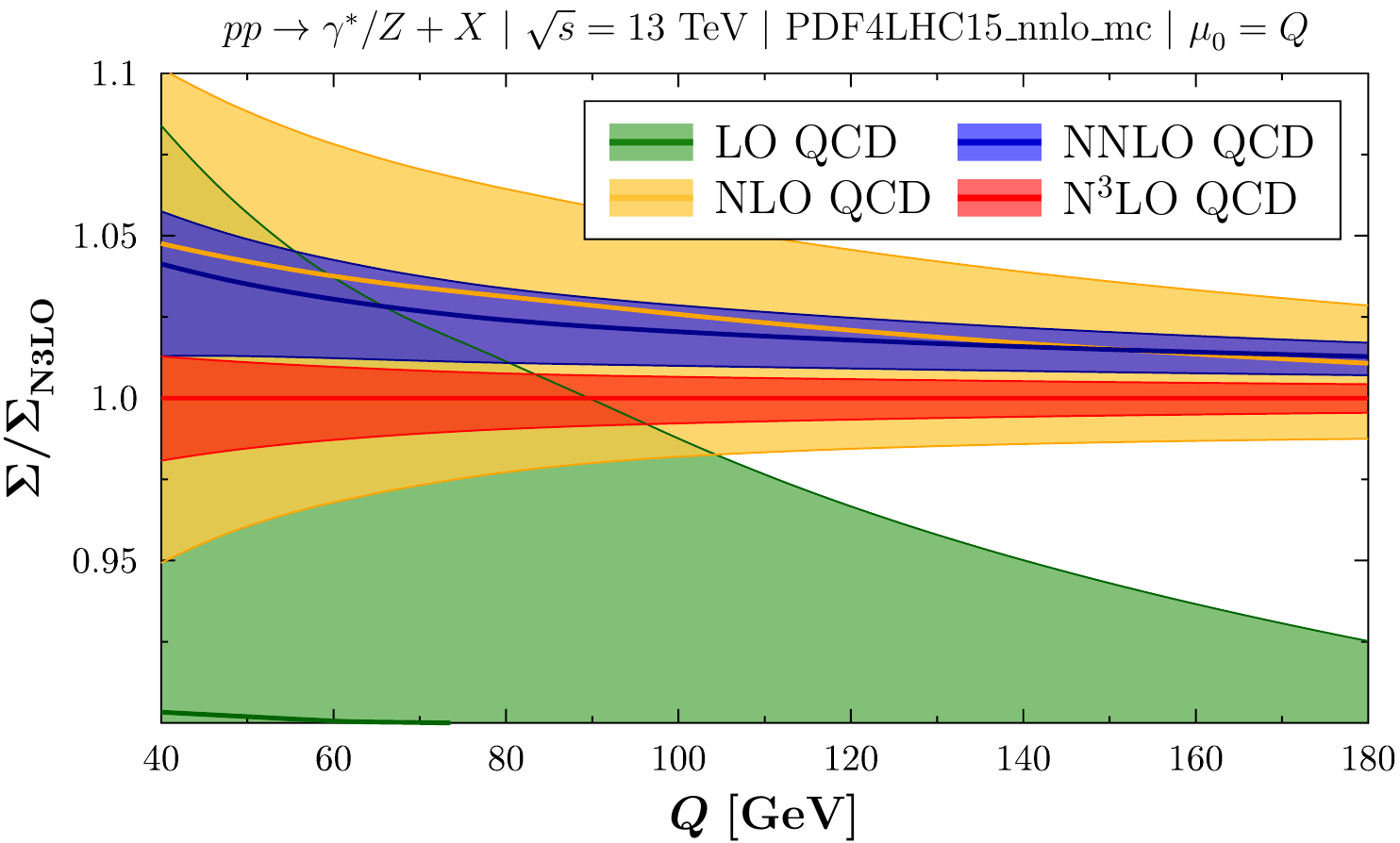} 
\includegraphics[width=0.48 \textwidth]{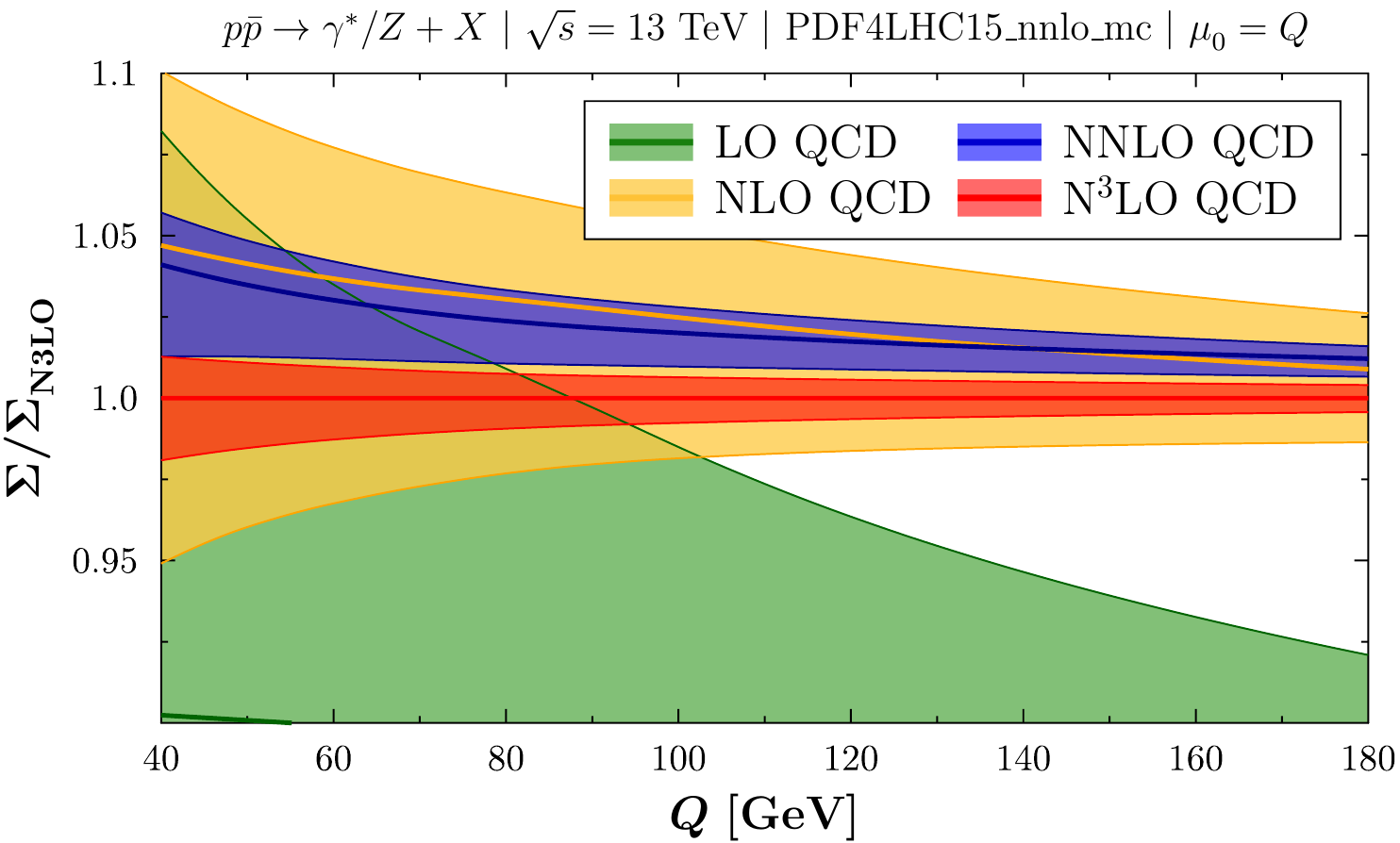} \\
\includegraphics[width=0.48 \textwidth]{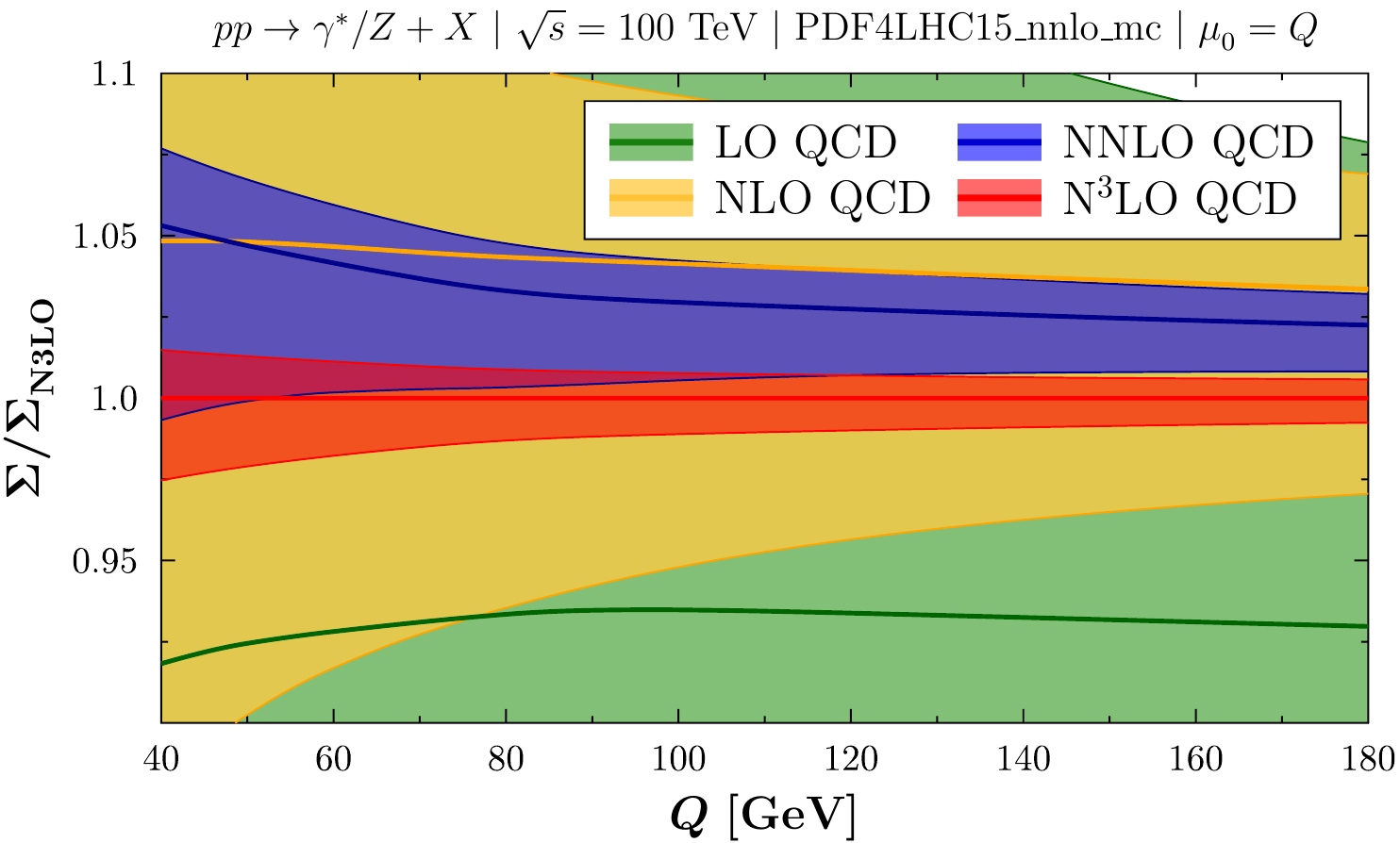} 
\includegraphics[width=0.48 \textwidth]{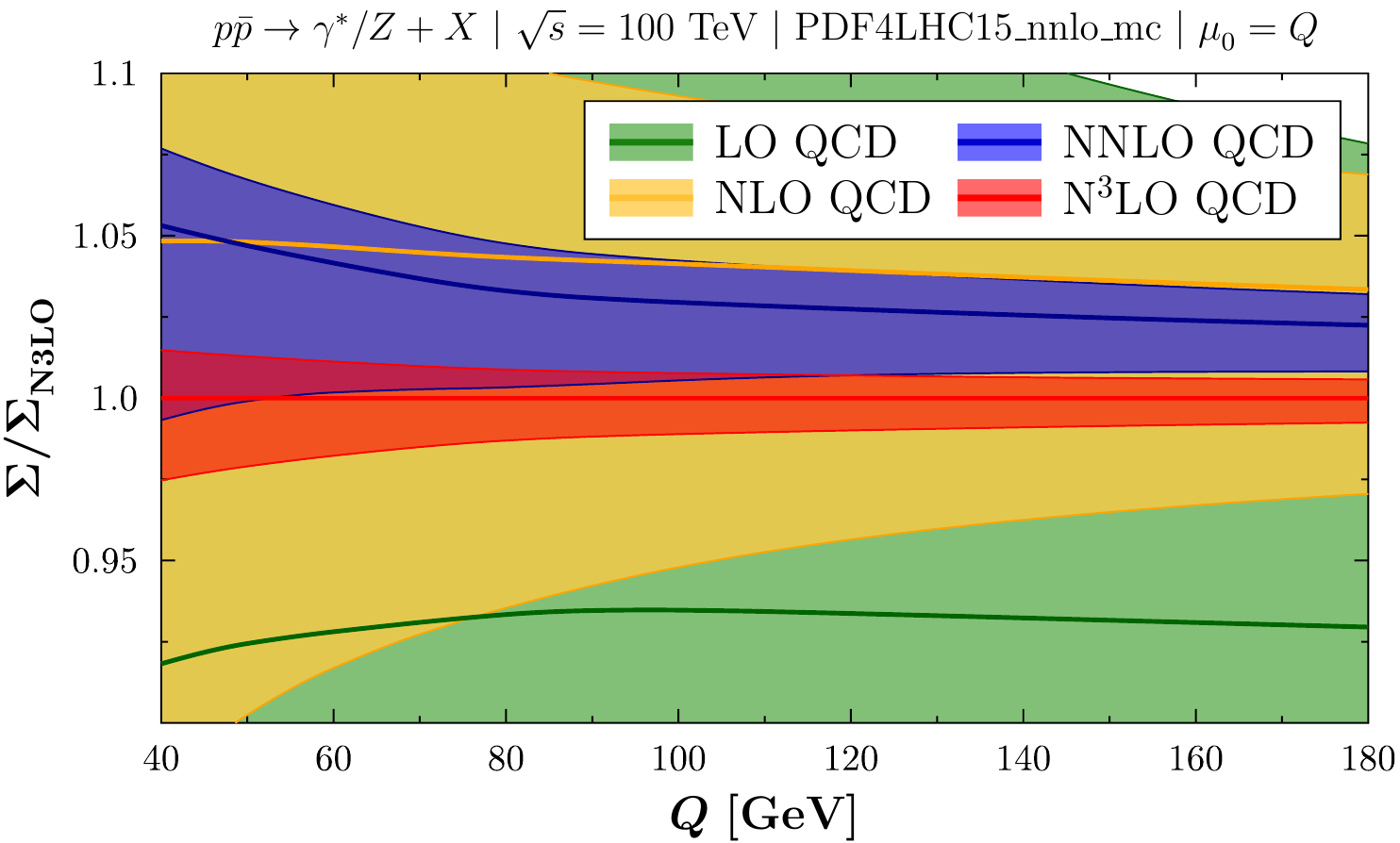} \\
\caption{\label{fig:Qvariation_DYn}Dependence of the neutral-current $\gamma/Z$ Drell-Yan cross sections on the invariant mass $Q$ of the lepton pair in the final state (in GeV) normalized to the N$^3$LO cross section at various center of mass energies for the $pp$ (left column) and $p\bar{p}$ collisions (right column). The bands indicate the 7-point scale uncertainty at each corresponding perturbative order.}
\end{center}
\end{figure}
\begin{figure}[!h]
\begin{center}
\includegraphics[width=0.48 \textwidth]{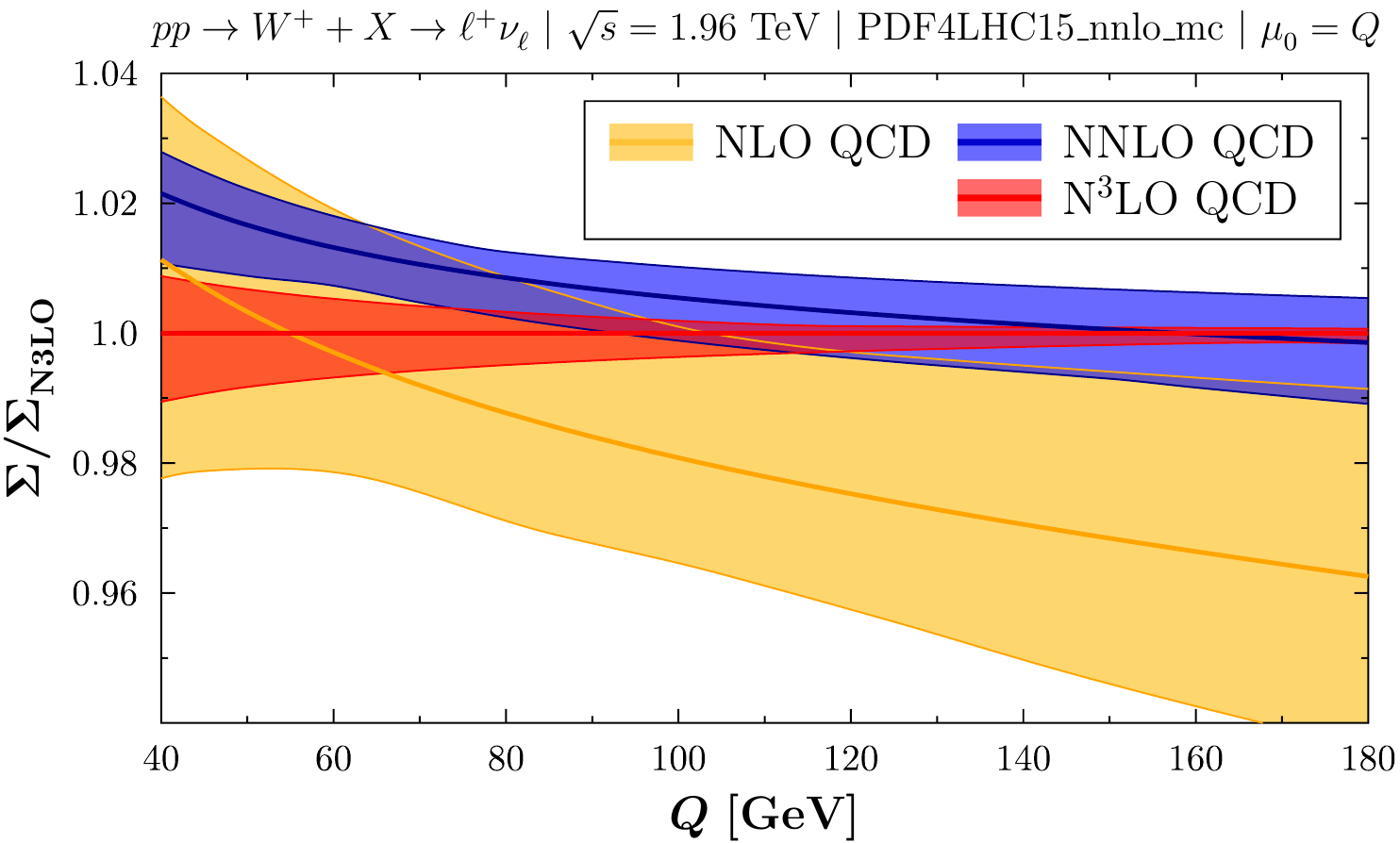} 
\includegraphics[width=0.48 \textwidth]{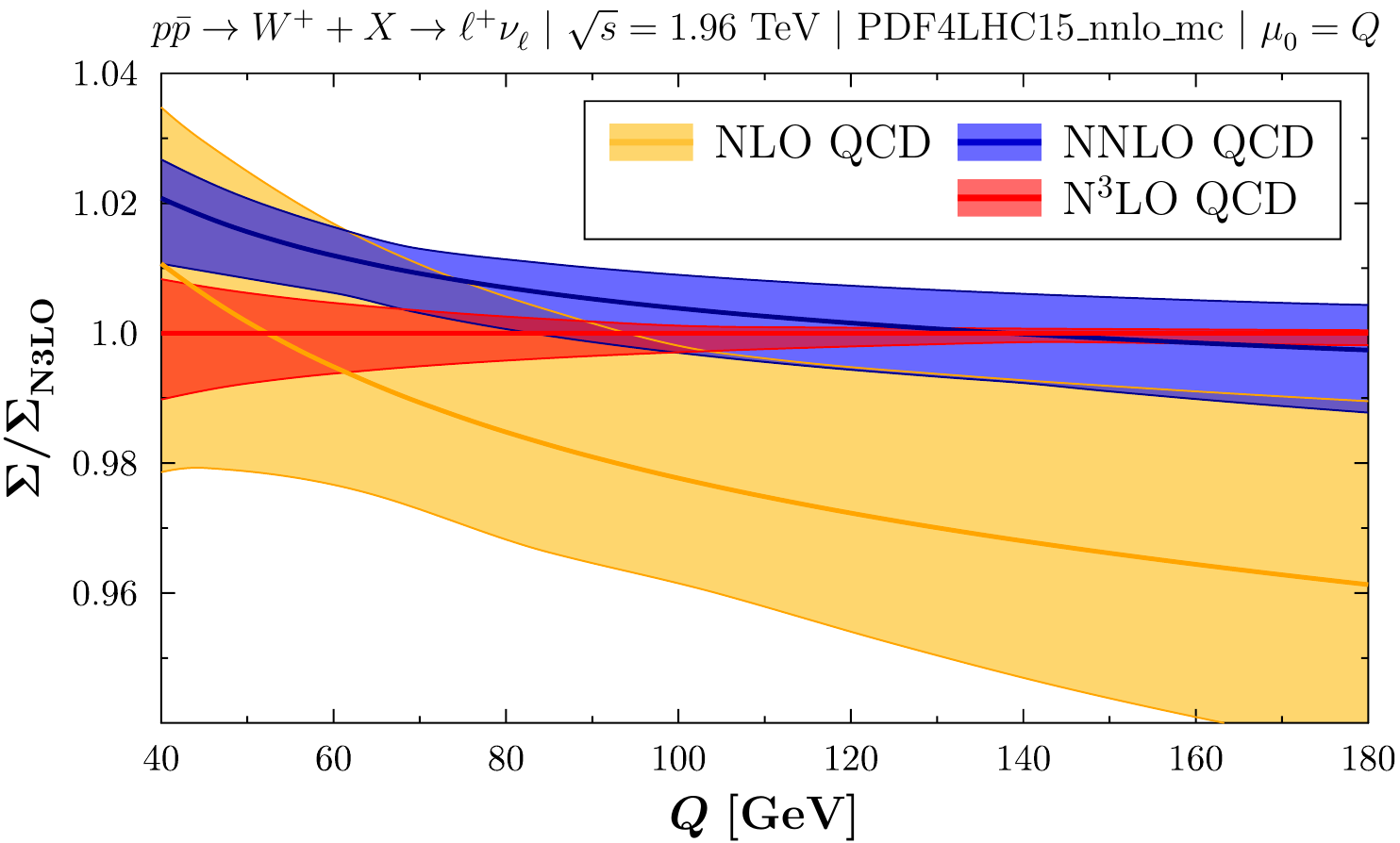} \\
\includegraphics[width=0.48 \textwidth]{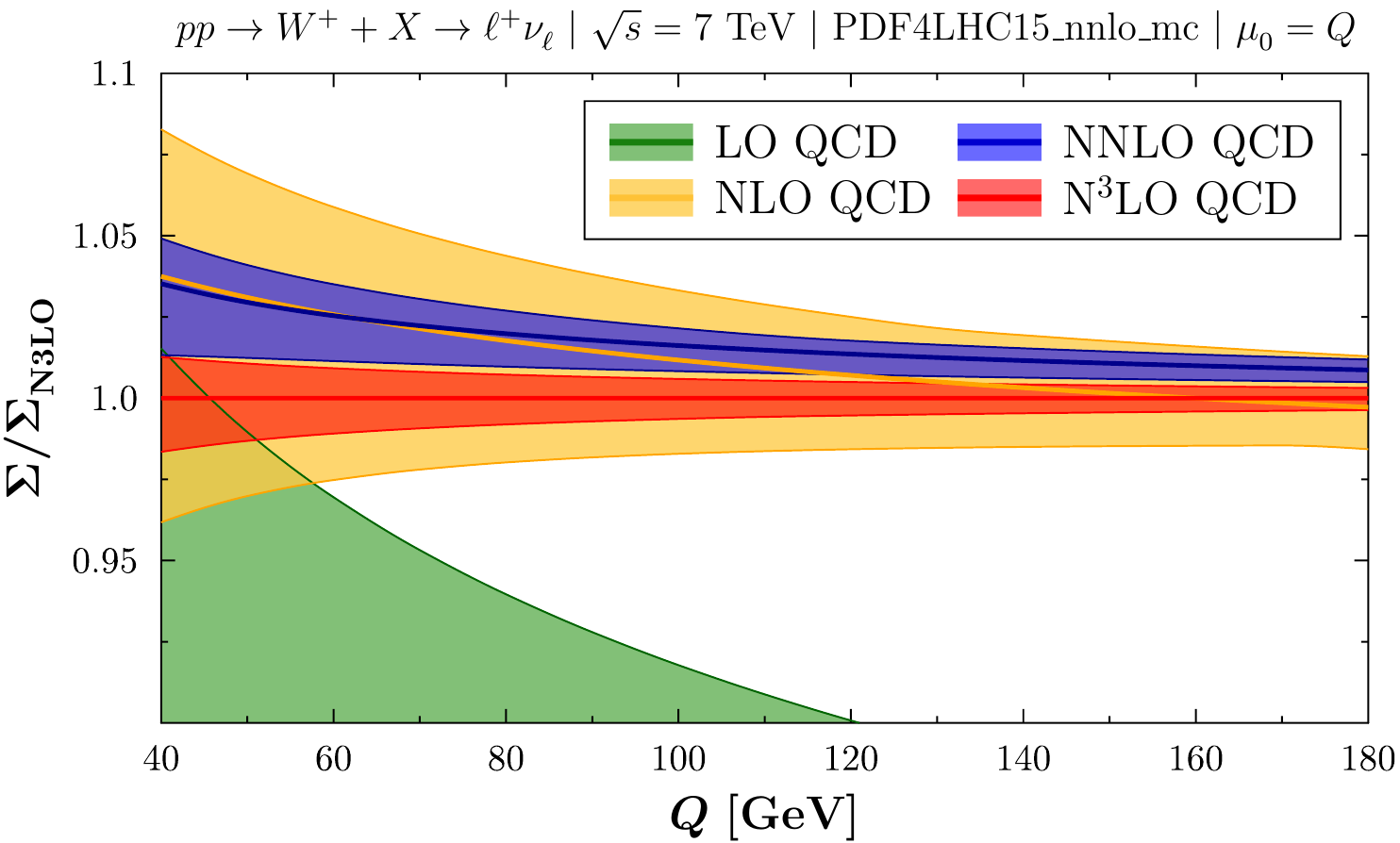} 
\includegraphics[width=0.48 \textwidth]{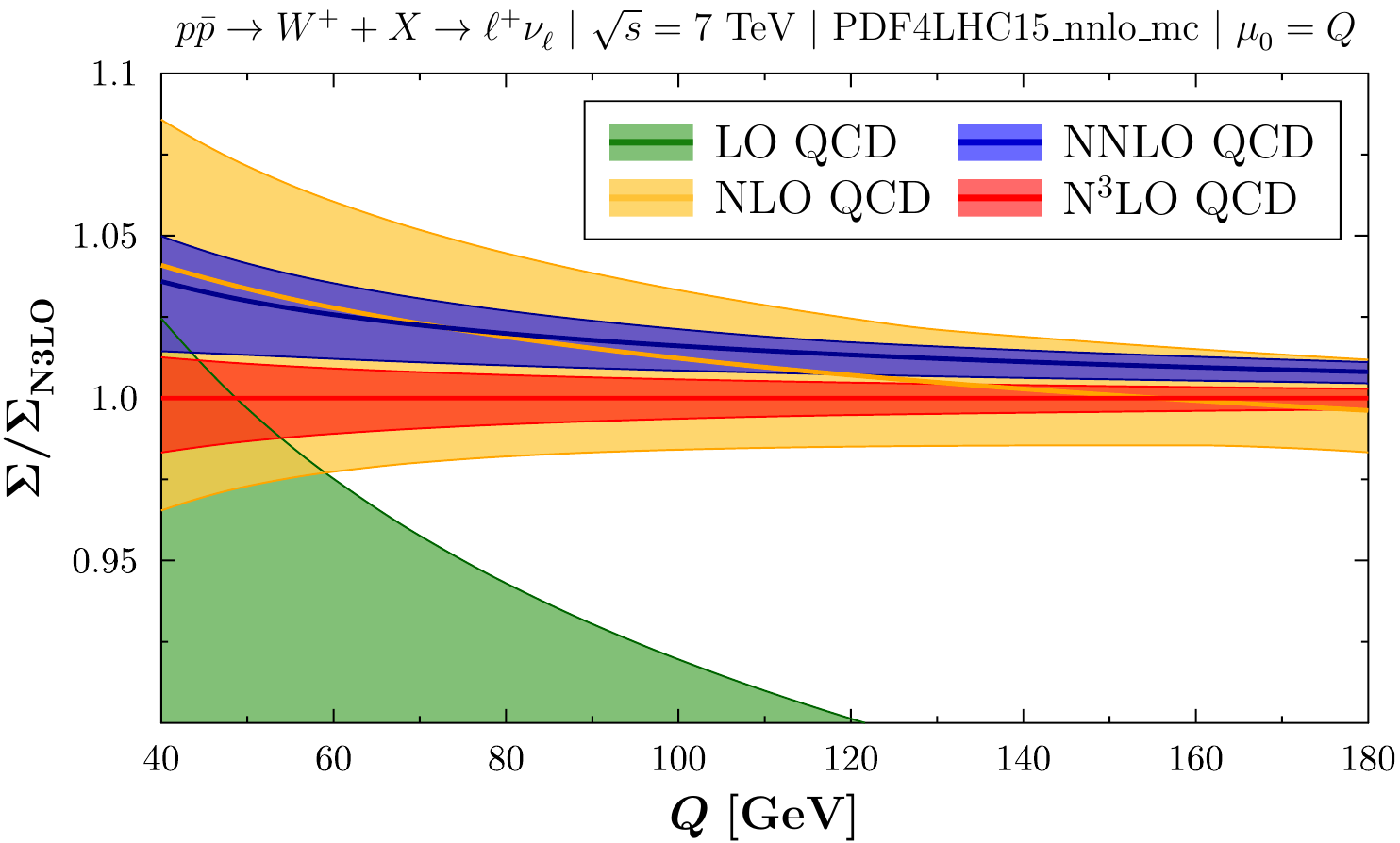} \\
\includegraphics[width=0.48 \textwidth]{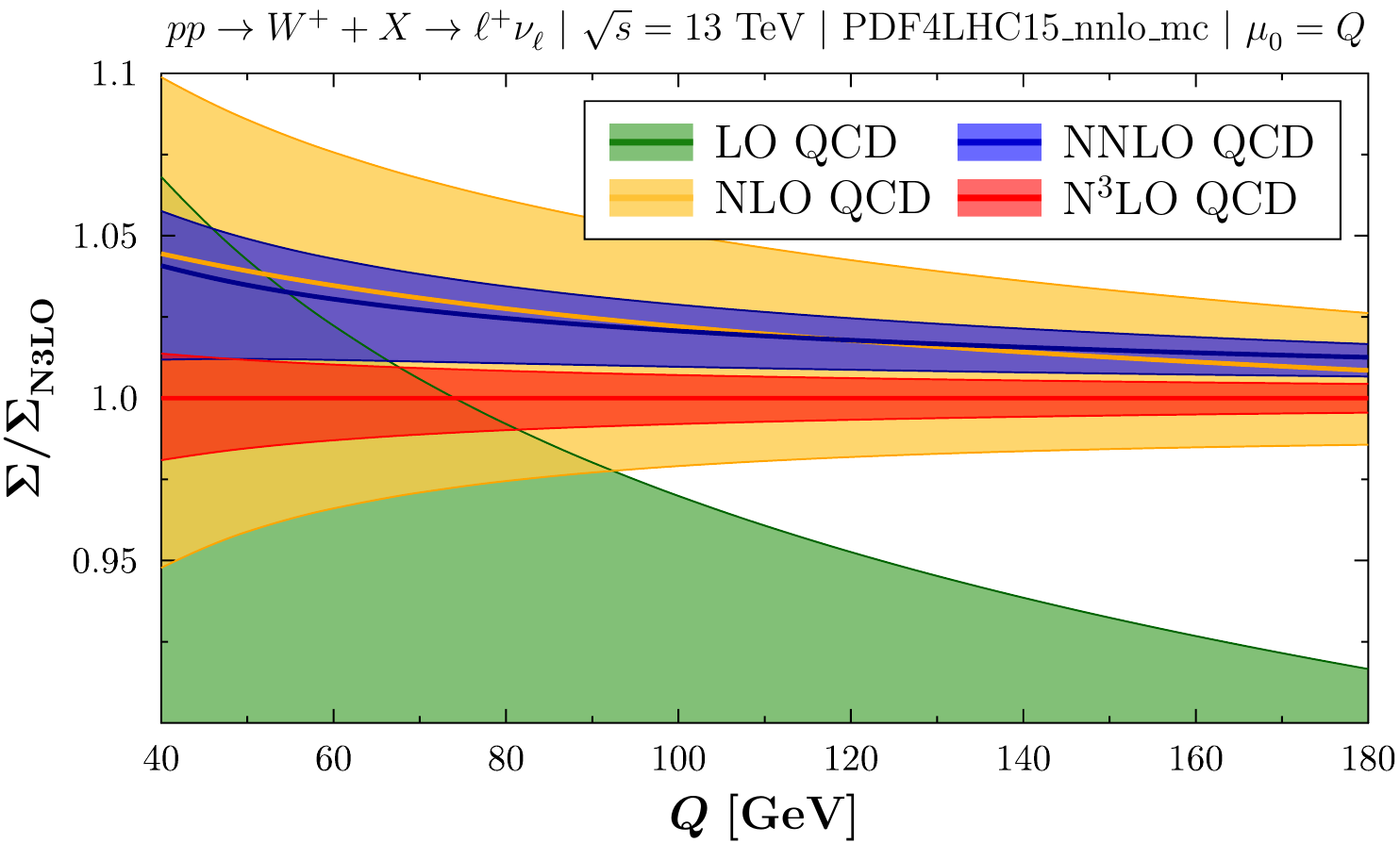} 
\includegraphics[width=0.48 \textwidth]{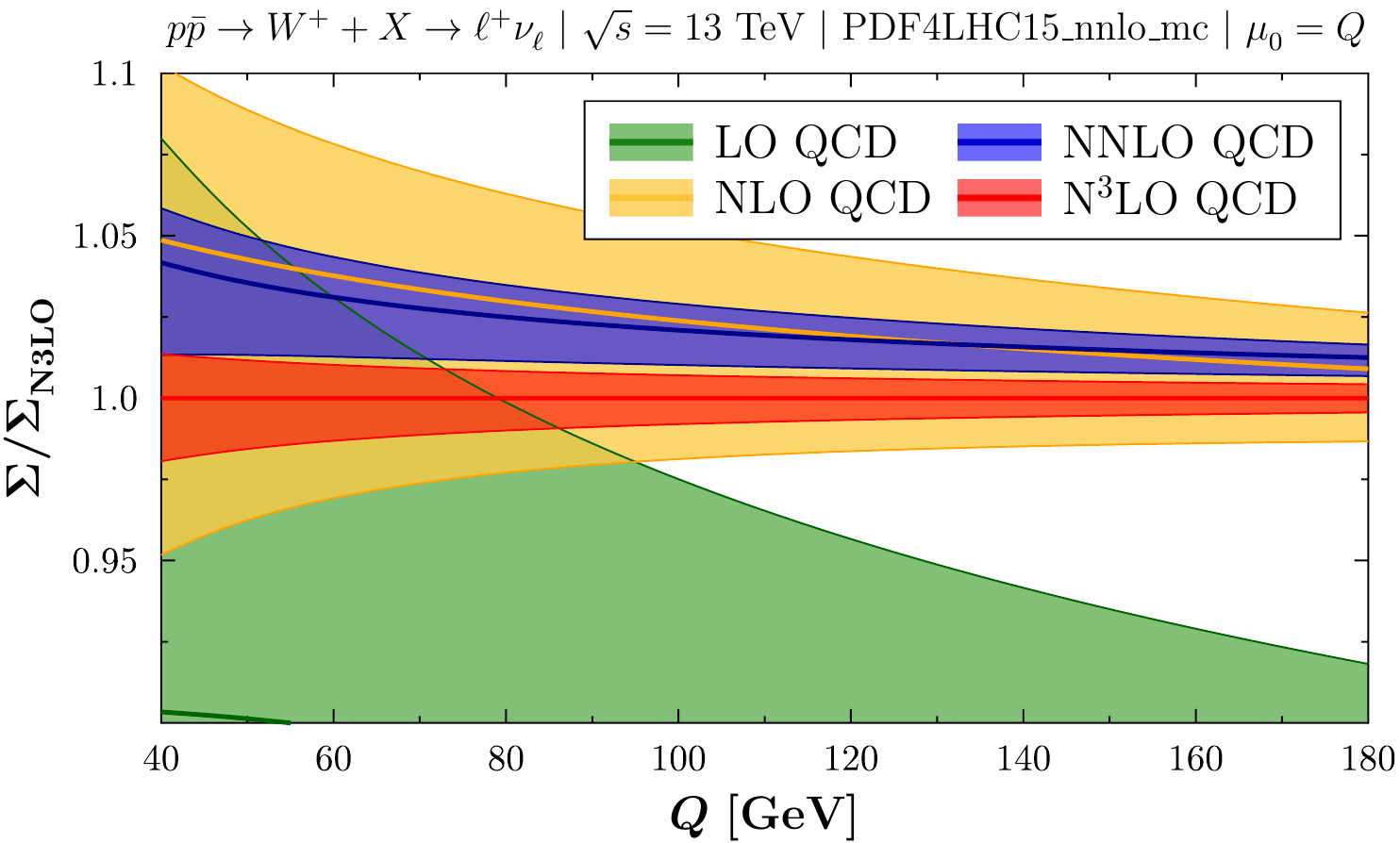} \\
\includegraphics[width=0.48 \textwidth]{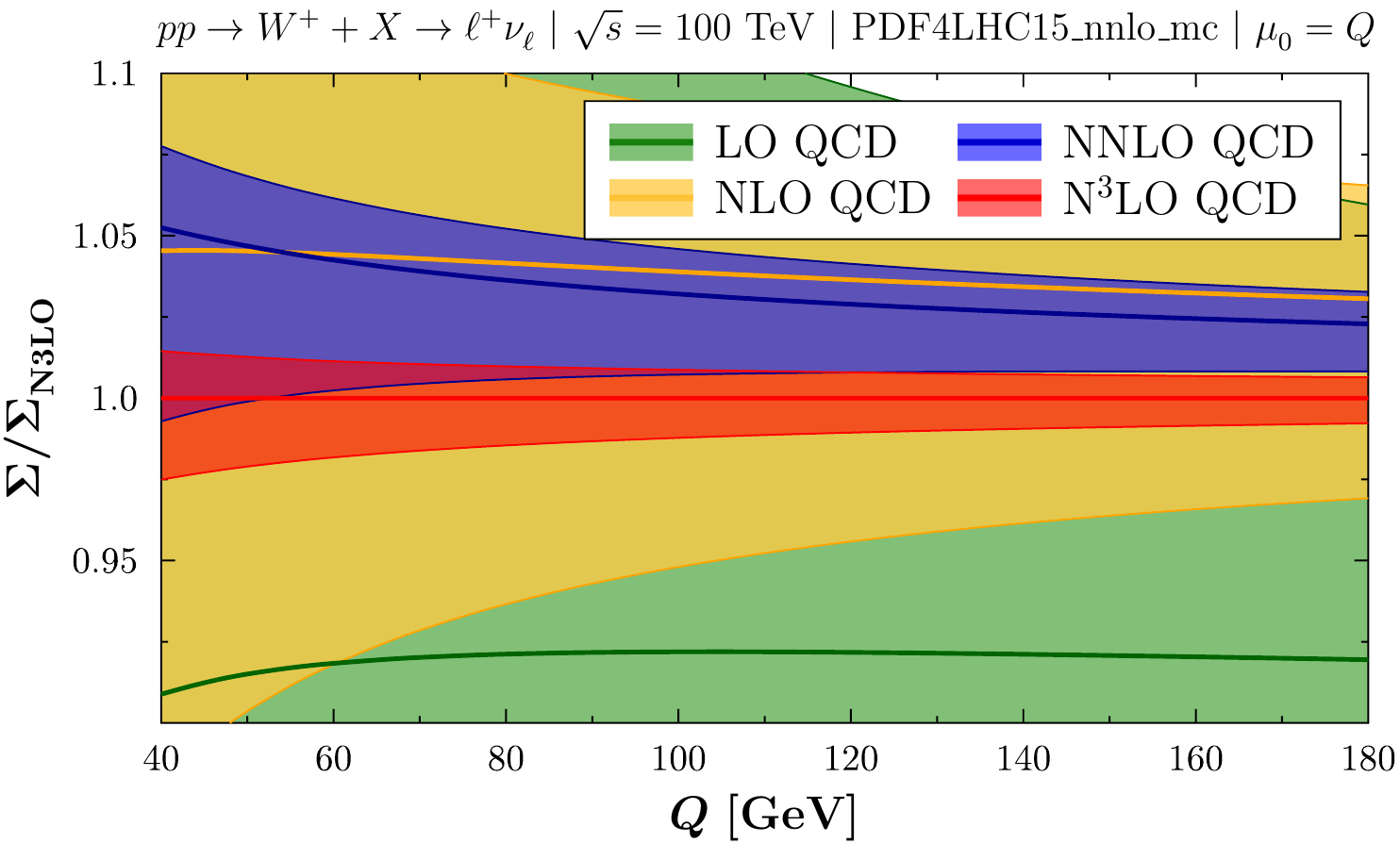} 
\includegraphics[width=0.48 \textwidth]{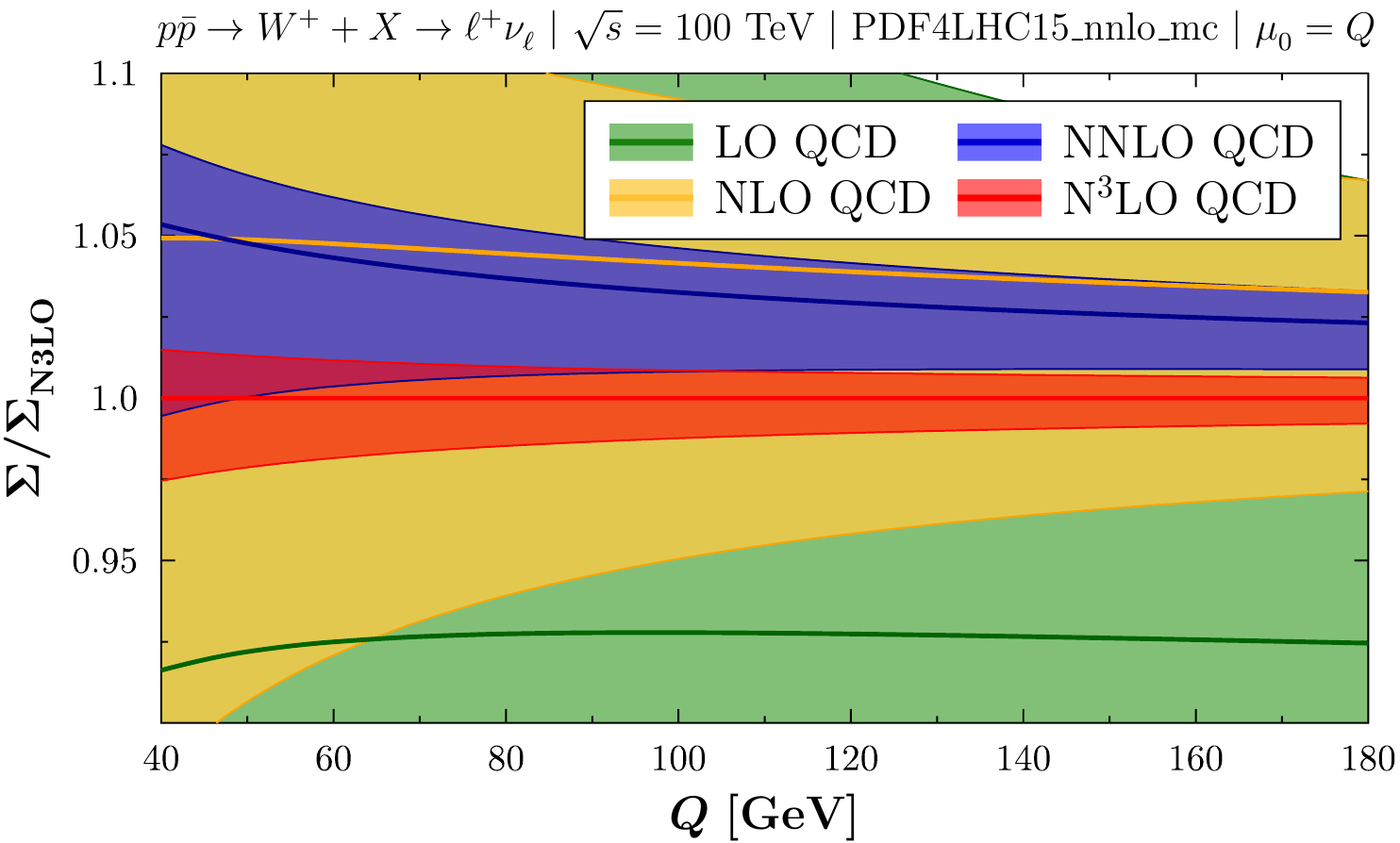} \\
\caption{\label{fig:Qvariation_DYp} Same as figure \ref{fig:Qvariation_DYn}, but for $W^+$ production.}
\end{center}
\end{figure}
\begin{figure}[!h]
\begin{center}
\includegraphics[width=0.48 \textwidth]{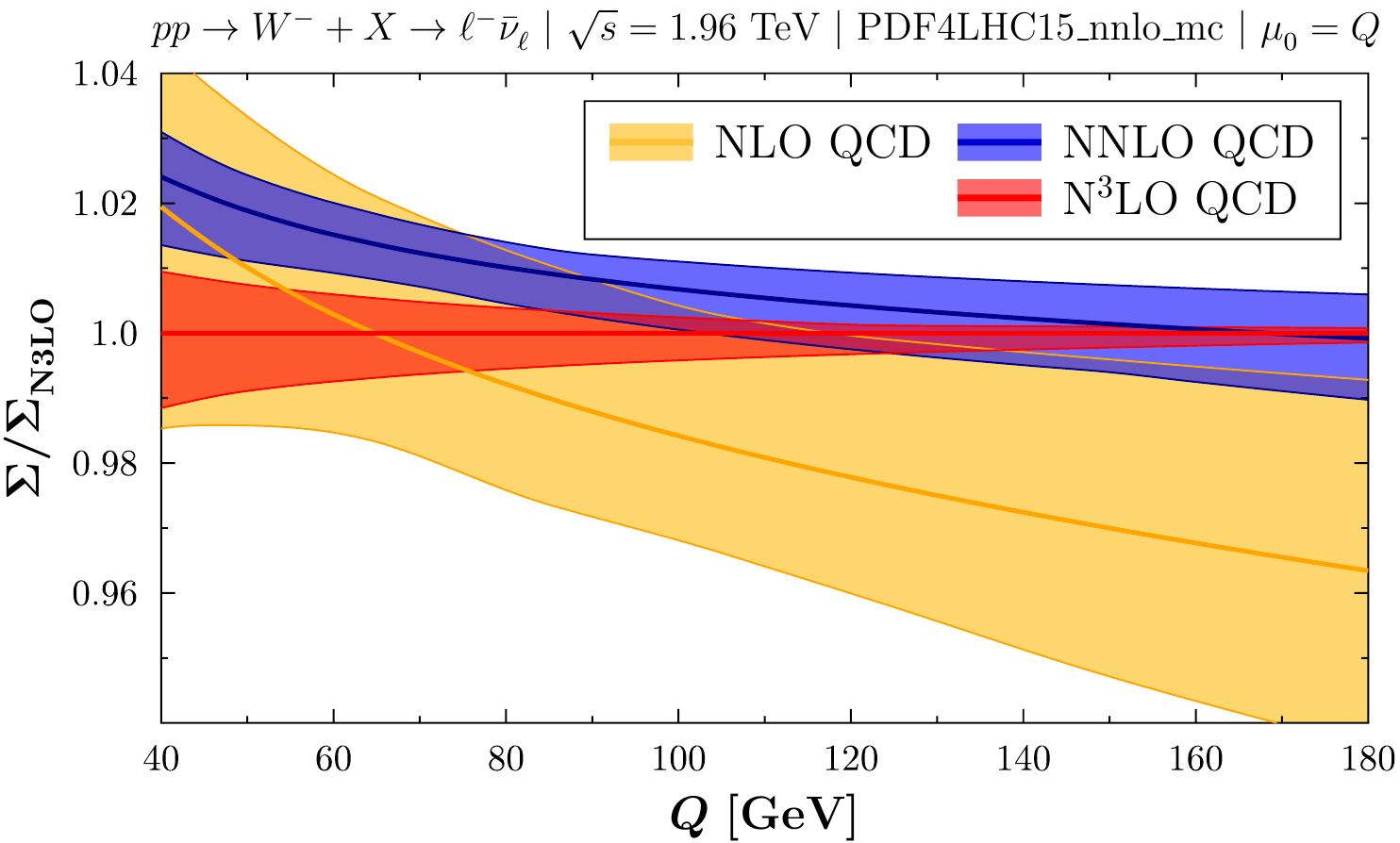} 
\includegraphics[width=0.48 \textwidth]{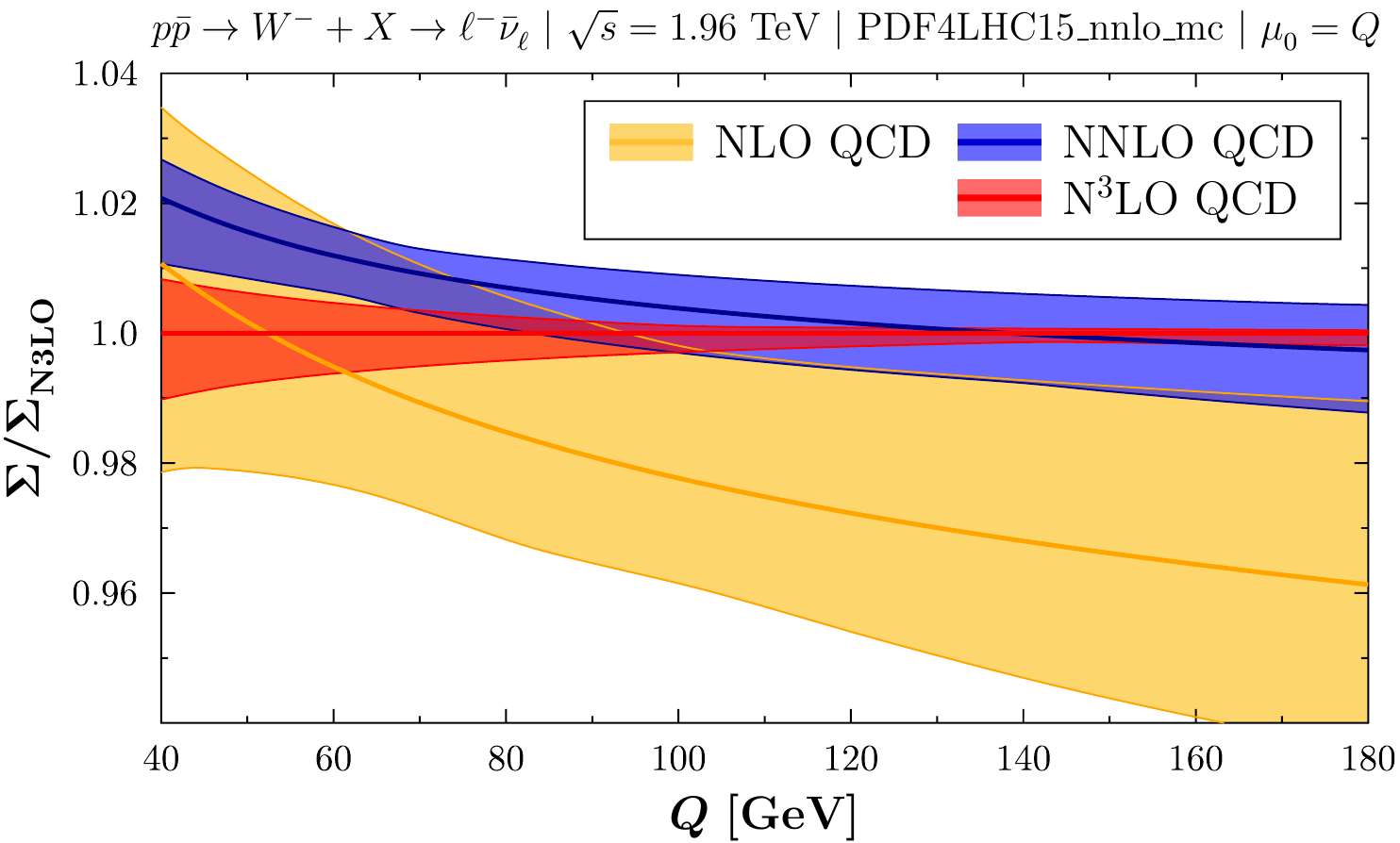} \\
\includegraphics[width=0.48 \textwidth]{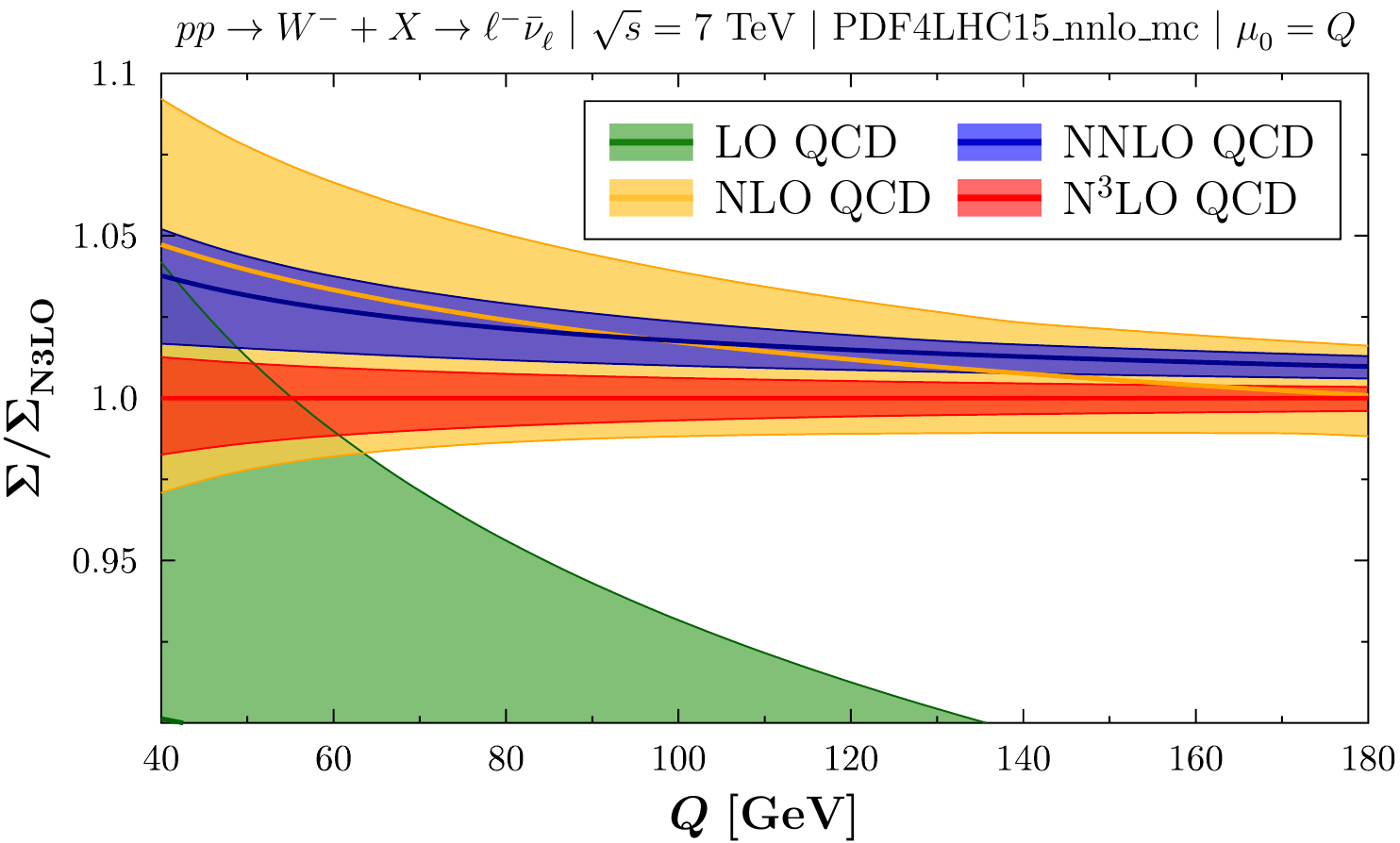} 
\includegraphics[width=0.48 \textwidth]{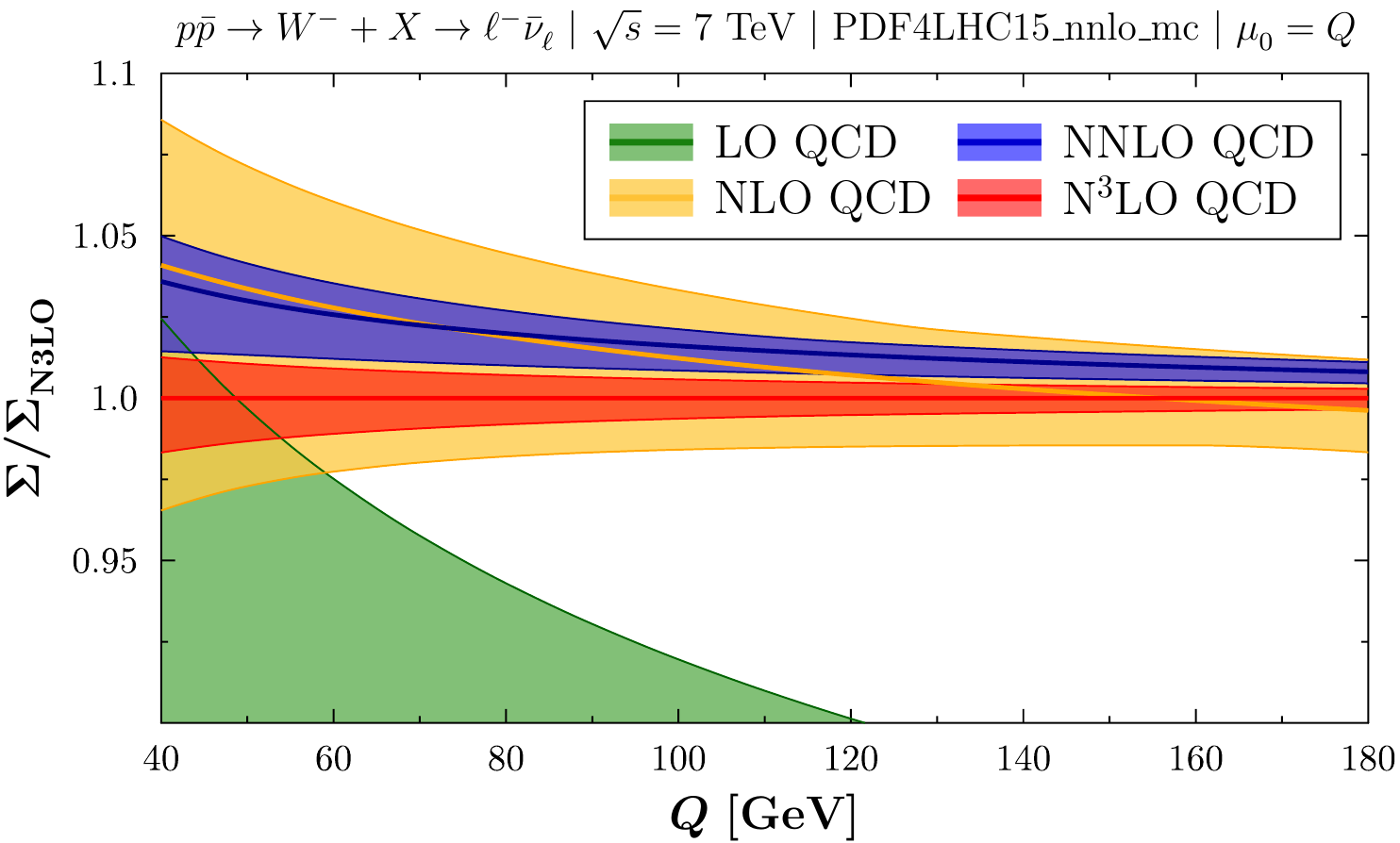} \\
\includegraphics[width=0.48 \textwidth]{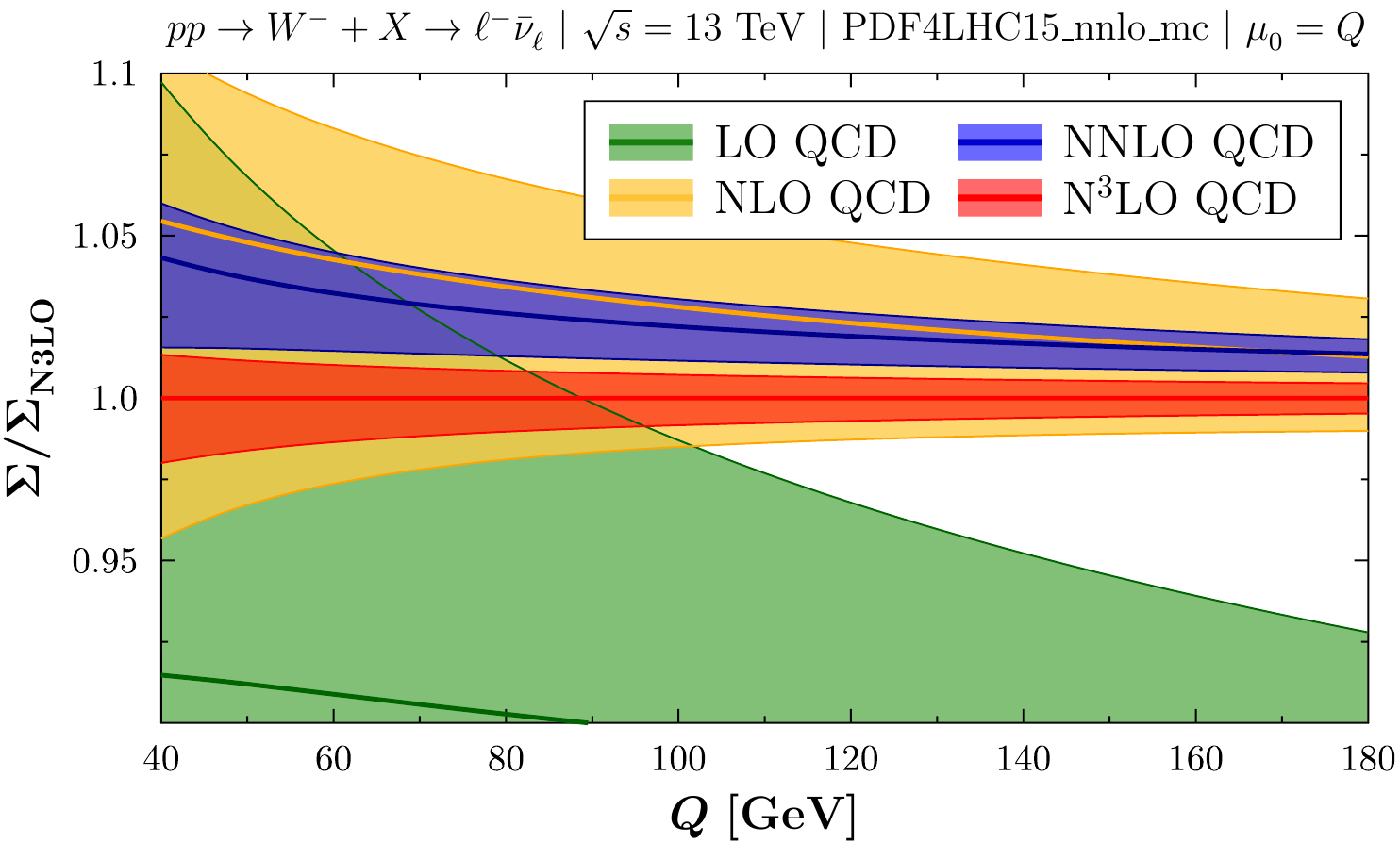} 
\includegraphics[width=0.48 \textwidth]{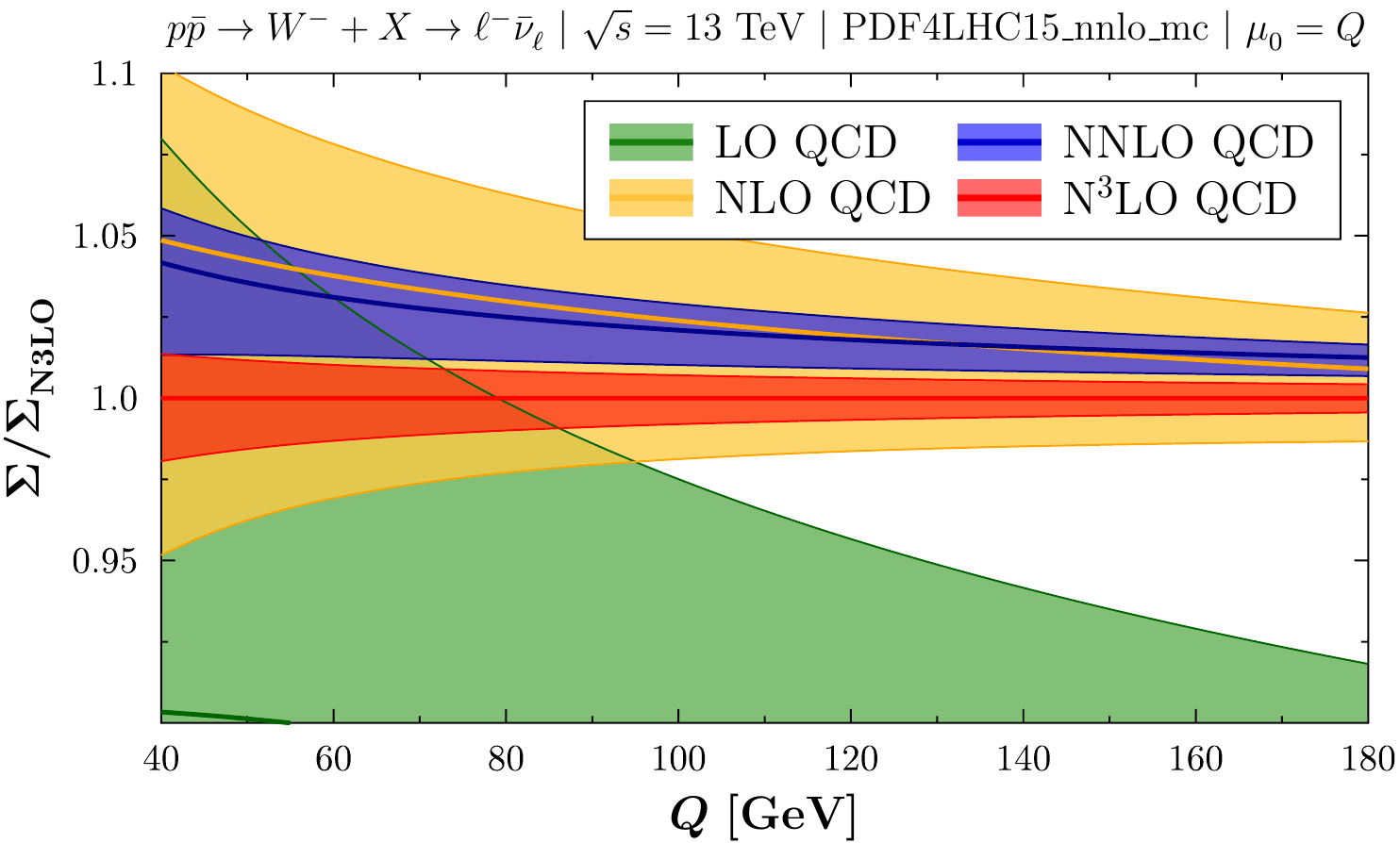} \\
\includegraphics[width=0.48 \textwidth]{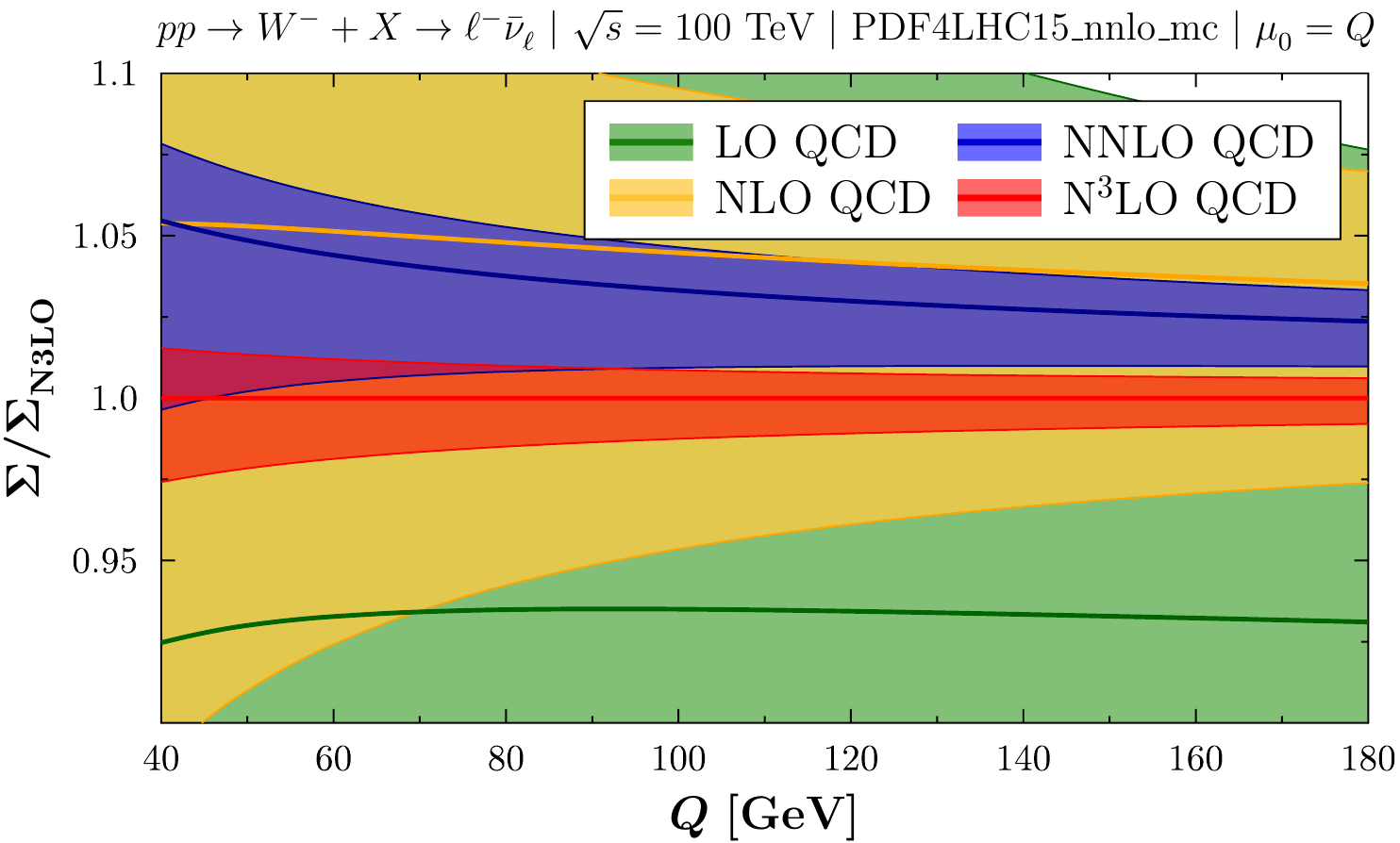} 
\includegraphics[width=0.48 \textwidth]{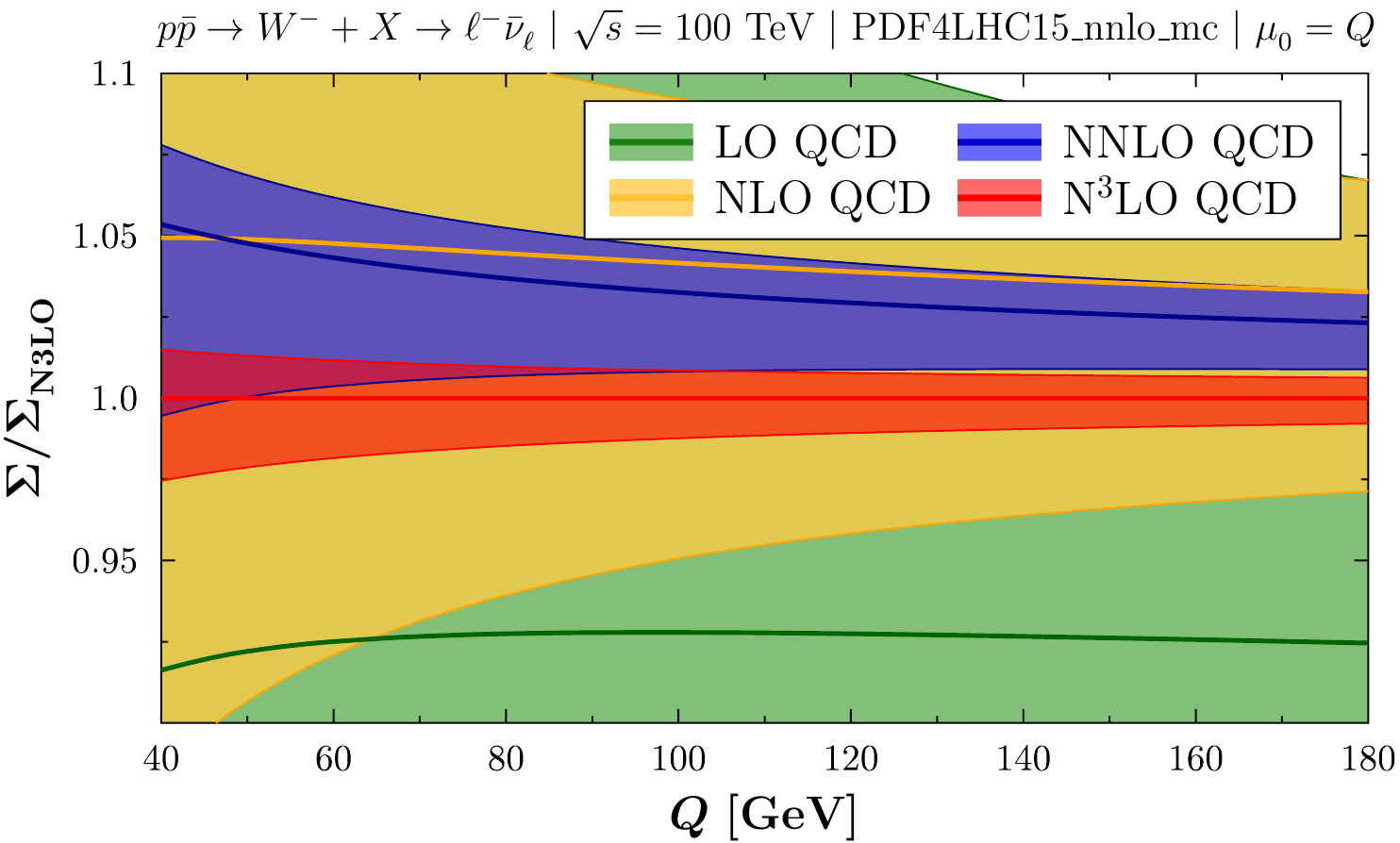} \\
\caption{\label{fig:Qvariation_DYm} Same as figure \ref{fig:Qvariation_DYn}, but for $W^-$ production.}
\end{center}
\end{figure}

We also present the scale dependence of Drell-Yan processes, i.e., the variation of the cross section when both the perturbative scales $\mu_F$ and $\mu_R$ are varied by a factor of two around the central scales $(\mu_F^0,\mu_R^0)$ while respecting the constraint
\beq
\frac{1}{2}\leq \frac{\mu_R}{\mu_F}\leq 2\,.
\eeq
Commonly, this is referred to as 7-point variation. The results are given in figs.~\ref{fig:Qvariation_DYn}, \ref{fig:Qvariation_DYp} and \ref{fig:Qvariation_DYm} where we show the invariant-mass distribution of the neutral- and charged-current Drell-Yan cross sections normalized to the N$^3$LO result as a function of the invariant mass $Q$ of the produced lepton pair with $40\textrm{ GeV}\le Q\le 180 \textrm{ GeV}$. 
For a 13 TeV $pp$ collider, we reproduce the results of refs.~\cite{Duhr:2020sdp,Duhr:2020seh,Duhr:2021vwj}. Just like in those references, we observe that for the range of invariant masses considered, the bands from the scale variation do not overlap between NNLO and N$^3$LO. 
We plot the scale dependence also for $\sqrt{s}=$1.96, 7, and 100 TeV, both for $pp$ and $p\bar{p}$ collisions. Even at 100 TeV, where, as expected, the scale variation bands are significantly wider than at 13 TeV, there is no overlap between the NNLO and N$^3$LO predictions at large $Q$ values.
At the Tevatron energy of 1.96 TeV, we note that the NNLO and N$^3$LO bands cross within the plotted range, and thus the non-overlapping bands appear at small values of $Q$.   
We do not observe any significantly different behavior between the $pp$ and $p\bar{p}$ colliders.

\begin{figure}[!h]
\includegraphics[width=\textwidth]{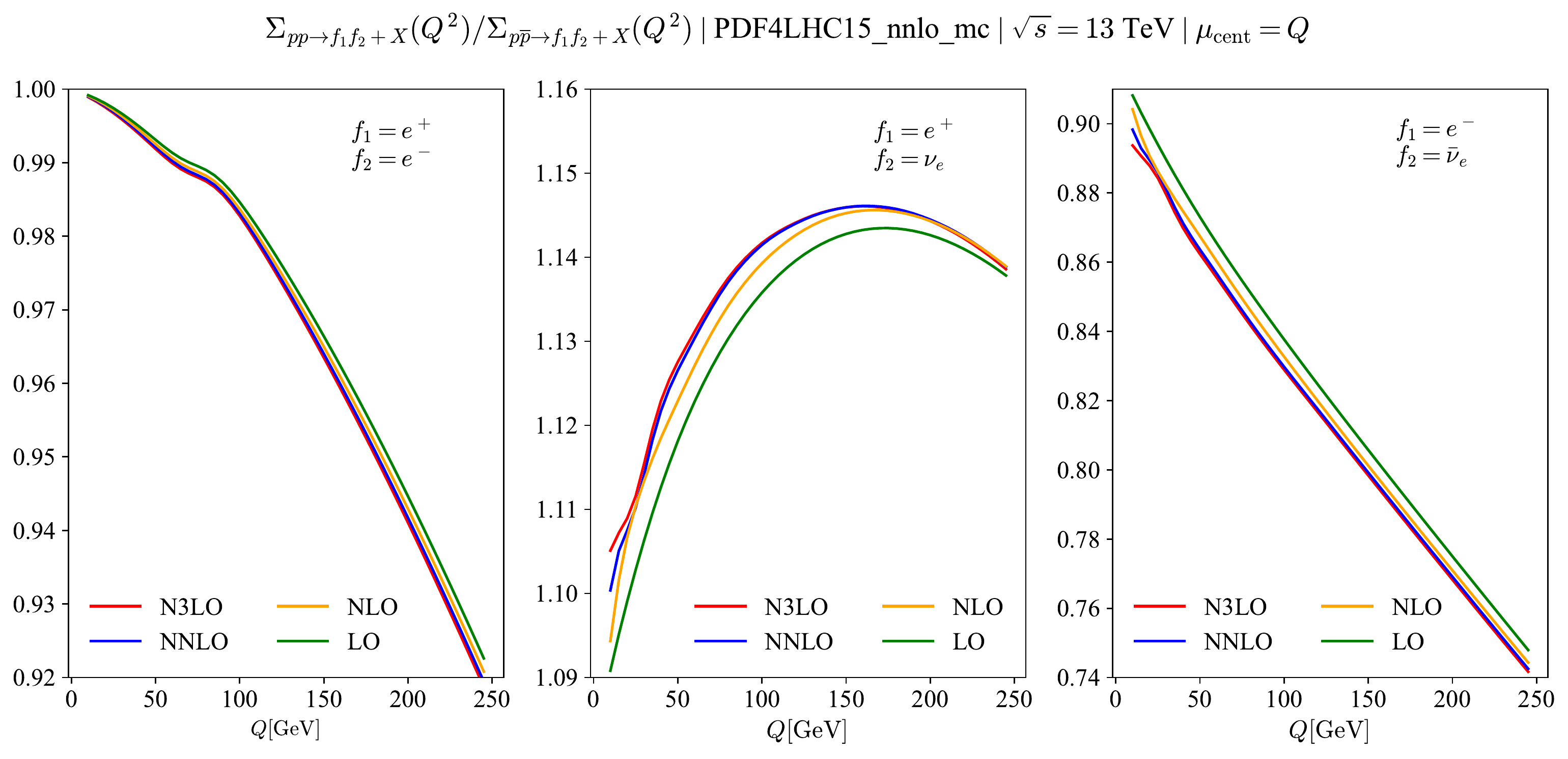}
\caption{\label{fig:ratioppppbar13} Ratio  of the proton-proton to the proton-anti-proton $\Sigma_{pp}^{\rm N^kLO}(Q^2)/\Sigma_{p\bar{p}}^{\rm N^kLO}(Q^2)$ production cross sections for Drell-Yan processes.}
\end{figure}

Figure~\ref{fig:ratioppppbar13} shows the ratio of the production cross section for a $pp$ and  $p\bar{p}$ collider for $\sqrt{s}=13$ TeV. The neutral- and charged-current processes show a distinct dependence on $Q$, but the higher-order corrections do not change this behavior qualitatively. Only for the production of an $e^+\nu_e$ pair, the $pp$ collider leads to higher values of the cross section than for the $p\bar{p}$ collisions, and this process shows non-monotonic dependence on the invariant mass $Q$.

\begin{figure}[!h]
\begin{center}
\includegraphics[width=0.48 \textwidth]{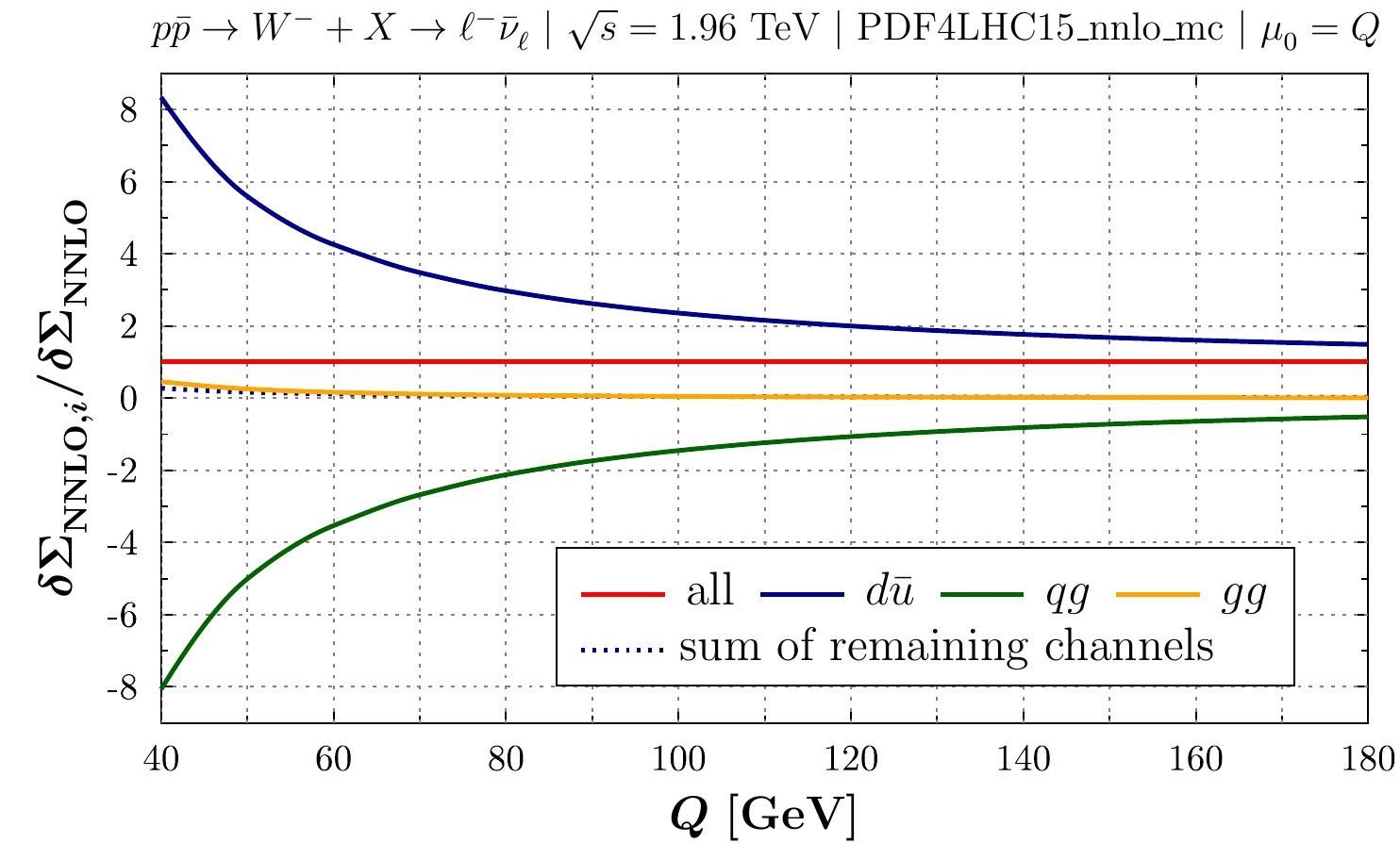} 
\includegraphics[width=0.48 \textwidth]{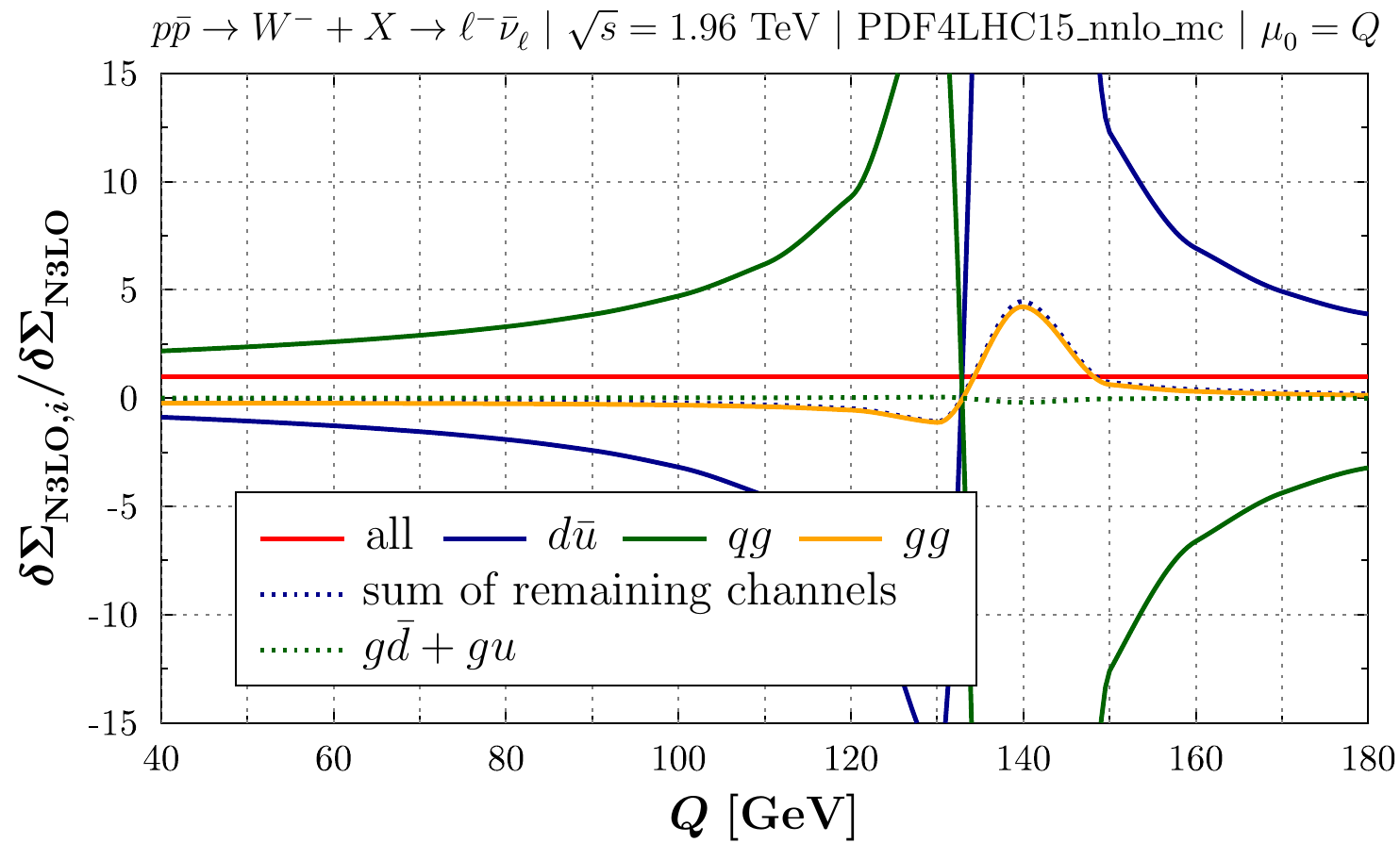} \\
\includegraphics[width=0.48 \textwidth]{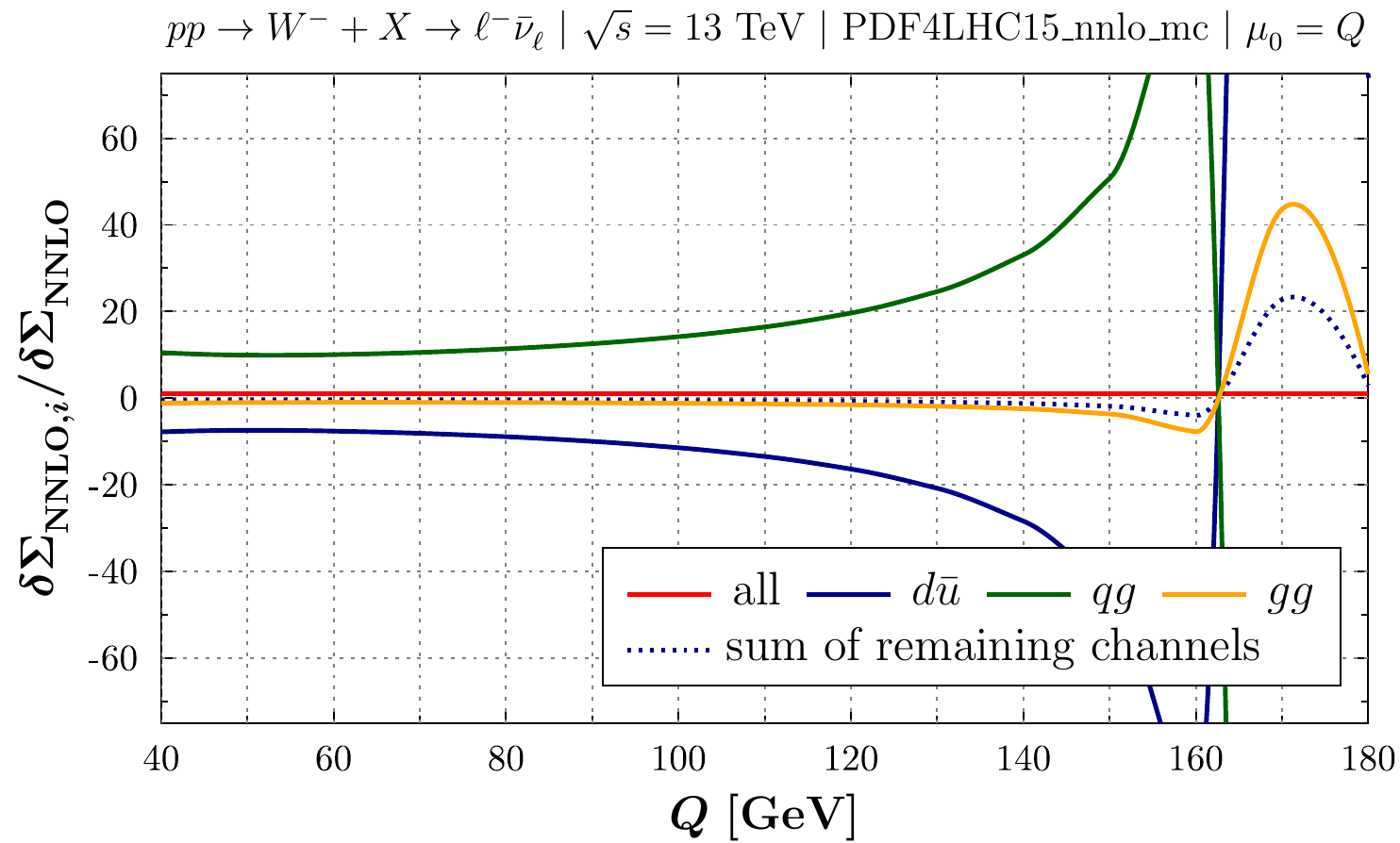} 
\includegraphics[width=0.48 \textwidth]{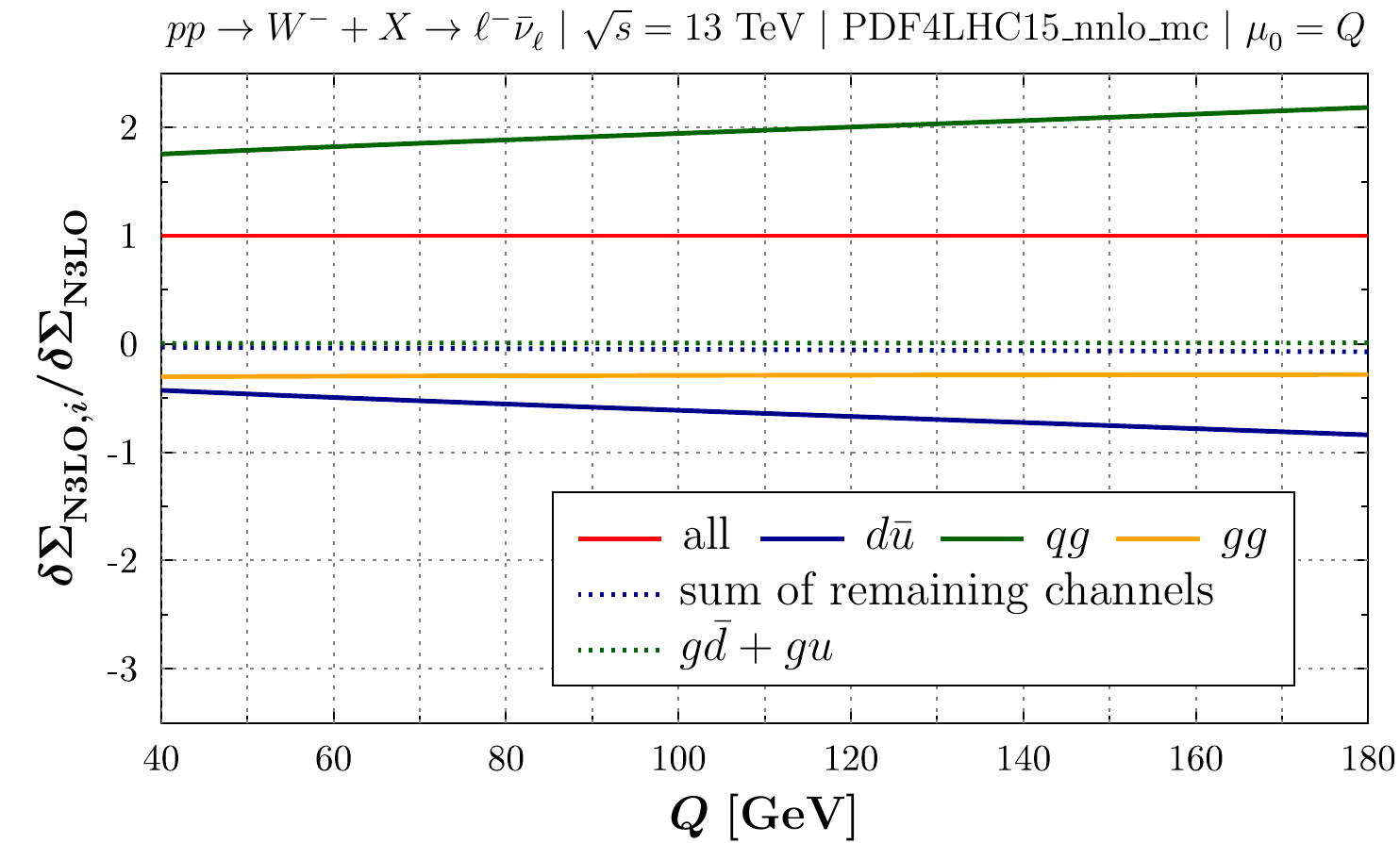} \\
\includegraphics[width=0.48 \textwidth]{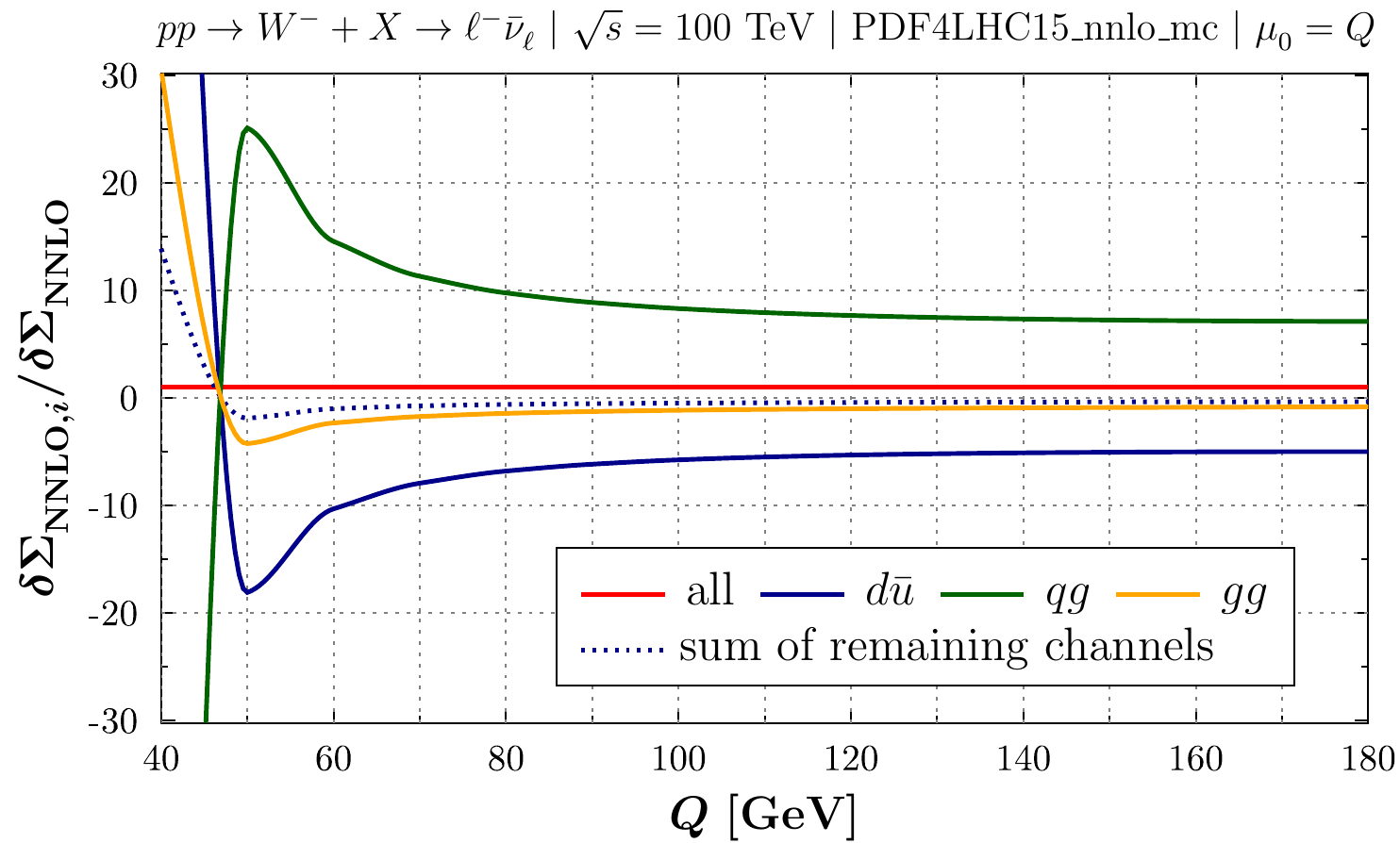} 
\includegraphics[width=0.48 \textwidth]{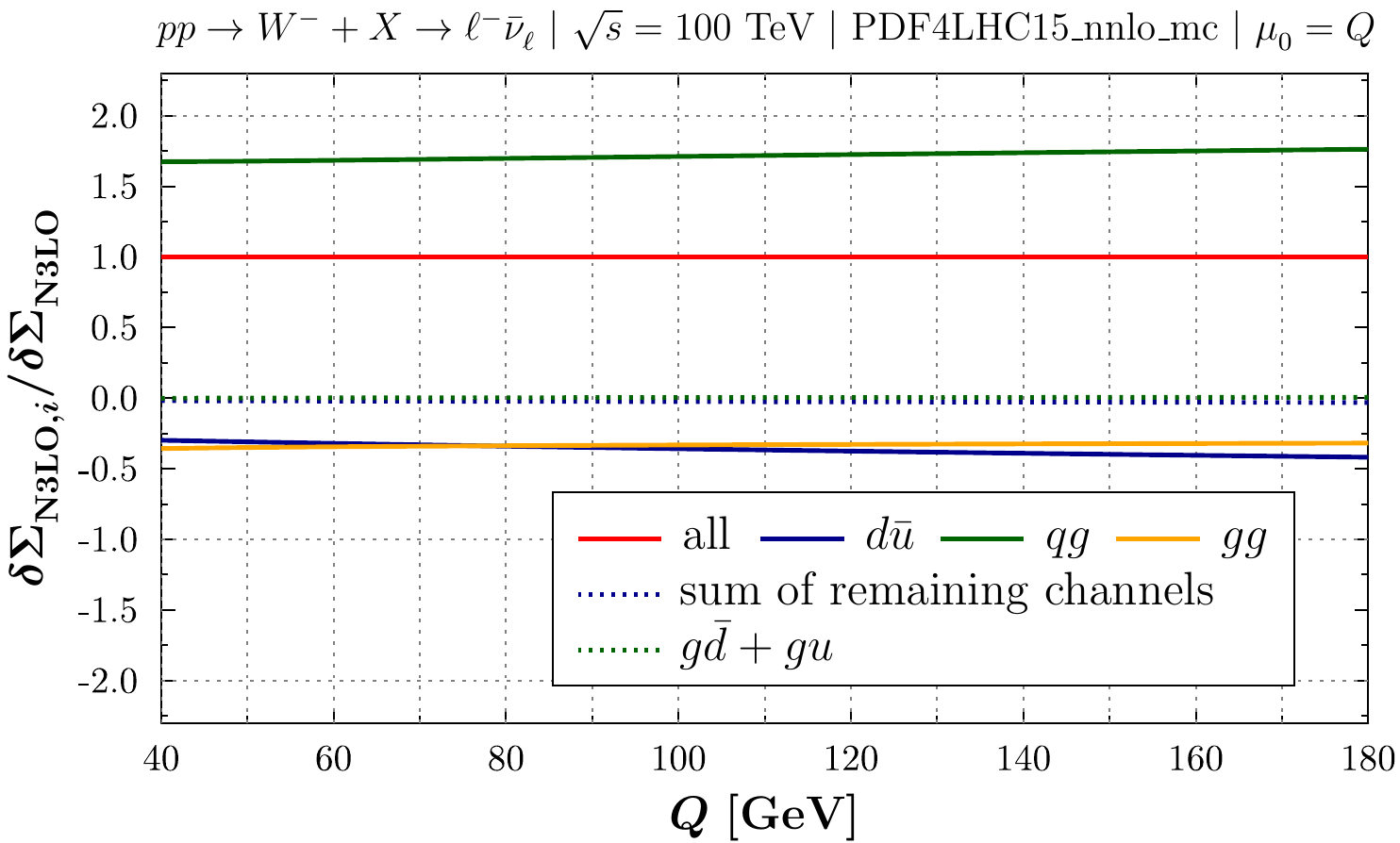} \\
\includegraphics[width=0.48 \textwidth]{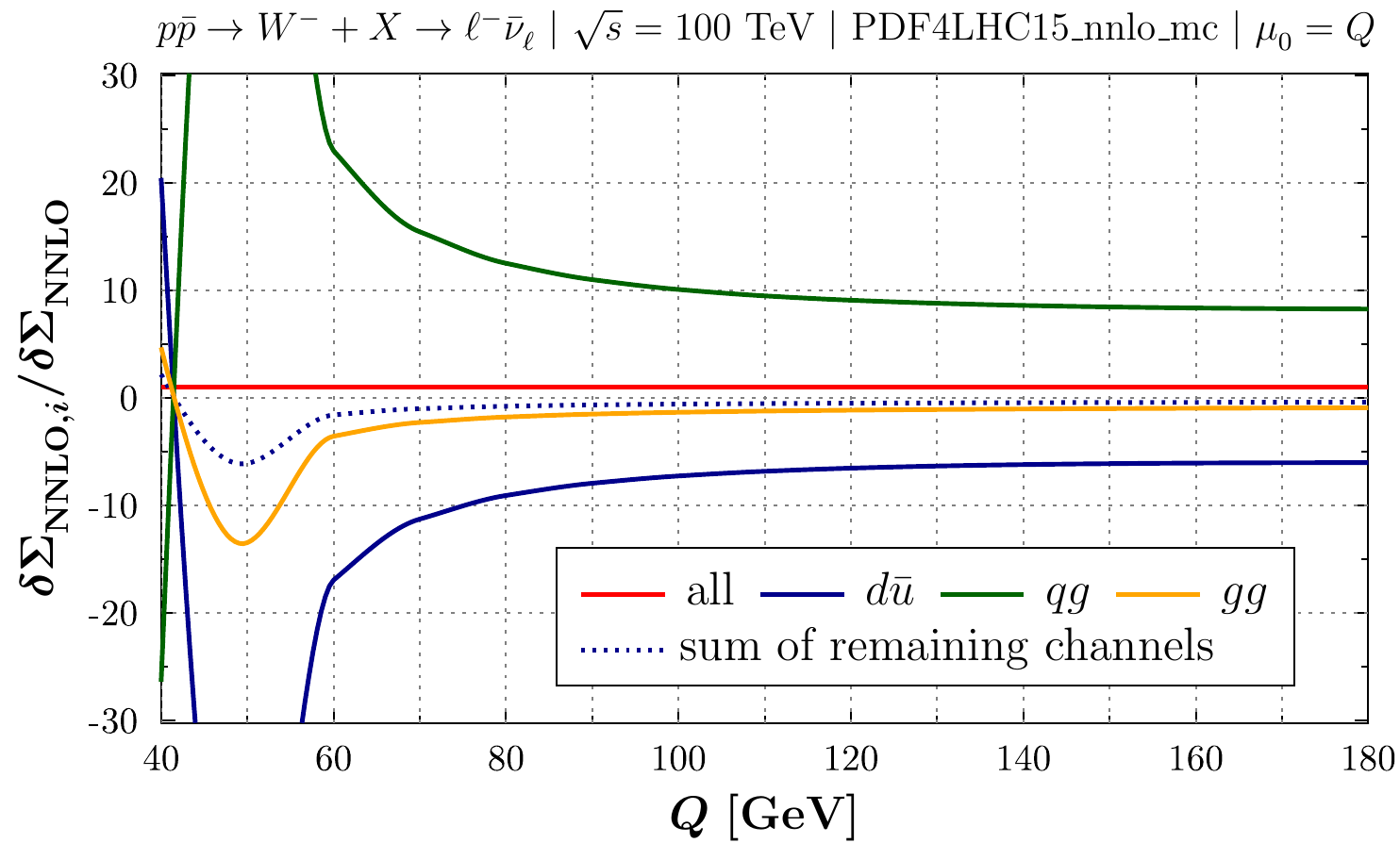} 
\includegraphics[width=0.48 \textwidth]{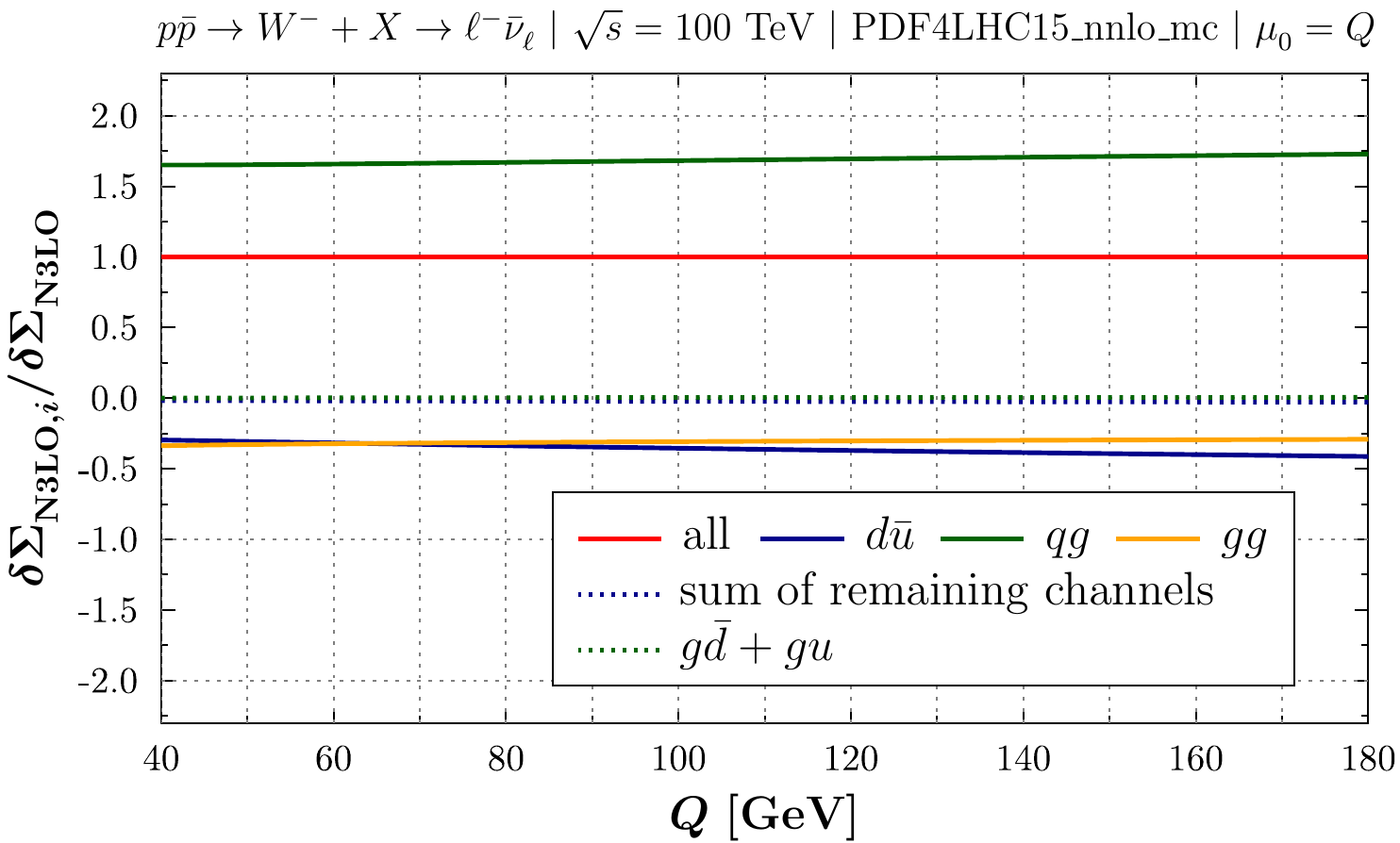} 

\caption{\label{fig:W-_channels} Individual channels contributing to the invariant mass distribution for the production of an $e^-\overline{\nu}_e$  pair at different colliders, normalized to the total correction at NNLO (left panels) and N$^3$LO (right panels).}
\end{center}
\end{figure}

\begin{figure}[!h]
\begin{center}
\includegraphics[width=0.48 \textwidth]{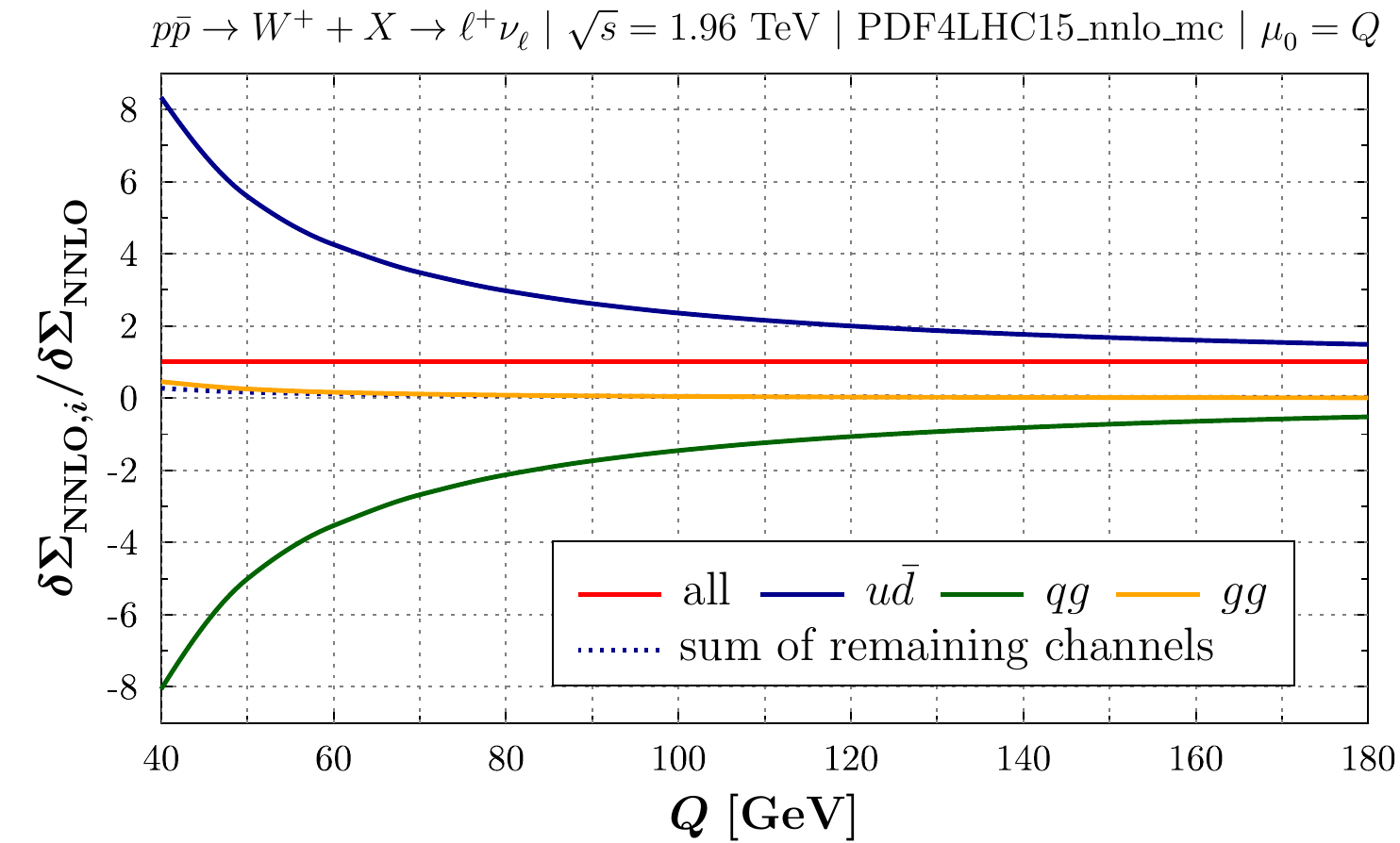} 
\includegraphics[width=0.48 \textwidth]{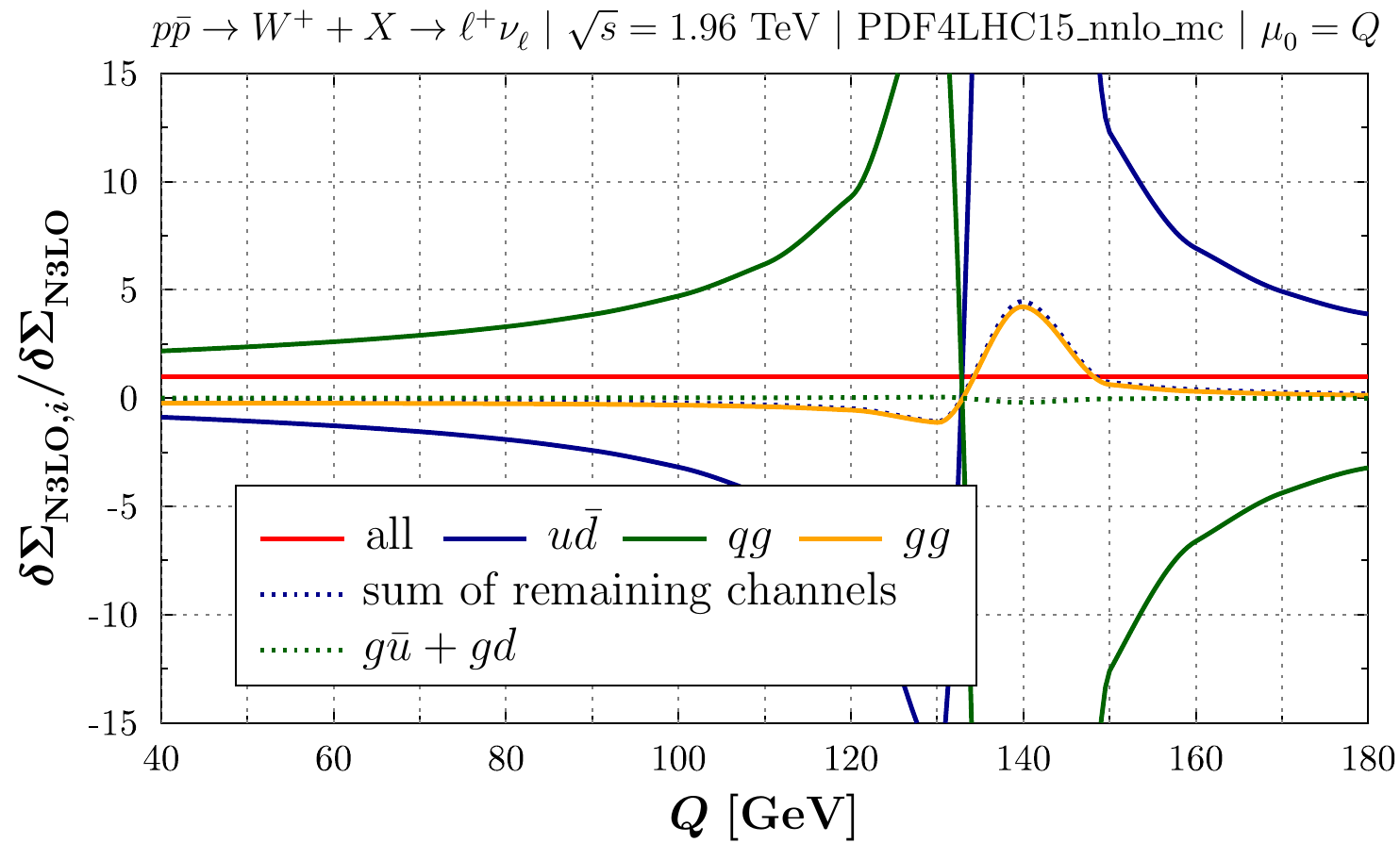} \\
\includegraphics[width=0.48 \textwidth]{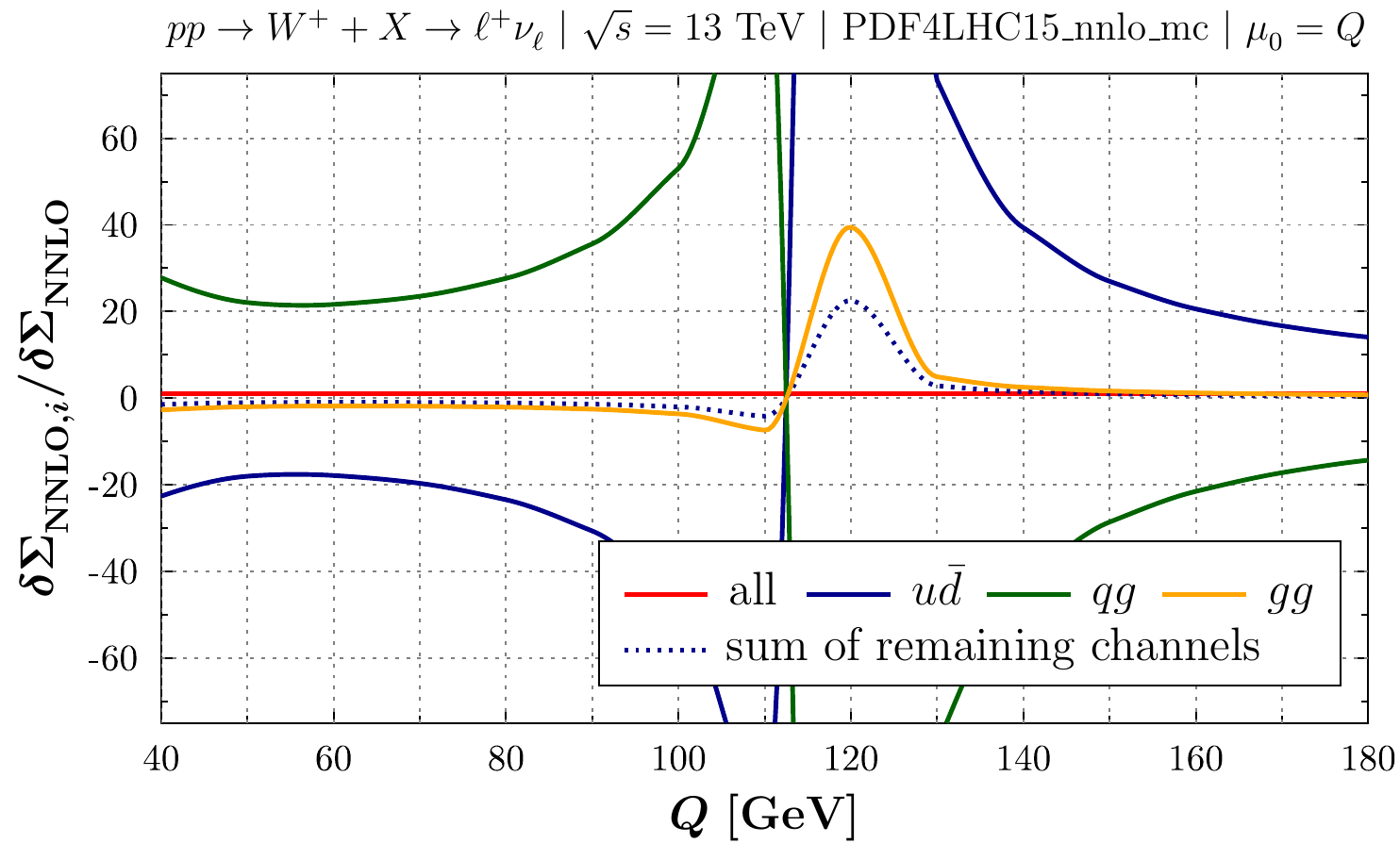} 
\includegraphics[width=0.48 \textwidth]{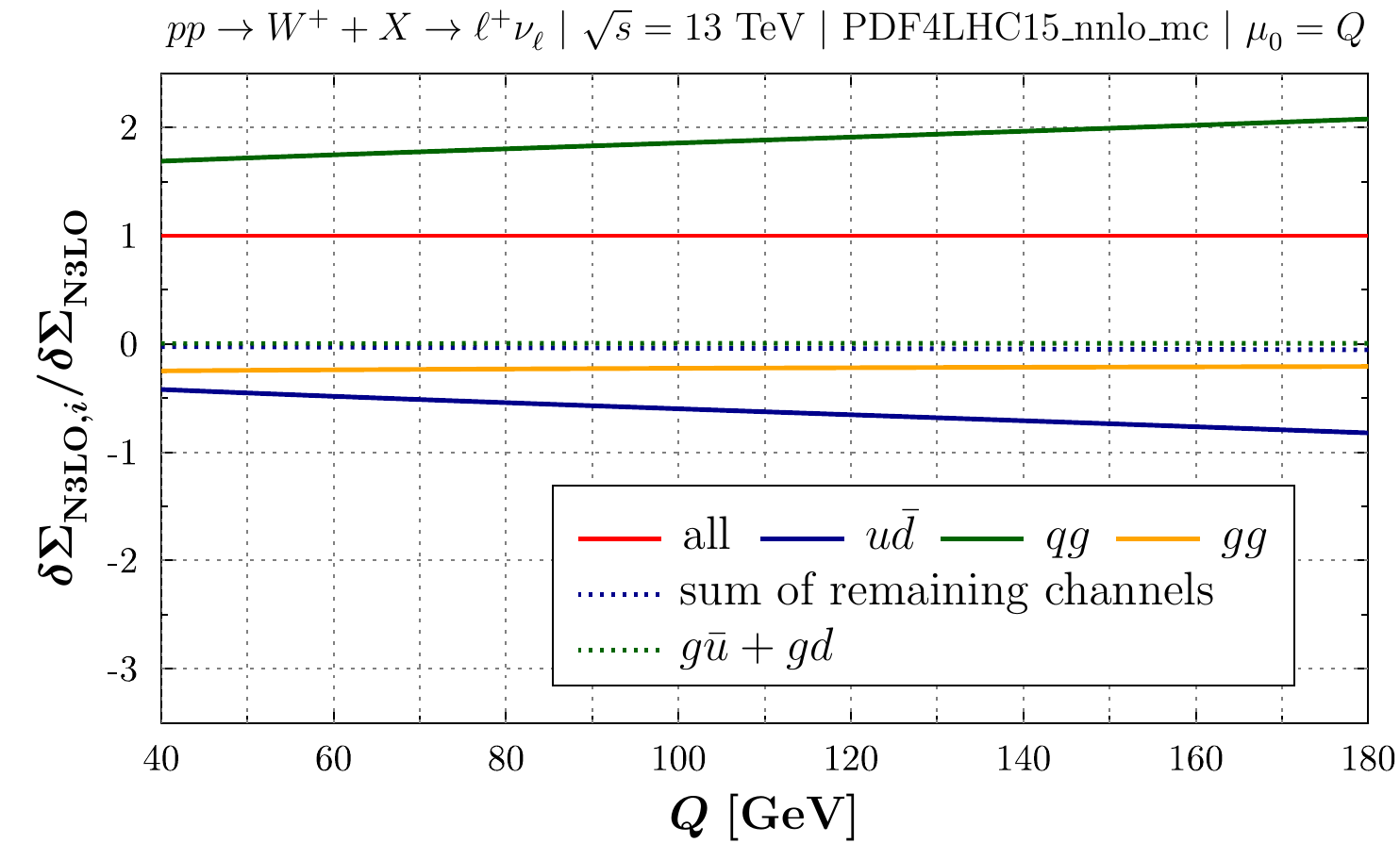} \\
\includegraphics[width=0.48 \textwidth]{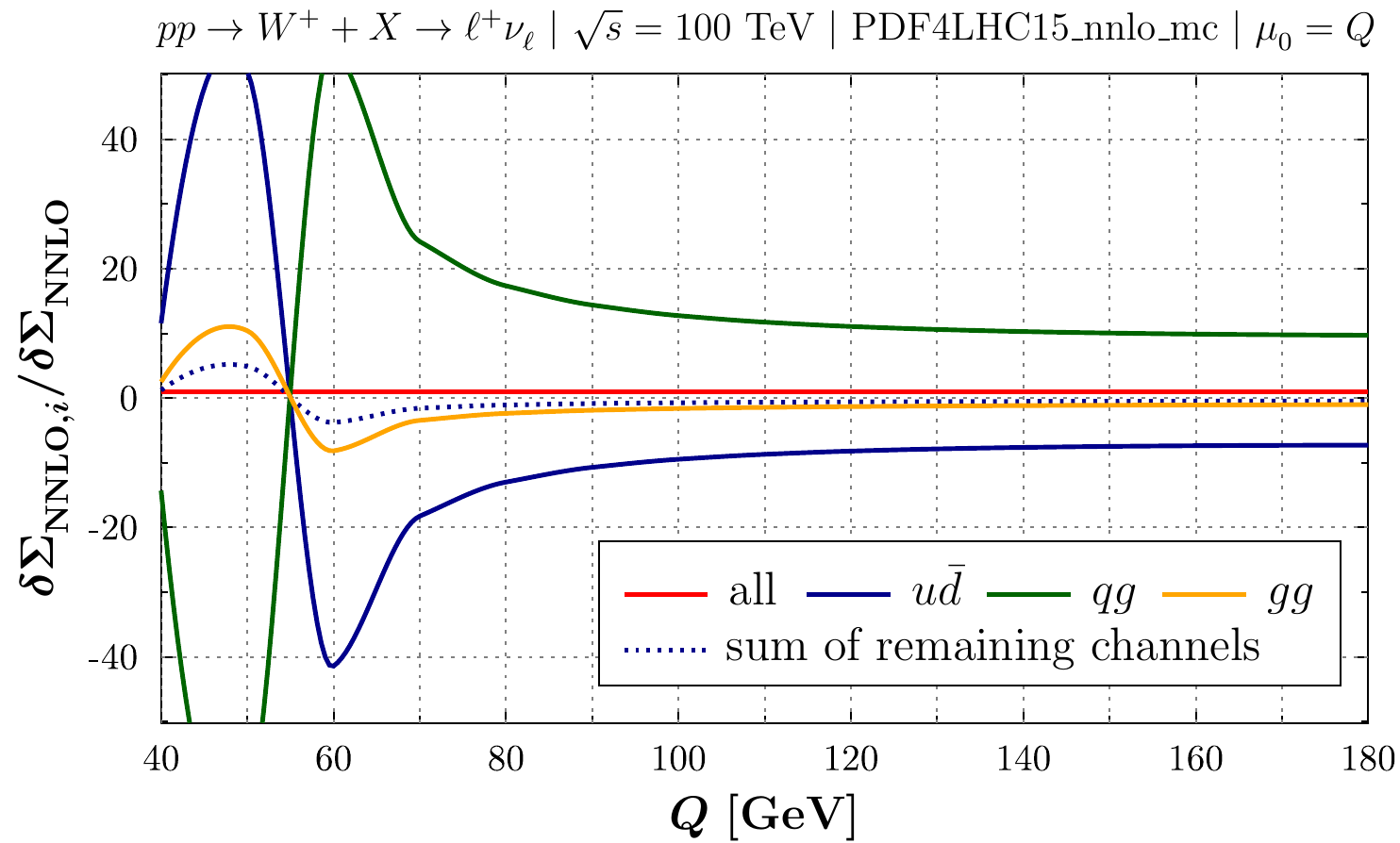} 
\includegraphics[width=0.48 \textwidth]{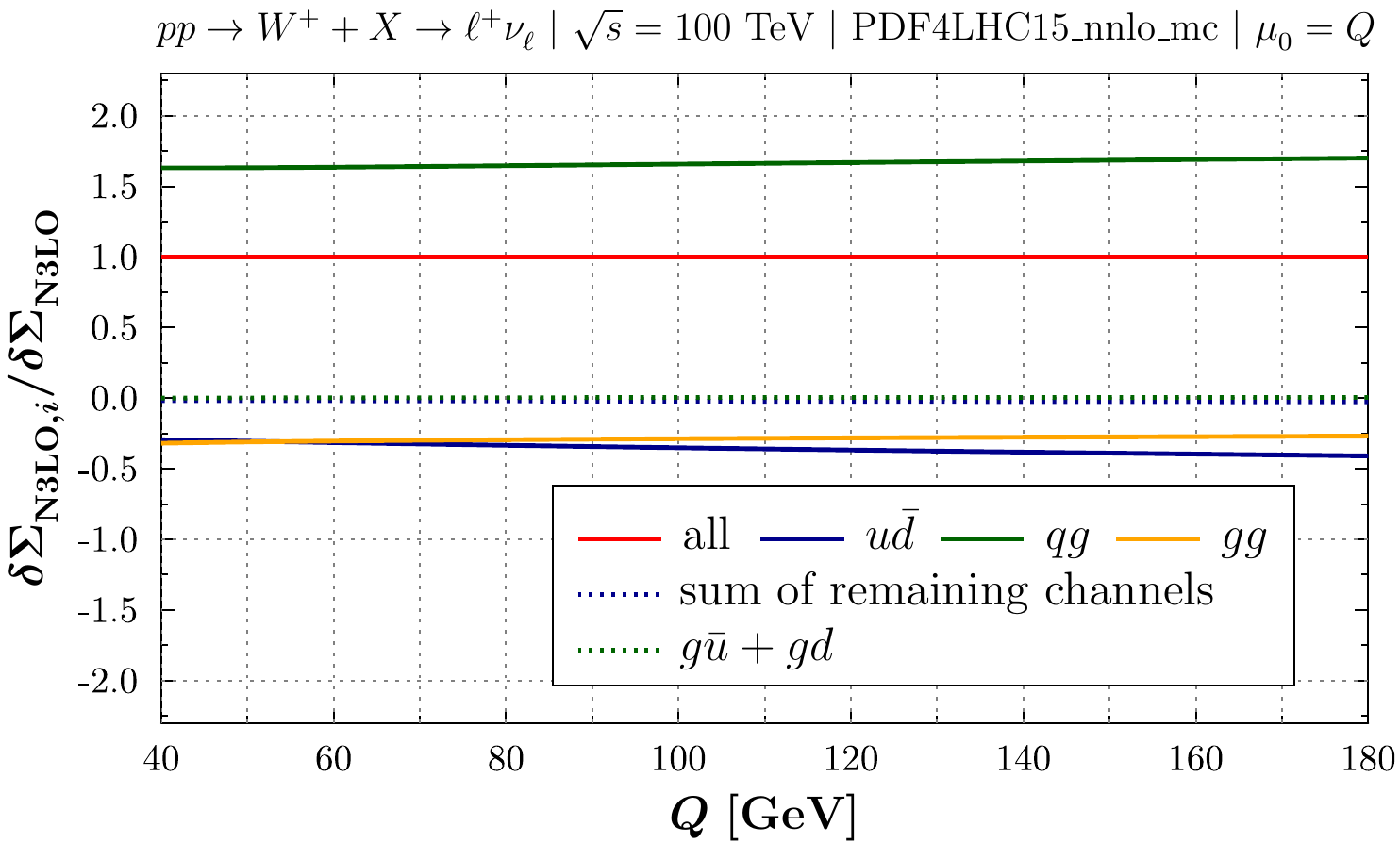} \\
\includegraphics[width=0.48 \textwidth]{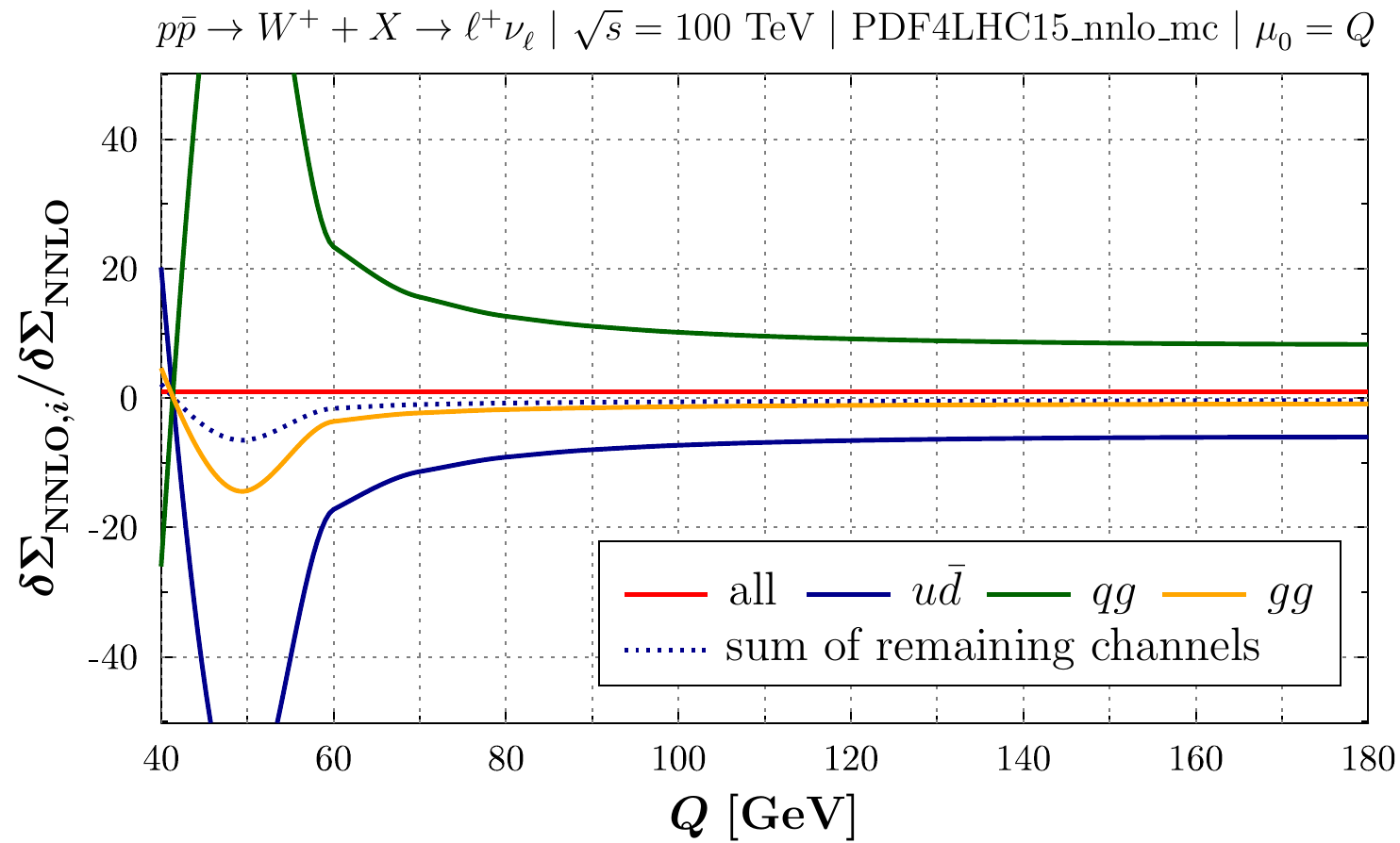} 
\includegraphics[width=0.48 \textwidth]{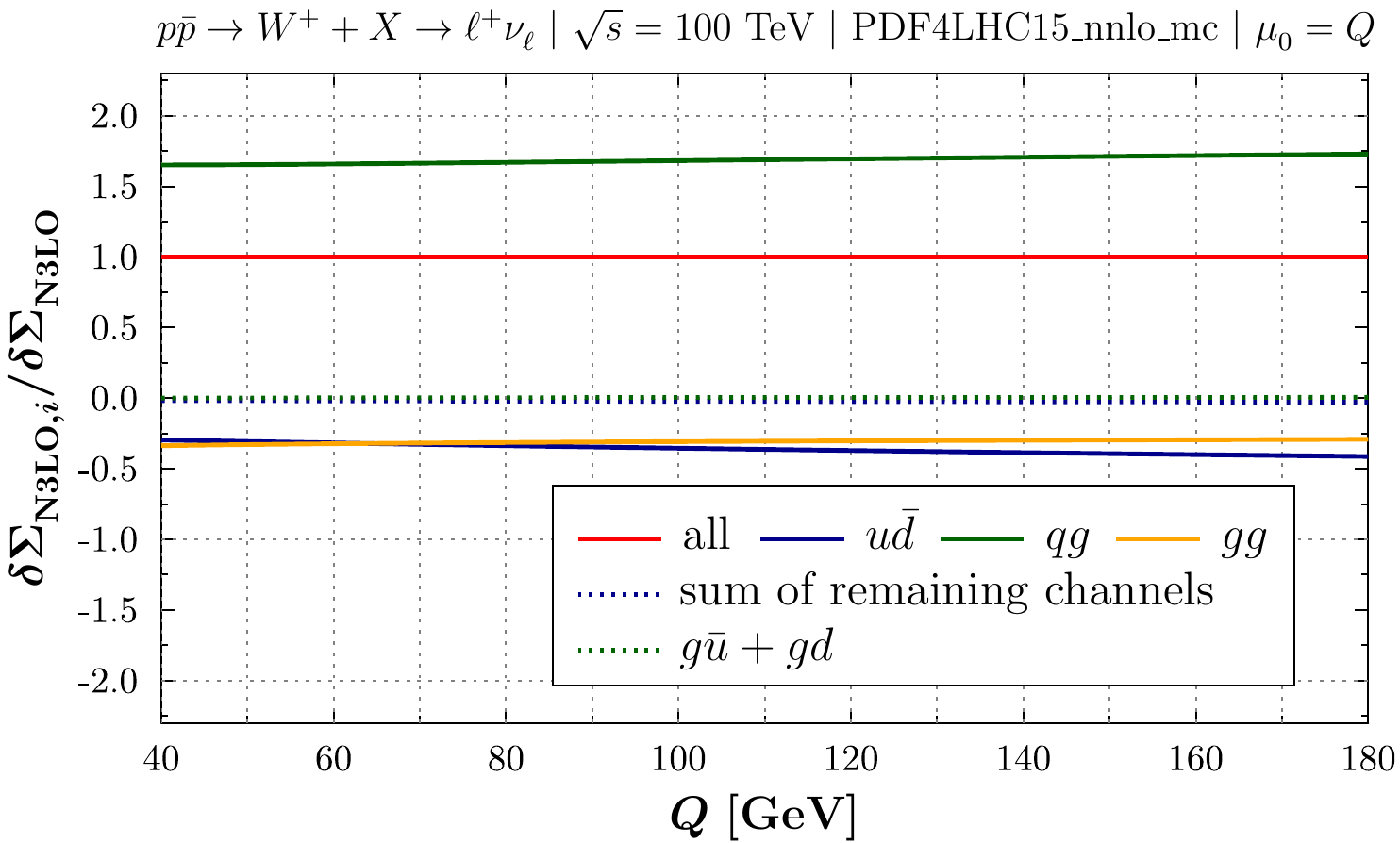} 

\caption{\label{fig:W+_channels} Individual channels contributing to the invariant mass distribution for the production of an $e^+\nu_e$  pair at different colliders normalized to the total correction at NNLO (left panels) and N$^3$LO (right panels).}
\end{center}
\end{figure}

\begin{figure}[!h]
\begin{center}
\includegraphics[width=0.48 \textwidth]{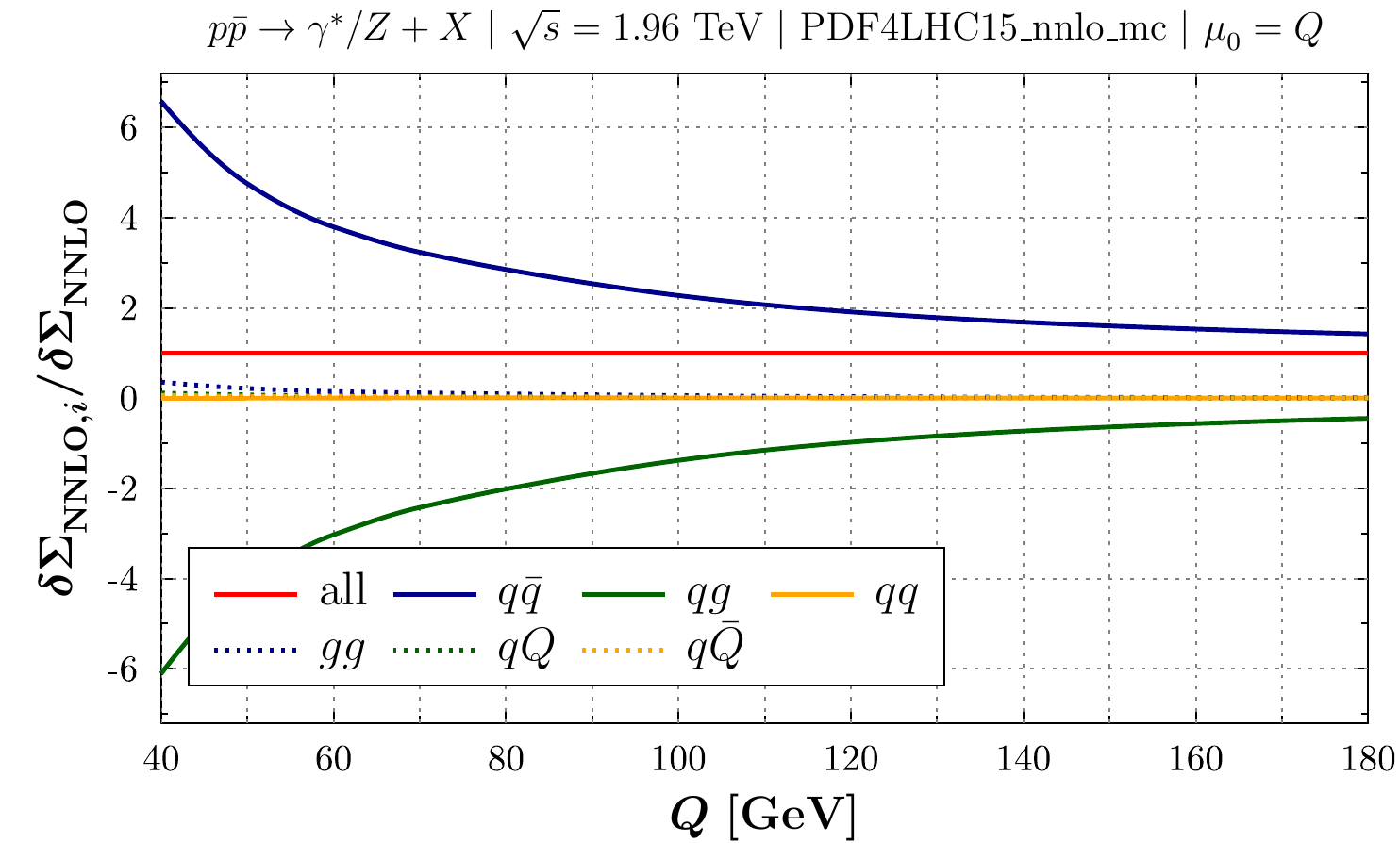} 
\includegraphics[width=0.48 \textwidth]{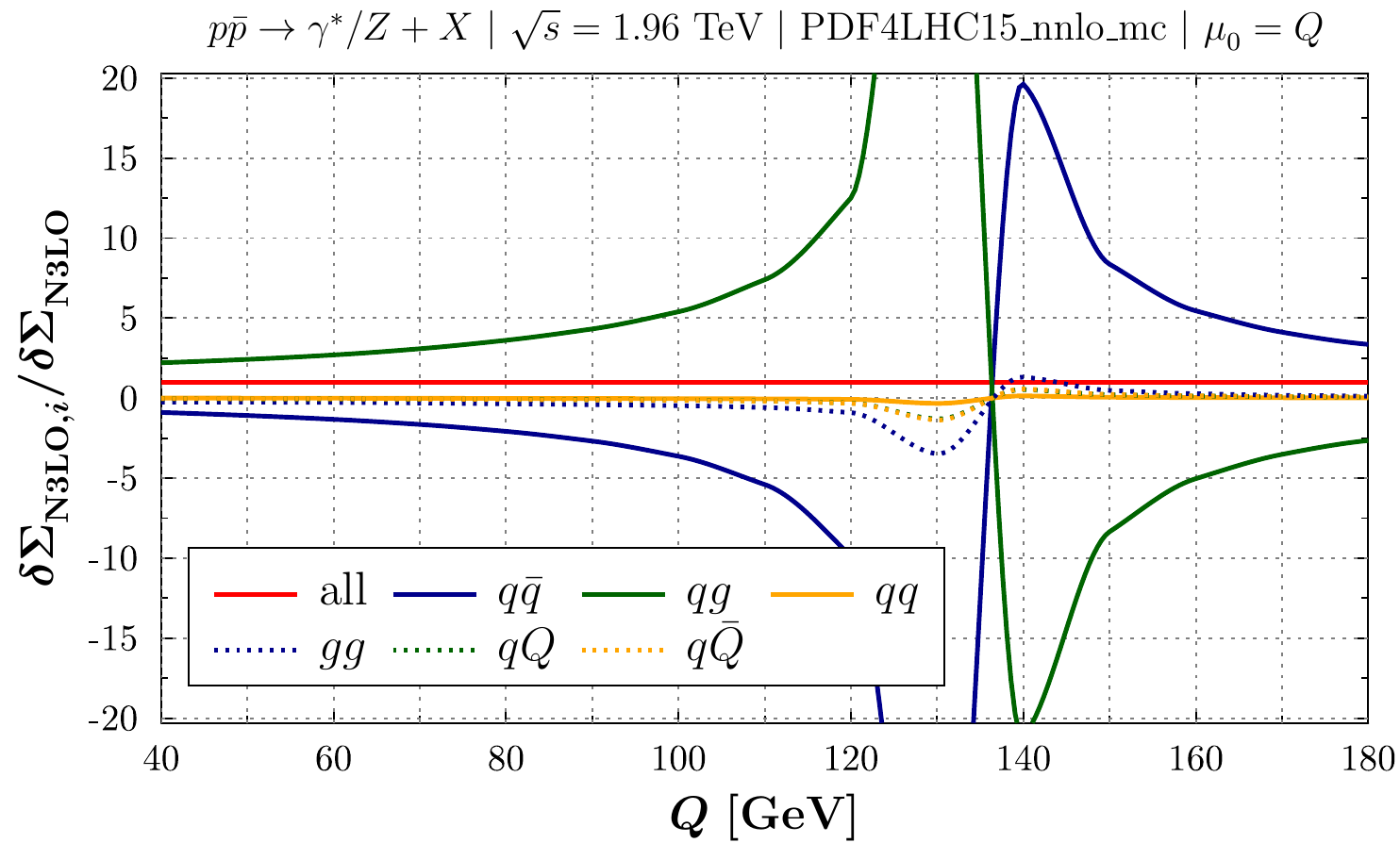} \\
\includegraphics[width=0.48 \textwidth]{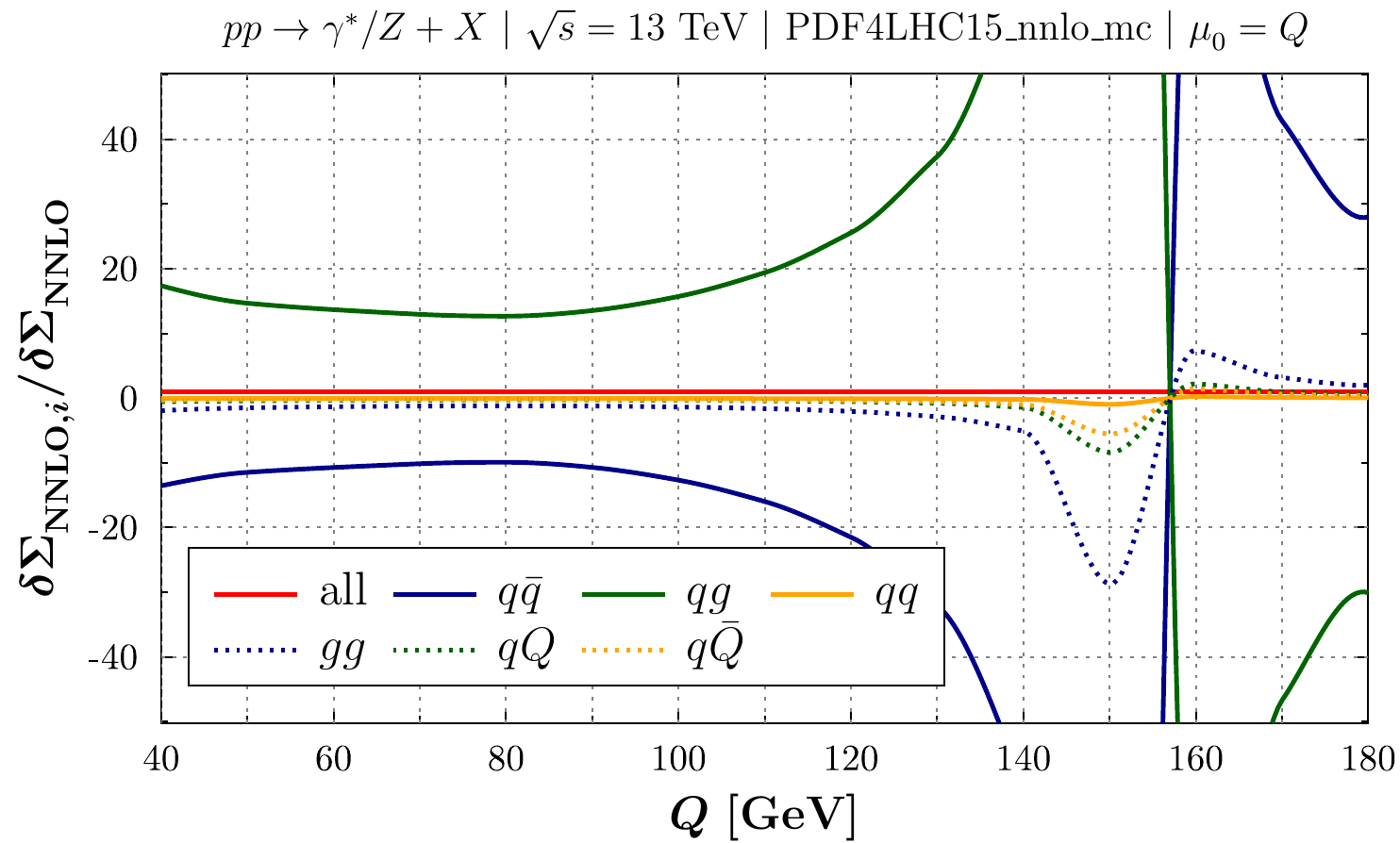} 
\includegraphics[width=0.48 \textwidth]{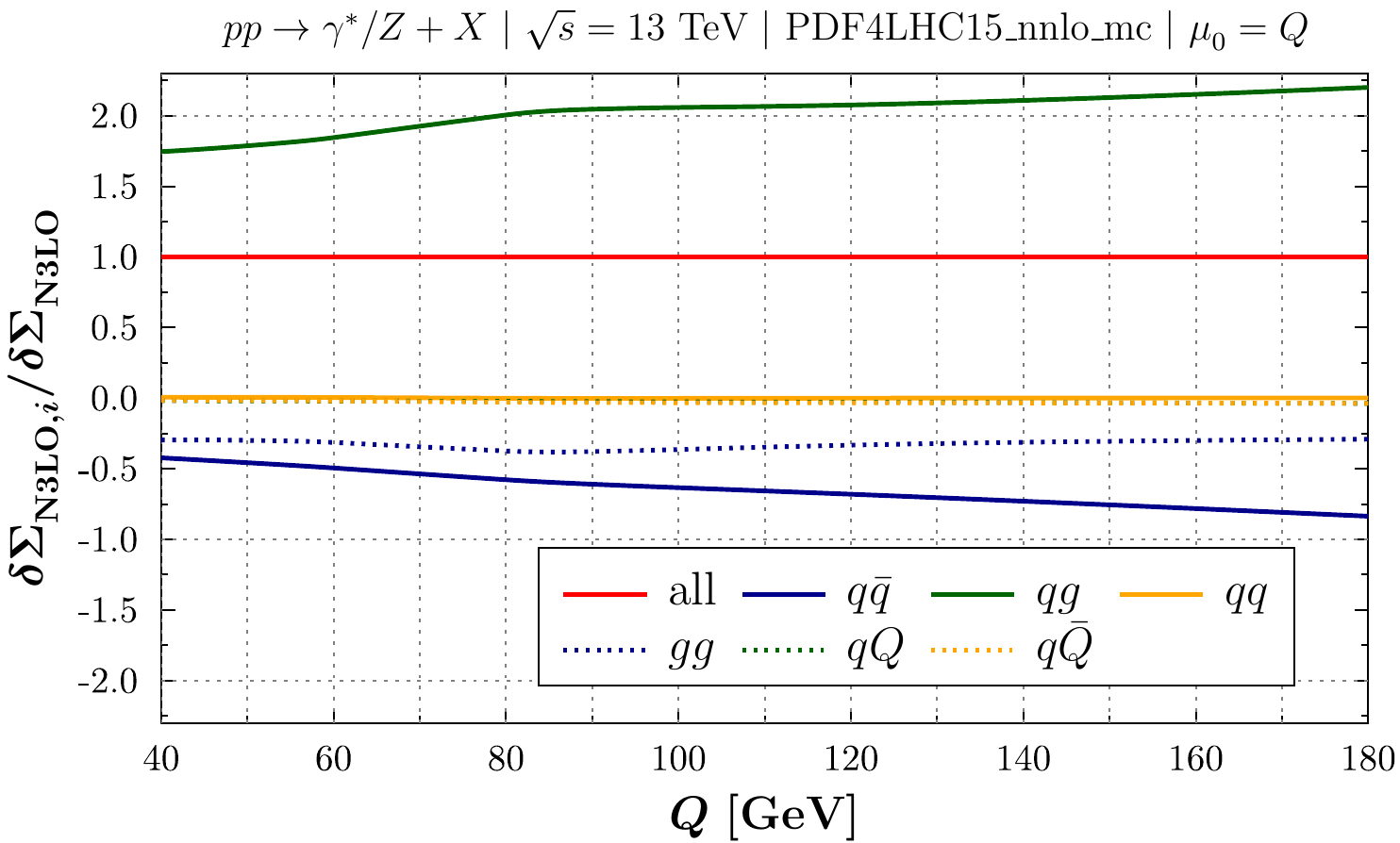} \\
\includegraphics[width=0.48 \textwidth]{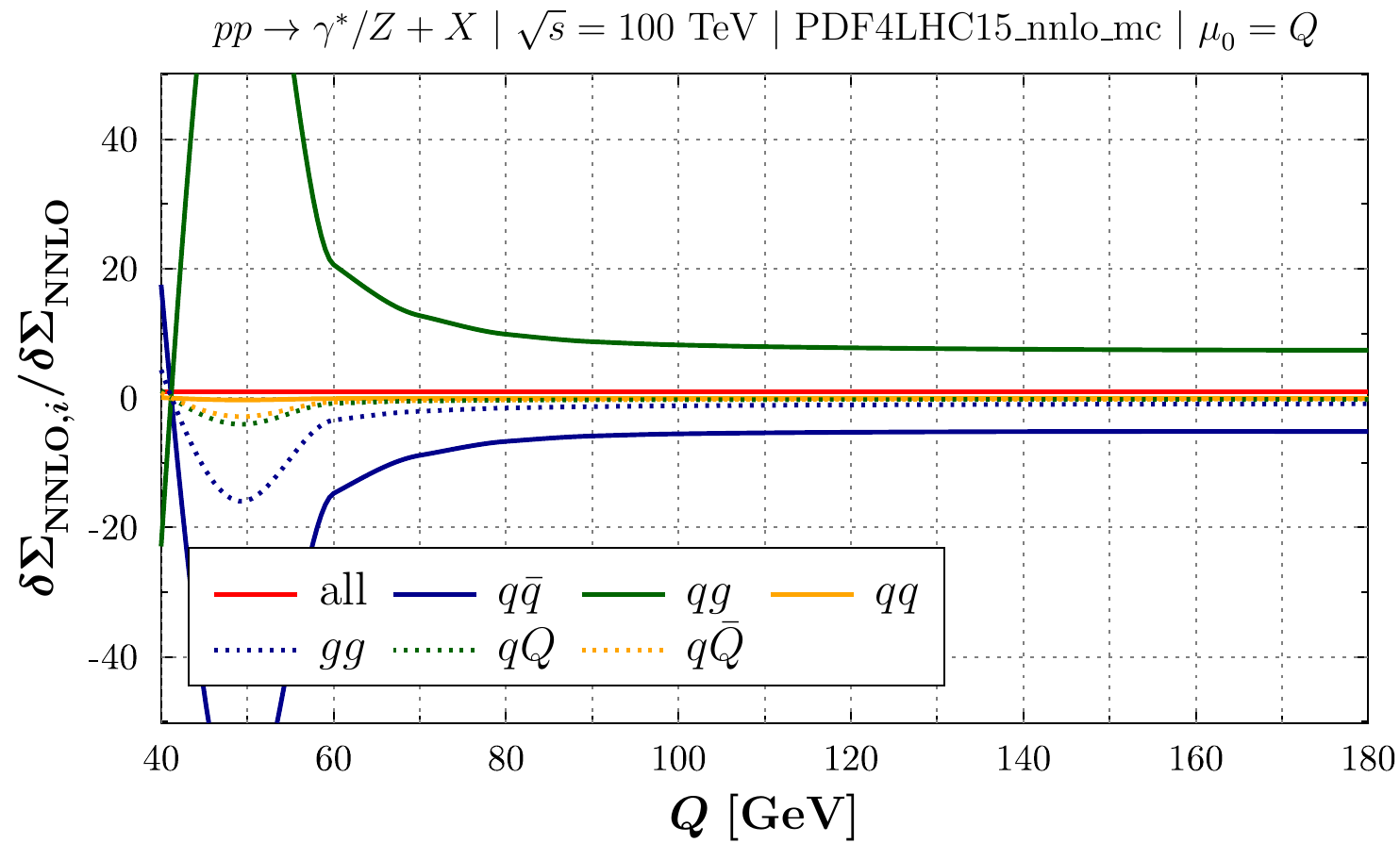} 
\includegraphics[width=0.48 \textwidth]{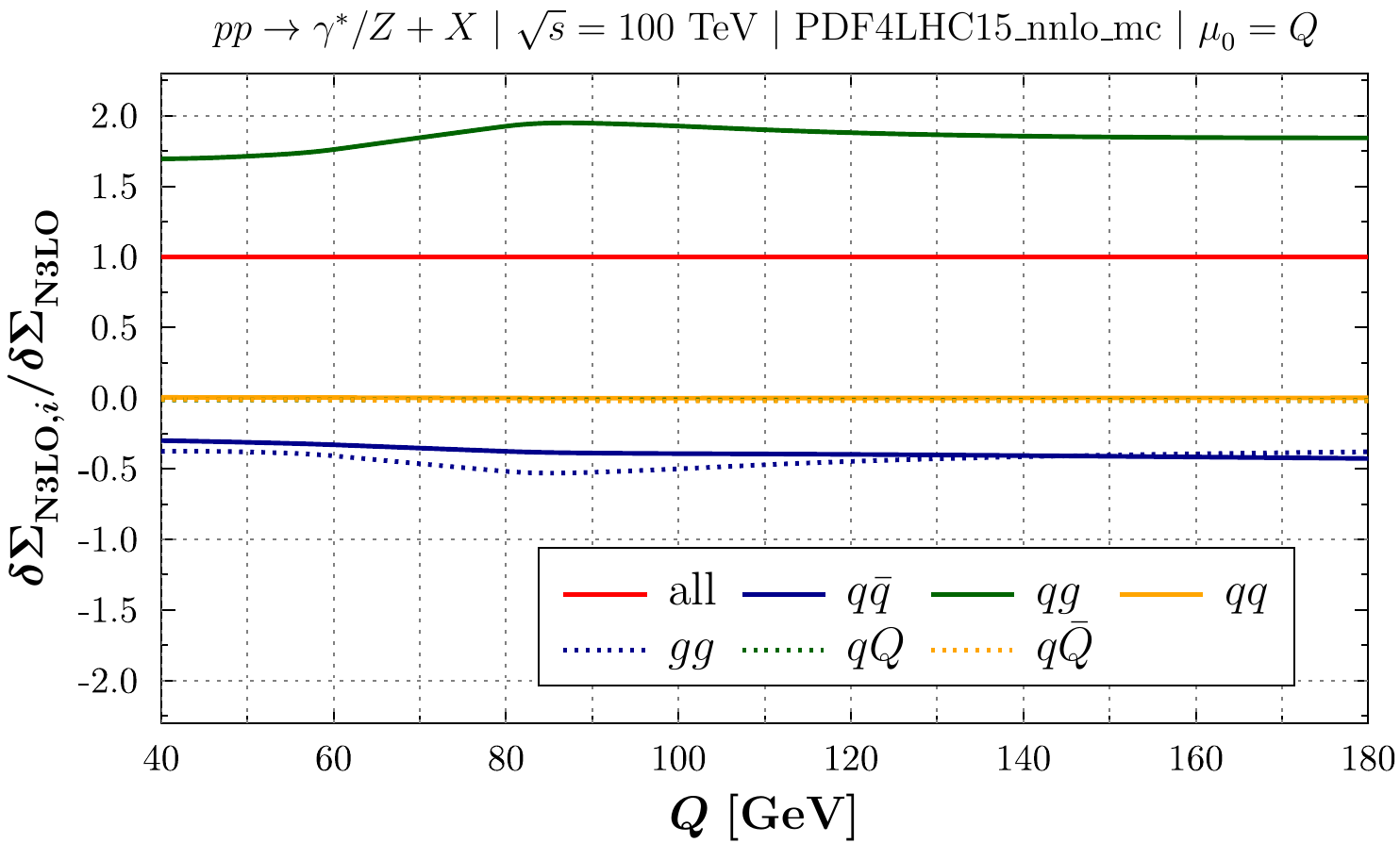} \\
\includegraphics[width=0.48 \textwidth]{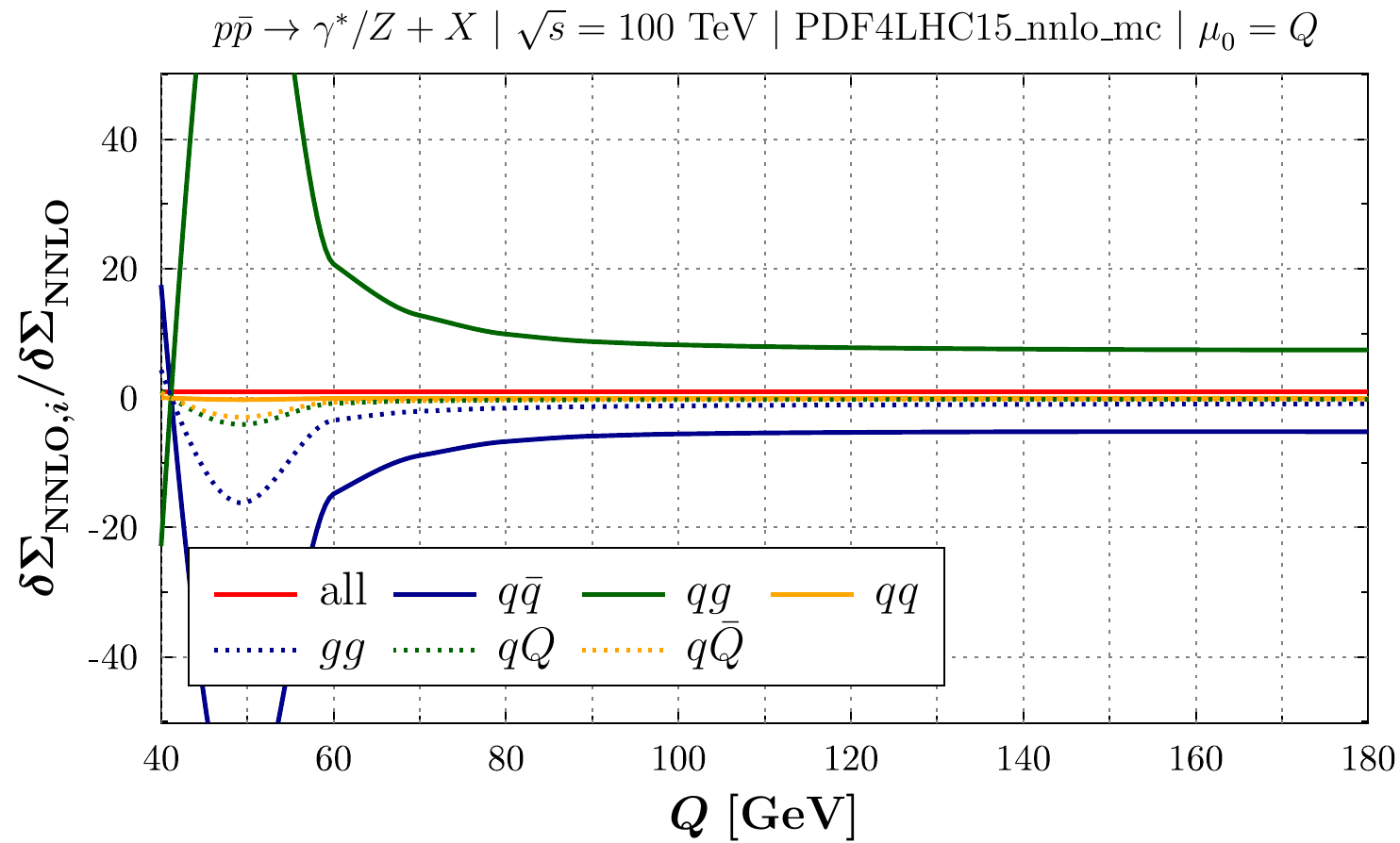} 
\includegraphics[width=0.48 \textwidth]{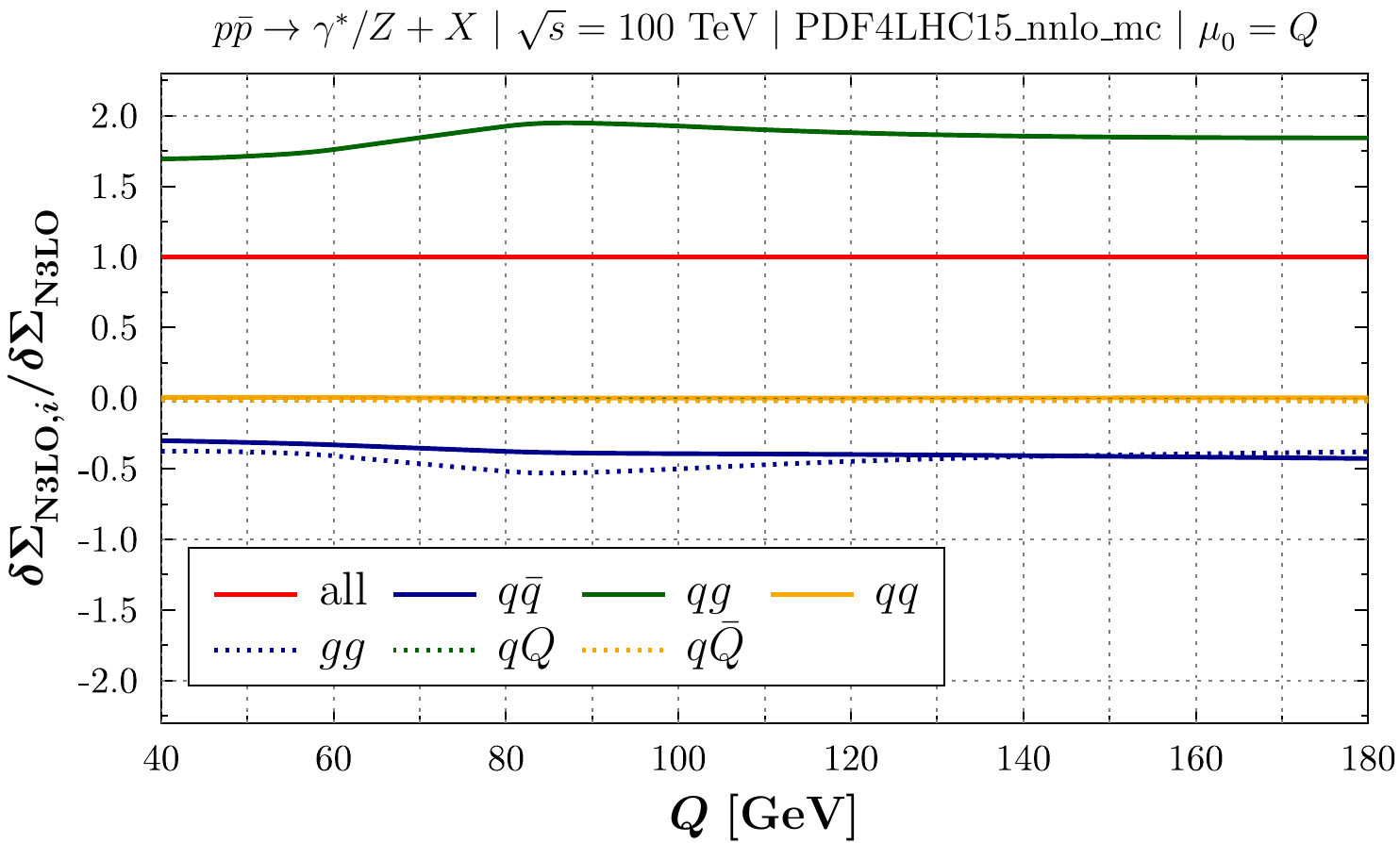} 

\caption{\label{fig:DY_channels} Individual channels contributing to the invariant mass distribution for the production of an $e^+e^-$  pair at different colliders normalized to the total correction at NNLO (left panels) and N$^3$LO (right panels).}
\end{center}
\end{figure}

To better understand the perturbative behavior, we plotted in figs.~\ref{fig:W-_channels},~\ref{fig:W+_channels} and~\ref{fig:DY_channels}  the individual channels contributing to the NNLO and N$^3$LO cross sections, for  1.96 TeV Tevatron, 13 TeV LHC, and 100 TeV $pp$ and $p\bar{p}$ colliders. 
We observe substantial, $\mathcal{O}(10)$, cancellations at NNLO for the LHC and 100 TeV collider over large range of $Q^2$. 
The cancellations are substantially less prominent at N$^3$LO, but nevertheless present.
They introduce an enhanced sensitivity to the PDFs associated with the cancelling channels. 
For example, if the combination of PDFs associated with a large, positive channel is slightly varied and the combination associated with a large, negative channel remains the same, the effect of the change in PDFs is enhanced in the prediction of the cross section.
If this effect is not accidental, the question arises if it is adequately taken into account at N$^3$LO, because our PDFs are currently extracted at NNLO.
Furthermore, the uncertainty of our knowledge of PDFs may give rise to varying cancellation patterns.
Note that the definition of the different partonic channels depends on the scheme used to define the PDFs (and all available NNLO PDF sets are defined in $\overline{\textrm{MS}}$-scheme), and so the pattern of cancellations may itself be scheme-dependent. 
If indeed this cancellation has an origin in our current definition of PDFs, we may ask if a more judicial choice of PDF scheme would give rise to an improved description.
The points raised above may be part of the answer to the question of why N$^3$LO corrections for Drell-Yan processes are partly not contained within NNLO scale uncertainties.

\subsection{PDF study for Drell-Yan and Higgs processes}
\label{sec:sec:pdfDYandHiggs}

Given the smallness of the scale uncertainties at N$^3$LO (and already at NNLO for Drell-Yan processes), the dependence of the cross sections on the PDFs becomes the dominant source of theoretical uncertainties at hadron colliders. It is therefore highly desirable to perform a comprehensive study of the various PDF sets available on the market. Indeed, since PDFs need to be extracted from experiment, their value and uncertainty depend on the experimental data used in the fit as well as on the fitting methodology used by the different groups.

Our default set is {\tt PDF4LHC15\textunderscore nnlo\textunderscore mc} (PDF4LHC15)~\cite{Butterworth:2015oua}, and in addition we have calculated predictions for Drell-Yan and Higgs production processes using the latest updates of the global sets available in 2022:
\begin{itemize}
\item The MSHT20 update of MMHT14: {\tt MSHT20nnlo\textunderscore as118} (MSHT20)~\cite{Bailey:2020ooq};
\item The CT18 update from the CTEQ-TEA collaboration, superseding the CT14 global analysis: {\tt CT18NNLO} (CT18)~\cite{Hou:2019efy};
\item The two last updates from the NNPDF collaboration: {\tt NNPDF31\textunderscore nnlo\textunderscore as\textunderscore 0118} (NNPDF~3.1)~\cite{NNPDF:2017mvq} and the latest set {\tt NNPDF40\textunderscore nnlo\textunderscore as\textunderscore 01180} (NNPDF~4.0)~\cite{NNPDF:2021njg};
\item The PDF4LHC21 update of PDF4LHC15, based on the statistical Monte-Carlo combination of the three global sets NNPDF 3.1, MSHT20, and CT18: \\ {\tt PDF4LHC21\textunderscore mc}~\cite{Ball:2022hsh};
\item The last update from the ABMP collaboration: {\tt ABMP16als118\textunderscore 5\textunderscore nnlo}\\ (ABMP16\textunderscore als118)~\cite{Alekhin:2017kpj}.
\end{itemize}
Each PDF set also includes a fitted value for the input strong coupling constant, $\alpha_S(m_Z)$. It is well known that the ABMP fitted value for this physical parameter is smaller than what is obtained in the other sets that we consider. This has in particular a strong impact on the predictions for Higgs production in gluon fusion. Since ABMP16 also contains PDF sets with alternative fitted values for $\alpha_S(m_Z)$, we have chosen to select the one which uses $\alpha_S(m_Z)=0.118$ as in the other sets. This greatly improves the agreement between e.g. PDF4LHC15 and ABMP16 predictions for the inclusive $gg\to H$ hadronic cross section at the LHC.

\begin{figure}[!t]
\begin{center}
\includegraphics[width=0.85\textwidth]{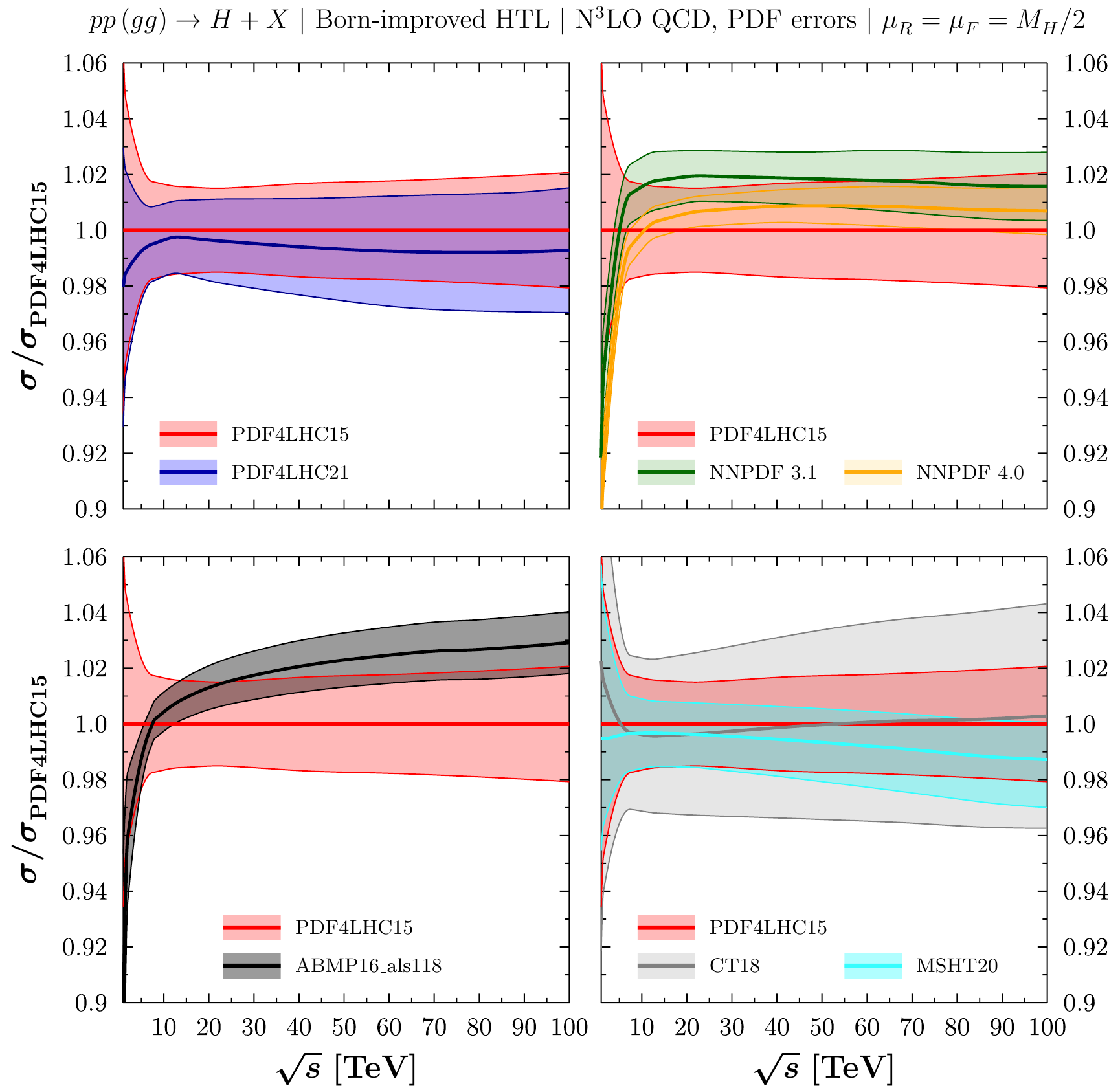}
\caption{\label{fig:PDF_ggH}
Dependence on the choice of the PDF set of the Born-improved HTL gluon fusion Higgs production cross section at N$^3$LO in QCD as a function of the proton-proton c.m. energy $\sqrt{s}$ in TeV. The results are normalized to the central PDF4LHC15 PDF set and the 68\% CL PDF uncertainties are represented by bands for all sets and calculated according to the prescription of ref.~\cite{Butterworth:2015oua}. The comparison is between PDF4LHC15 and: PDF4LHC21 (upper left panel); NNPDF 3.1 and NNPDF 4.0 (upper right panel); ABMP16\textunderscore als118 (lower left panel); CT18 and MSHT 20 (lower right panel).}
\end{center}
\end{figure}

\begin{figure}[!t]
\begin{center}
\includegraphics[width=0.85\textwidth]{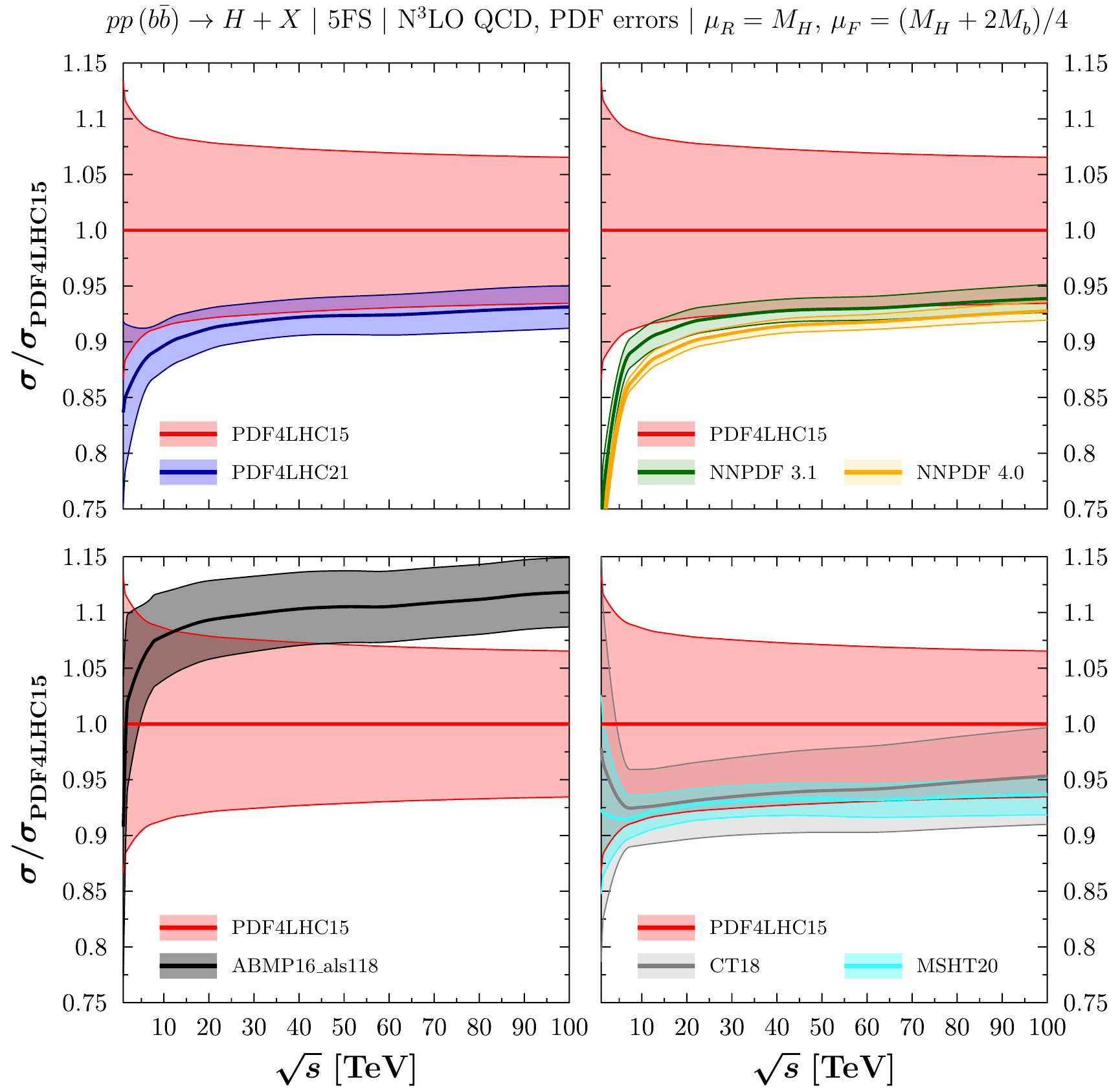}
\caption{\label{fig:PDF_bbH}
Same as in fig.~\ref{fig:PDF_ggH}, but for bottom-quark-fusion Higgs production in the five-flavor-scheme.}
\end{center}
\end{figure}

\begin{figure}[!t]
\begin{center}
\includegraphics[width=0.85\textwidth]{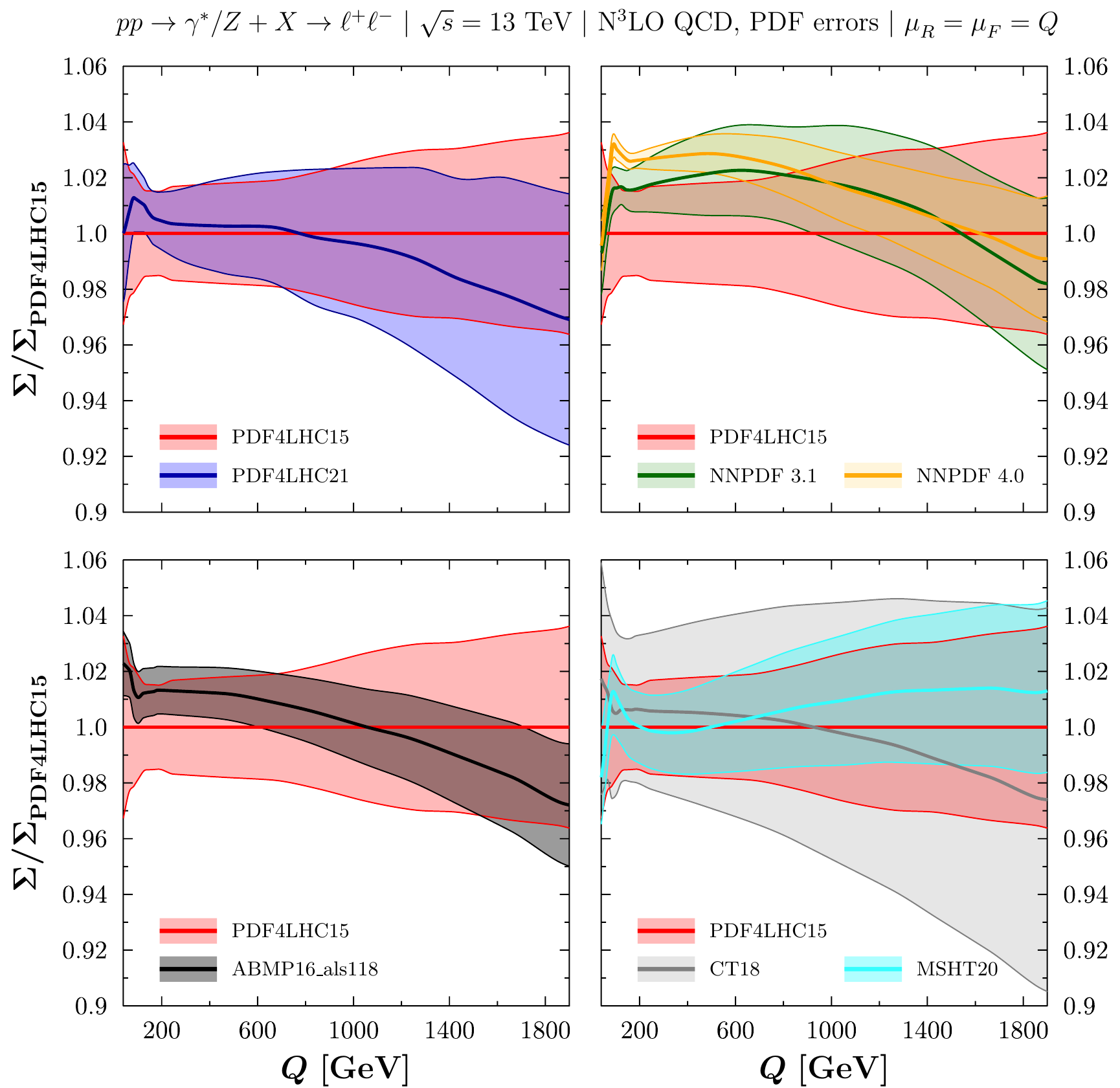}
\caption{\label{fig:PDF_NCDY}
Dependence on the choice of the PDF set of the neutral-current Drell-Yan differential cross section at N$^3$LO in QCD as a function of the invariant mass $Q$ of the final-state lepton pair, at the 13 TeV LHC. The results are normalized to the central PDF4LHC15 PDF set and the 68\% CL PDF uncertainties are represented by bands for all sets and calculated according to the prescription of ref.~\cite{Butterworth:2015oua}. The comparison is between PDF4LHC15 and: PDF4LHC21 (upper left panel); NNPDF 3.1 and NNPDF 4.0 (upper right panel); ABMP16\textunderscore als118 (lower left panel); CT18 and MSHT 20 (lower right panel).}
\end{center}
\end{figure}

\begin{figure}[!t]
\begin{center}
\includegraphics[width=0.85\textwidth]{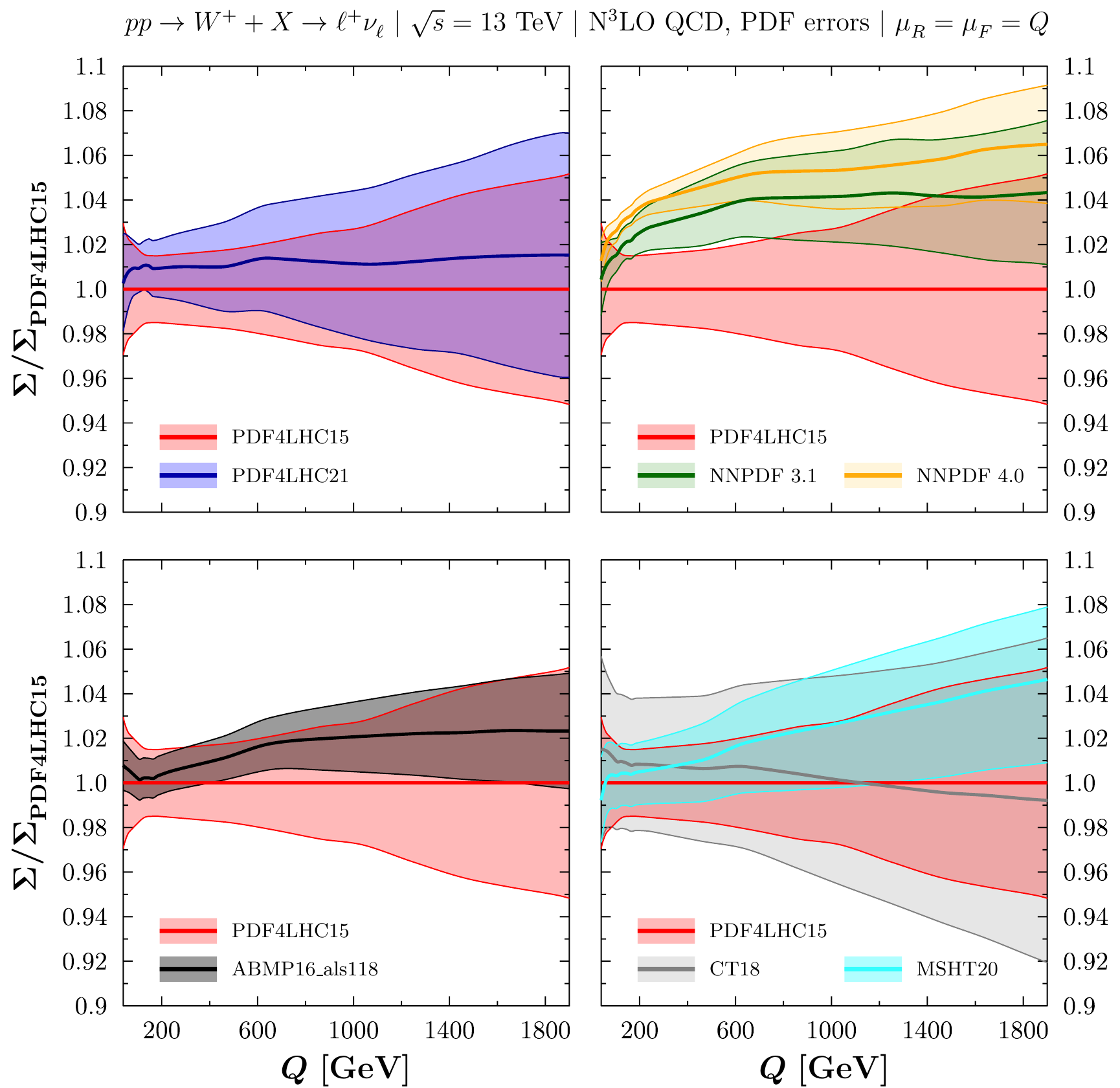} 
\caption{\label{fig:PDF_CCDY_Wplus}
Same as in fig.~\ref{fig:PDF_NCDY}, but for charged-current $W^+$ Drell-Yan differential cross section.}
\end{center}
\end{figure}

\begin{figure}[!t]
\begin{center}
\includegraphics[width=0.85\textwidth]{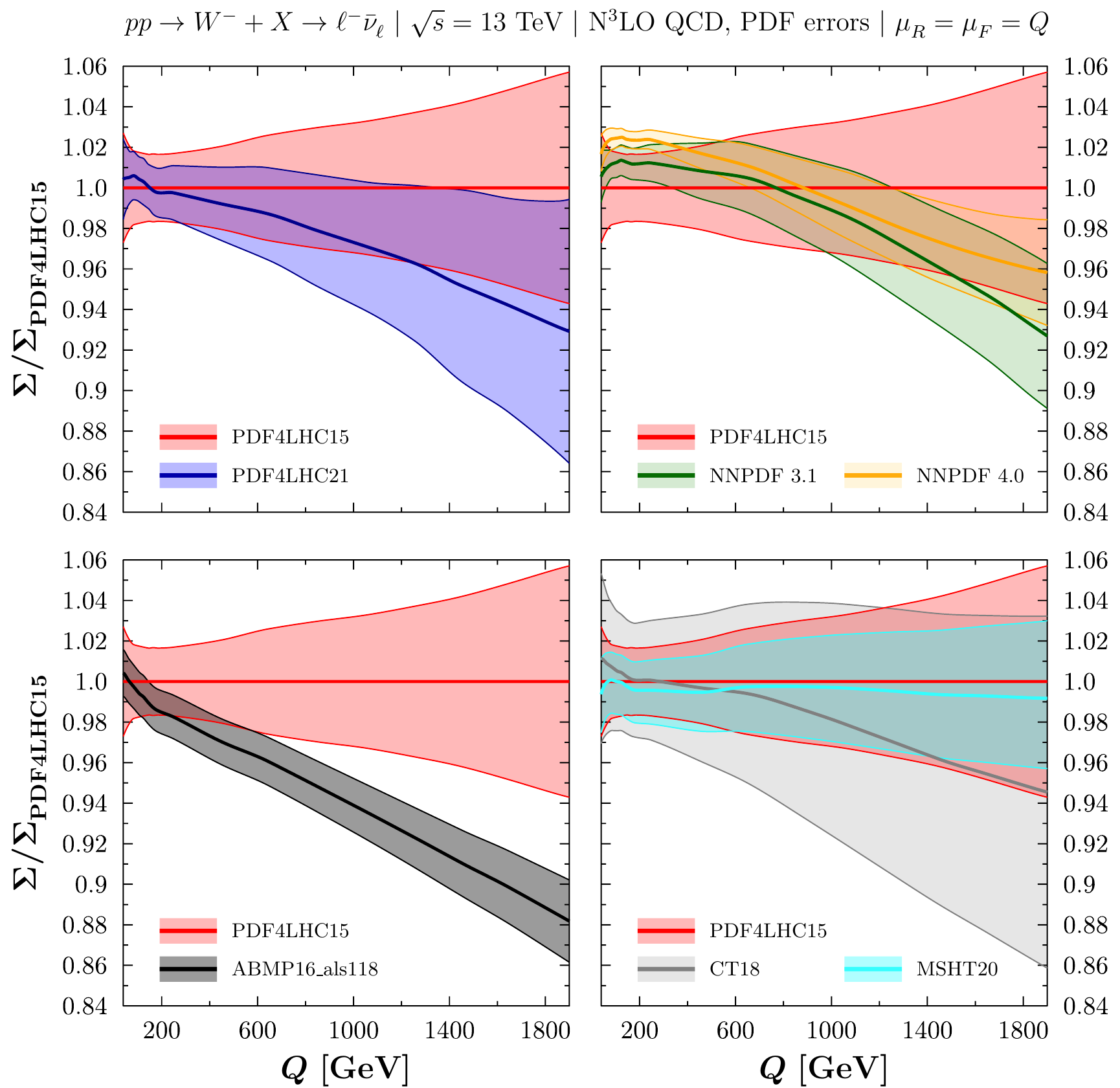} 
\caption{\label{fig:PDF_CCDY_Wminus}
Same as in fig.~\ref{fig:PDF_NCDY}, but for charged-current $W^-$ Drell-Yan differential cross section.}
\end{center}
\end{figure}

In figs.~\ref{fig:PDF_ggH}, \ref{fig:PDF_bbH}, \ref{fig:PDF_NCDY}, \ref{fig:PDF_CCDY_Wplus}, and~\ref{fig:PDF_CCDY_Wminus}, we show the dependence of the Higgs and Drell-Yan cross sections on the choice of the PDF set. We normalize all our results to the central PDF4LHC15 set. The first panel shows a comparison to the PDF4LHC21 set, the second panel shows a comparison to the NNPDF~3.1 and NNPDF~4.0 sets. For all these sets the PDF uncertainties have been computed using the Monte-Carlo replica method adapted to produce a 68\% CL uncertainty, following the prescription of ref.~\cite{Butterworth:2015oua}. The lower panels show results with ABMP16\textunderscore als118,  CT18, and MSHT20 sets. The PDF uncertainties for these sets have been evaluated using the Hessian prescription rescaled to 68\% CL ($1\sigma$,  assuming a Gaussian distribution) when needed (this is the case for CT18 which provides error sets for 90\% CL Hessian errors). Note that the PDF uncertainty for the MSHT20 PDF set is asymmetric. We observe that various sets show overall rather good agreement with one another (up to some deviations which we will highlight below), and that CT18 always provides the largest PDF uncertainty amongst the newest sets.
Nonetheless, the systematic bias introduced by a specific choice of PDF set can be substantial, and we observe large variations in the width of the error bands for different sets.

In the case of gluon fusion Higgs production, all sets are in agreement, as seen in fig.~\ref{fig:PDF_ggH}, and apart from CT18, all newer sets give smaller PDF uncertainties compared to PDF4LHC15. This process is more sensitive to the input value of $\alpha_S(m_Z)$ than to the PDFs themselves. Indeed, while we observe a good agreement between ABMP16\textunderscore als118 and PDF4LHC15, this would not have been the case between PDF4LHC15 and the nominal ABMP16 PDF set which has a lower fitted value of $\alpha_S(m_Z)$.

There is a spectacular reduction of the PDF uncertainties for bottom-quark fusion Higgs production as seen in fig.~\ref{fig:PDF_bbH}, with all new sets giving substantially smaller uncertainties compared to the nominal PDF4LHC15 set. The central value also shifted noticeably.  This reflects a strong improvement in the determination of the bottom-quark PDF in the newest sets. For c.m. energies above 50 TeV, ABMP16\textunderscore als118 and PDF4LHC15 display a $1\sigma$ disagreement, while for low c.m. energies below 10 TeV there is also at least $1\sigma$ deviation between both NNPDF sets and PDF4LHC15.

The study of the Drell-Yan processes is very interesting, as they are standard candle processes at hadron colliders. In the case of neutral Drell-Yan production, displayed in fig.~\ref{fig:PDF_NCDY}, PDF4LHC21 has smaller uncertainties at values of the invariant mass $Q$ below 600~GeV, but larger uncertainties for $Q>600$~GeV. All sets but CT18 display a similar downward trend when going to larger values of $Q$, and more interestingly the newest NNPDF~4.0 set is in disagreement with PDF4LHC15 at the $1\sigma$ level for $Q<600$~GeV, and in particular around the $Z$ peak where it also disagrees with NNPDF~3.1. In the case of charged-current $W^+$ production, displayed in fig.~\ref{fig:PDF_CCDY_Wplus}, both NNPDF sets show significant differences, at the $1\sigma$ level, for $Q$ values between 200 and 600~GeV. The deviation extends up to $Q$ values of around 1250~GeV for NNPDF~4.0. The case of $W^-$ production cross sections, shown in fig.~\ref{fig:PDF_CCDY_Wminus}, displays significant differences in large $Q^2$ ranges between the older set and the newer updates. The NNPDF and PDF4LHC sets show a significant shift toward the ABMP16\textunderscore als118 results for this process. The latter, in particular, is in  significant disagreement with PDF4LHC15 predictions for $Q>600$~GeV, beyond the $1\sigma$ level. Again NNPDF~4.0 is in a $1\sigma$ disagreement with PDF4LHC15 (and PDF4LHC21) for low $Q$ values below 200~GeV and in particular around the $W$ peak.   This behavior signals a systematic trend in all Drell-Yan processes that calls for further investigations to understand why these differences exist in this particular set.
 


 
\section{\boldmath Inclusive associated $VH$ production at N$^3$LO}
\label{sec:vh}

In this section, we present for the first time results for the fully inclusive cross section at N$^3$LO for the production of a Higgs boson in association with a weak boson $V = Z, W^{\pm}$ in QCD with $N_f=5$ massless flavors. We only include Drell-Yan-like contributions, i.e. the diagrams where the Higgs boson couples directly to the intermediate vector boson.\footnote{In particular, we neglect contributions to $ZH$ production from box diagrams where the Higgs boson couples to a heavy fermion loop. The LO analysis can be found in ref.~\cite{Kniehl:1990iva}, the NLO QCD corrections to $g g\to Z H$ in the heavy-top quark limit can be found in ref.~\cite{Altenkamp:2012sx} while the NLO QCD corrections (two-loop corrections) including the top-quark dependence have been computed recently, see refs.~\cite{Davies:2020drs,Chen:2020gae,Alasfar:2021ppe,Wang:2021rxu,Chen:2022rua,Degrassi:2022mro}. We also neglect the top-quark radiated contributions to $VH$ calculated in ref.~\cite{Brein:2011vx}, known to be very small for inclusive rates.} The NLO QCD corrections were evaluated in refs.~\cite{Han:1991ia,Baer:1992vx,Ohnemus:1992bd}, while the NNLO QCD corrections were given in ref.~\cite{Brein:2003wg} for the inclusive cross section and have been implemented in the code {\tt vh@nnlo}~\cite{Brein:2012ne,Harlander:2018yio}. The NNLO QCD corrections have also been extended to exclusive cross sections and fully differential observables in refs.~\cite{Ferrera:2011bk,Ferrera:2014lca}, and later also to include the Higgs decays in refs.~\cite{Ferrera:2013yga,Campbell:2016jau,Caola:2017xuq,Ferrera:2017zex}. An NNLO QCD analysis with a resolved jet was performed in refs.~\cite{Majer:2020kdg,Gauld:2021ule}. The NLO electroweak corrections are also known \cite{Ciccolini:2003jy,Denner:2011id} and are rather seizable.

The Drell-Yan-like contribution comes from the virtual vector boson $V^*$ exchanged in the $s$-channel and its subsequent splitting into Higgs and on-shell $V$ boson. We can obtain this cross section by integrating the invariant-mass distribution in eq.~\eqref{eq:master} over the virtuality of the intermediate off-shell vector boson:
\begin{align}
  \sigma_{N_1 N_2 \to VH} = & \int_{(m_H +m_V)^2}^s dQ^2\, \frac{d\sigma_{N_1 N_2\to V^*}}{dQ^2}\, \frac{Q^4}{(Q^2-m_V^2)^2}\,\frac{m_V^2}{48\pi^2 v^2}\nonumber\\
  \times & \sqrt{\lambda\left(1,\frac{m_V^2}{Q^2},\frac{m_H^2}{Q^2}\right)} \left[\lambda\left(1,\frac{m_V^2}{Q^2},\frac{m_H^2}{Q^2}\right) + 12 \frac{m_V^2}{Q^2}\right],\label{eq:VHinclusive}
\end{align}
with the K\"all\'en function
\beq
\lambda(a,b,c)=(a-b-c)^2-4 b c.
\eeq
The integration over the virtuality can be performed numerically within {\tt n3loxs} (in fact, it can be done for arbitrary invariant mass ranges, see appendix~\ref{app:binned}). 

\subsection{Inclusive cross sections and scale uncertainties}

\begin{table}

\begin{centering}
\begin{tabular}{|c|c|c|c|c|c|c|c|}
\hline 
Process & $\sigma^{\textrm{LO}}$ [pb] & $\sigma^{\textrm{NLO}}$ [pb] & $\textrm{K}^{\textrm{NLO}}$ & $\sigma^{\textrm{NNLO}}$ [pb] & $\textrm{K}^{\textrm{NNLO}}$ & $\sigma^{\textrm{N$^3$LO}}$ [pb] & $\textrm{K}^{\textrm{N$^3$LO}}$\tabularnewline
\hline 
\hline 
$W^{+}H$ & $0.753_{-3.49\%}^{+2.70\%}$ & $0.886_{-1.27\%}^{+1.54\%}$ & $1.18$ & $0.891_{-0.29\%}^{+0.24\%}$ & $1.18$ & $0.883_{-0.34\%}^{+0.29\%}$ & $1.17$\tabularnewline
\hline 
$W^{-}H$ 
& $0.480_{-3.63\%}^{+2.79\%}$ 
& $0.562_{-1.23\%}^{+1.49\%}$ 
& $1.17$ 
& $0.564_{-0.29\%}^{+0.24\%}$ 
& $1.17$ 
& $0.558_{-0.36\%}^{+0.31\%}$ 
& $1.16$\tabularnewline
\hline 
$ZH$ & $0.673_{-3.44\%}^{+2.66\%}$ & $0.788_{-1.23\%}^{+1.48\%}$ & $1.17$ & $0.792_{-0.27\%}^{+0.22\%}$ & $1.18$ & $0.785_{-0.32\%}^{+0.28\%}$ & $1.17$\tabularnewline
\hline 
\end{tabular}
\par\end{centering}
\caption{\label{tab:WZH} Total cross section for associated Higgs production up to the third-order in perturbation theory at a 13 TeV proton-proton collider for a fixed central scale choice $\mu_0=m_V+m_H$. The PDF set {\tt PDF4LHC15\textunderscore nnlo\textunderscore mc} has been used for all predictions, and the K-factors, defined as $\textrm{K}^{\textrm{N}^k \textrm{LO}} = \sigma^{\textrm{N}^k \textrm{LO}}/\sigma^{\textrm{LO}}$ for $k=1,2,3$, are also given. The quoted uncertainties (in percent) are calculated using a 7-point scale variation around the central scale.}
\end{table}

We now present phenomenological results for the LHC. Unless stated otherwise, we use the same setup as for the results in section~\ref{sec:pheno}. In Table~\ref{tab:WZH} we show the total cross section at the LHC with a c.m. energy $\sqrt{s}=13$ TeV. Additional results for other c.m. energies are provided in appendix~\ref{app:vhadditional}. Uncertainties have been computed using a 7-point scale variation by varying the scales around a fixed central scale $\mu_{0} = m_H+m_V$. K-factors, defined as the ratio between higher-order QCD predictions and LO predictions, $\textrm{K}^{\textrm{N}^k \textrm{LO}} = \sigma^{\textrm{N}^k \textrm{LO}}/\sigma^{\textrm{LO}}$ for $k=1,2,3$, are also given. 
We note that the third-order corrections are larger than the NNLO correction, and that the N$^3$LO predictions have a slightly larger scale uncertainty (albeit very small, just like at NNLO already).
This behavior is analogous to NNLO corrections in the Drell-Yan channels~\cite{Duhr:2020sdp,Duhr:2021vwj}, which are based on the same hadronic production cross sections.

\begin{figure}
\includegraphics[width=\textwidth]{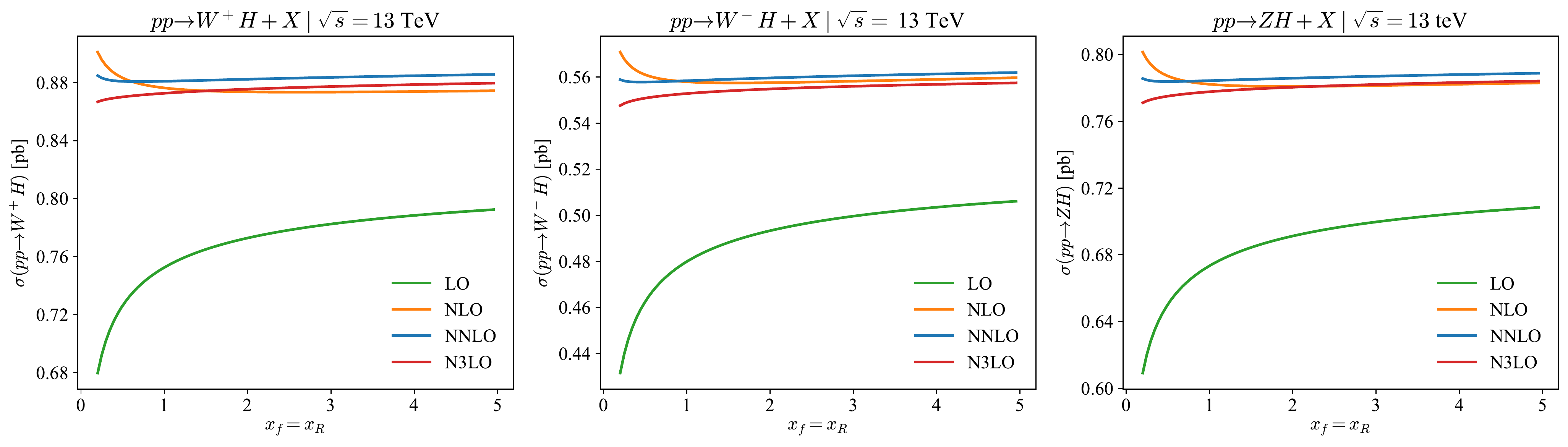}
\caption{\label{fig:WpmZscale} Scale dependence for the associated Higgs production for a fixed central scale choice $\mu_0=m_V+m_H$.
We vary the factorization and renormalization scale in a correlated fashion by varying the parameters $x_F$ and $x_R$ simultaneously (see main text).
}
\end{figure}

Figure~\ref{fig:WpmZscale} shows the dependence of the associated Higgs production on the renormalization and factorization scales set to a common value and varied simultaneously. 
We defined $x_F = \frac{\mu_F}{\mu_0}$ and $x_R= \frac{\mu_R}{\mu_0}$, with $\mu_0 = m_V +m_H$. 
Starting from NNLO, we  observe that the scale dependence is relatively flat. The corrections for  $W^\pm$ and $Z$ exhibit similar dependence. 
Coincidentally, the N$^3$LO correction almost entirely cancels with the NNLO contribution for large values of $x$.

\begin{figure}
  \centering
  \includegraphics[width=0.48\textwidth]{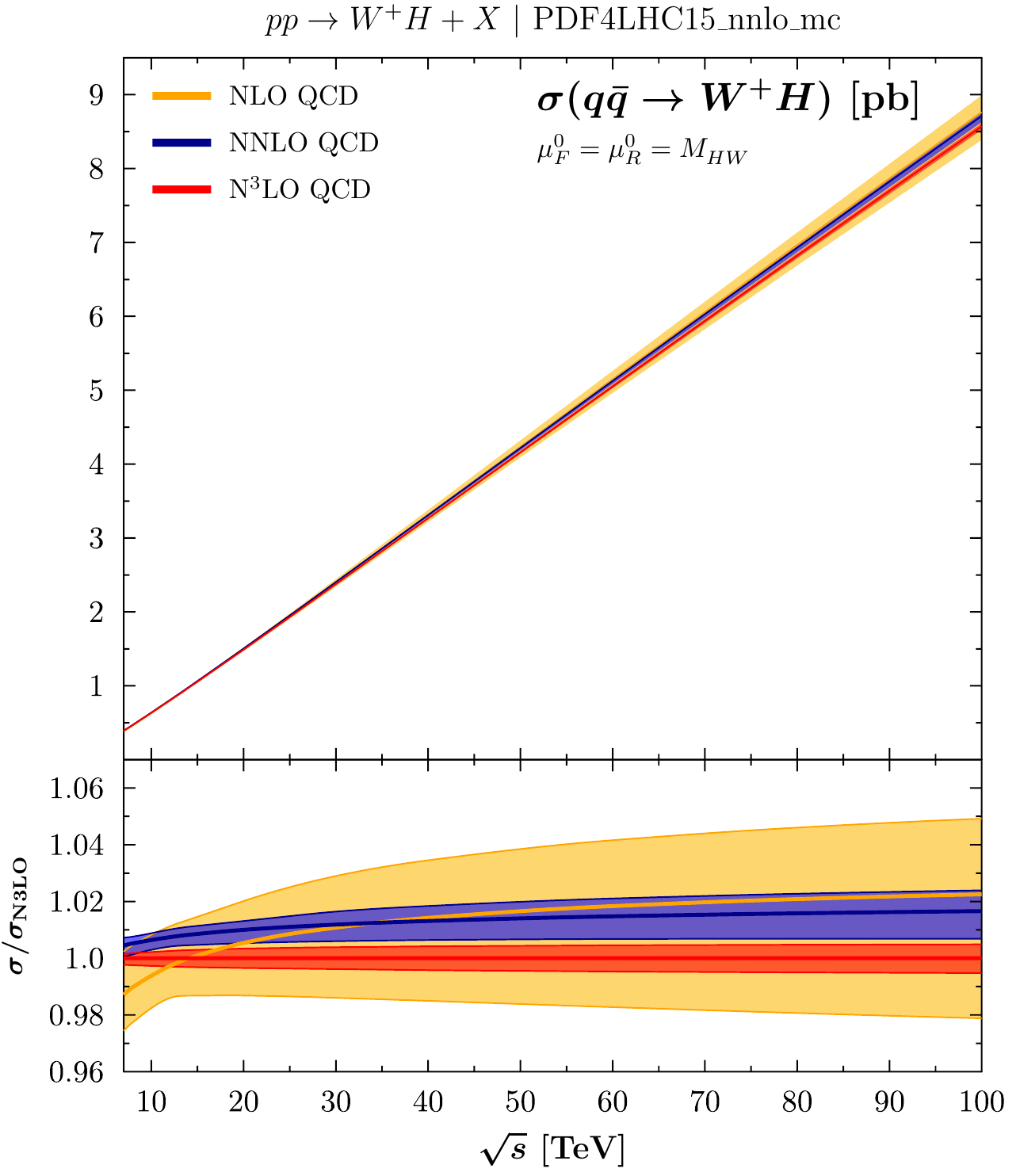}
  \includegraphics[width=0.48\textwidth]{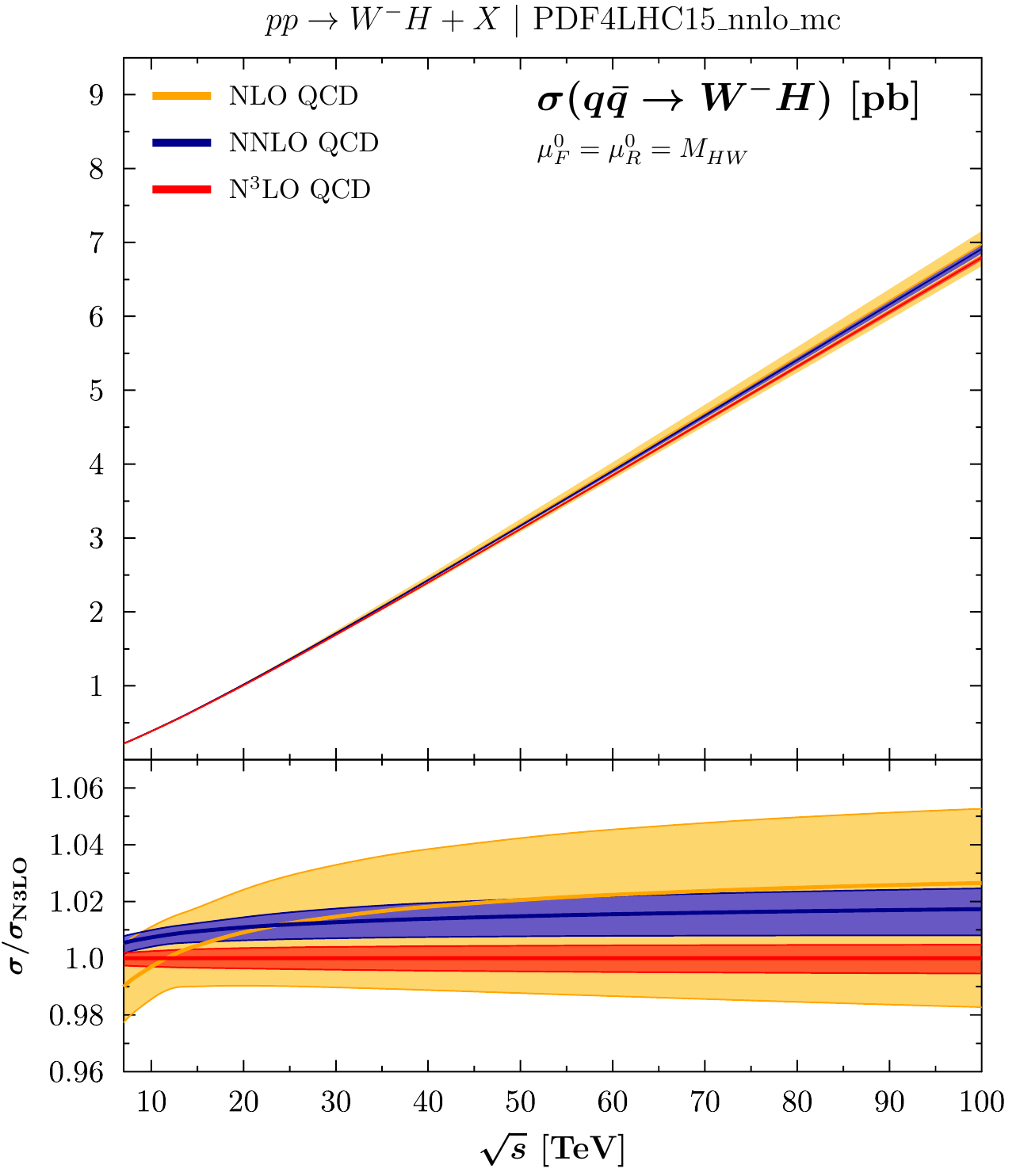}\\
  \includegraphics[width=0.48\textwidth]{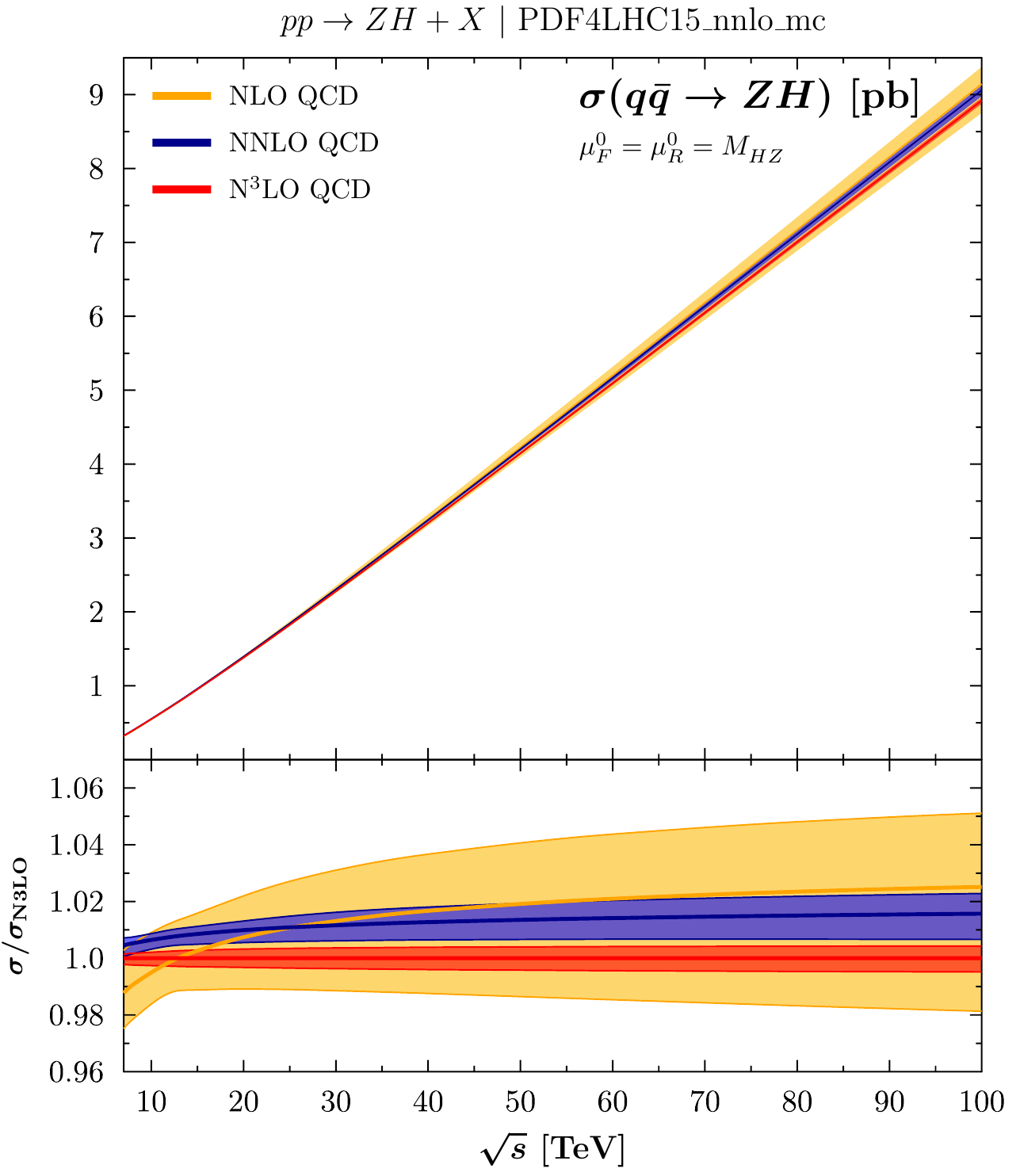}
  \caption{\label{fig:WpmZscale_energy} Inclusive cross sections for the associated Higgs production with a massive gauge boson (in pb) at a proton-proton collider as a function of the c.m. energy (in TeV), up to N$^3$LO in QCD including the 7-point scale uncertainty. All cross sections are calculated with the {\tt PDF4LHC15\textunderscore nnlo\textunderscore mc} PDF set. The lower panels display the ratio to the central N$^3$LO prediction. Upper row: $W^\pm H$ predictions. Lower row: $ZH$ predictions.
  }
\end{figure}

Table~\ref{tab:WZHdynamical} contains the total cross sections and the corresponding K-factors for Higgs associated production using a dynamical central scale which is defined in terms of the invariant mass of the $VH$ pair, $M_{VH}$. Computations are performed again for the a 13 TeV $pp$ (LHC) collider using the \verb+PDF4LHC15_nnlo_mc+ PDF set for all predictions. Typically, when comparing Tables~\ref{tab:WZH} and \ref{tab:WZHdynamical}, both scale choices lead to comparable values of the cross section, with the dynamical scale giving slightly smaller scale uncertainties. Similar to the fixed scale choice, the scale uncertainty is not decreasing when going from NNLO to N$^3$LO and stays quite the same. We also present in fig.~\ref{fig:WpmZscale_energy} the predictions for the inclusive cross section as a function of the c.m. energy at a proton-proton collider, including the 7-point scale uncertainties. We noted before that the NNLO and N$^3$LO bands do not overlap for the Drell-Yann processes. As we calculate DY-type contributions to associated Higgs production, this behavior is expected to follow what has already been observed in DY processes.
\begin{table}

\begin{centering}
\begin{tabular}{|c|c|c|c|c|c|c|c|}
\hline 
Process & $\sigma^{\textrm{LO}}$ [pb] & $\sigma^{\textrm{NLO}}$ [pb] & $\textrm{K}^{\textrm{NLO}}$ & $\sigma^{\textrm{NNLO}}$ [pb] & $\textrm{K}^{\textrm{NNLO}}$ & $\sigma^{\textrm{N$^3$LO}}$ [pb] & $\textrm{K}^{\textrm{N$^3$LO}}$\tabularnewline
\hline 
\hline 
$W^{+}H$
  & $0.758_{-3.13\%}^{+2.43\%}$ 
  & $0.883_{-1.20\%}^{+1.38\%}$ 
  & $1.16$ 
  & $0.891_{-0.34\%}^{+0.28\%}$ 
  & $1.18$
  & $0.884_{-0.30\%}^{+0.27\%}$ 
  & $1.17$\tabularnewline
    \hline 
$W^{-}H$
  & $0.484_{-3.26\%}^{+2.50\%}$
  & $0.560_{-1.23\%}^{+1.34\%}$ 
  & $1.16$ 
  & $0.564_{-0.34\%}^{+0.27\%}$ 
  & $1.17$
  & $0.559_{-0.33\%}^{+0.30\%}$ 
  & $1.16$\tabularnewline
\hline 
$ZH$
  & $0.678_{-3.11\%}^{+2.40\%}$ 
  & $0.786_{-1.16\%}^{+1.33\%}$ 
  & $1.16$ 
  & $0.792_{-0.32\%}^{+0.25\%}$ 
  & $1.17$
  & $0.786_{-0.29\%}^{+0.26\%}$
  & $1.16$\tabularnewline
\hline 
\end{tabular}
\par\end{centering}
\caption{\label{tab:WZHdynamical}
Same as in Table~\ref{tab:WZH}, but for a dynamical central scale choice $\mu_0=M_{VH}$.
}
\end{table}

\subsection{\boldmath Study of the PDF sets and associated PDF, PDF+$\alpha_S$ and PDF-TH uncertainties}

\begin{figure}
\centering
\includegraphics[width=0.85\textwidth]{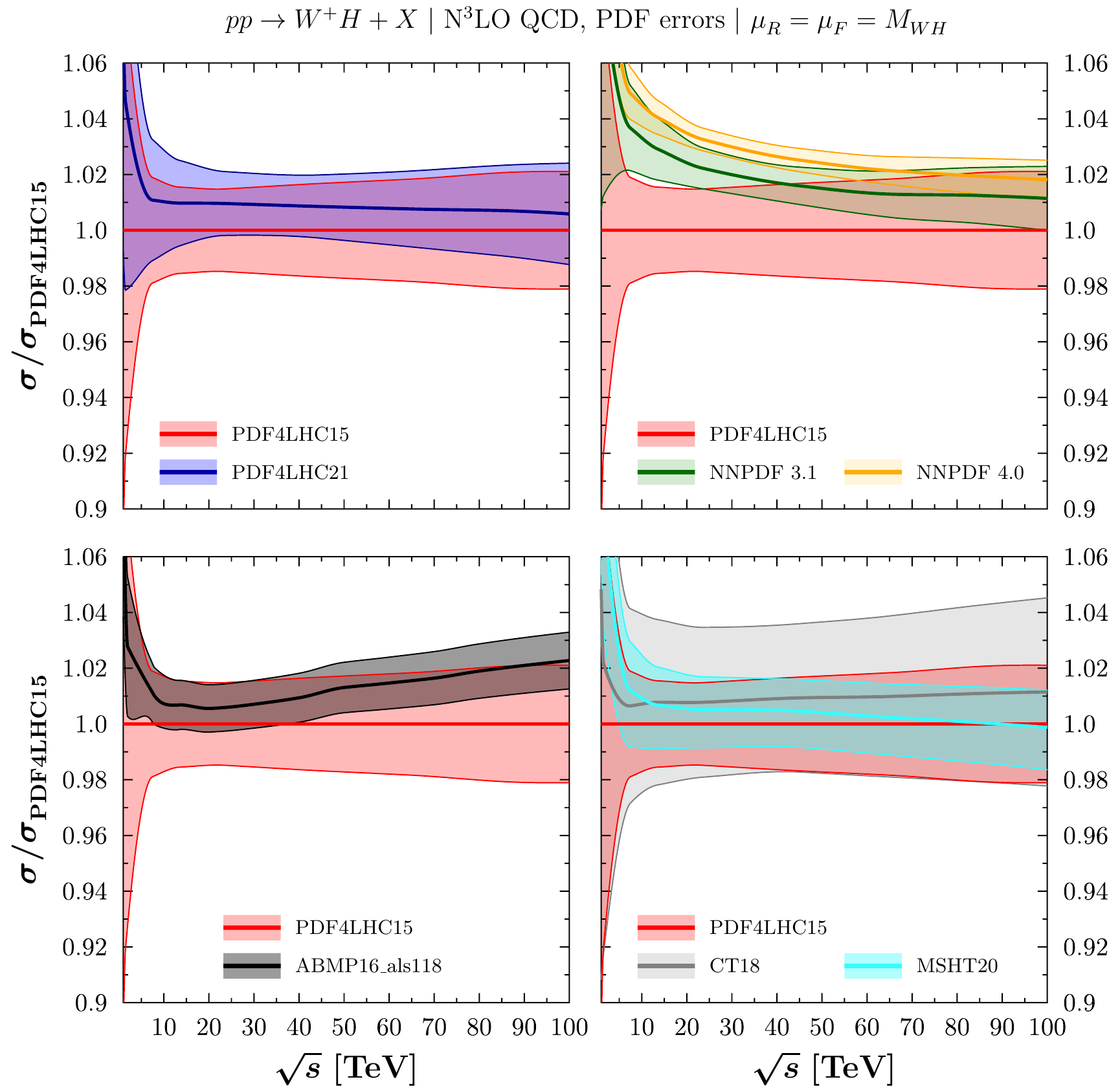}
\caption{\label{fig:PDF_WplusH} Dependence on the choice of the PDF set of the inclusive cross section at N$^3$LO in QCD, for the associated Higgs production with a $W^+$ boson at a proton-proton collider using a dynamical central scale, as a function of the proton-proton c.m. energy $\sqrt{s}$ in TeV. The results are normalized to the central PDF4LHC15 PDF set and the 68\% CL PDF uncertainties are represented by bands for all sets and calculated according to the prescription of ref.~\cite{Butterworth:2015oua}. The comparison is between PDF4LHC15 and: PDF4LHC21 (upper left panel); NNPDF 3.1 and NNPDF 4.0 (upper right panel); ABMP16 als118 (lower left panel); CT18 and MSHT 20 (lower right panel).}
\end{figure}

\begin{figure}
\centering
\includegraphics[width=0.85\textwidth]{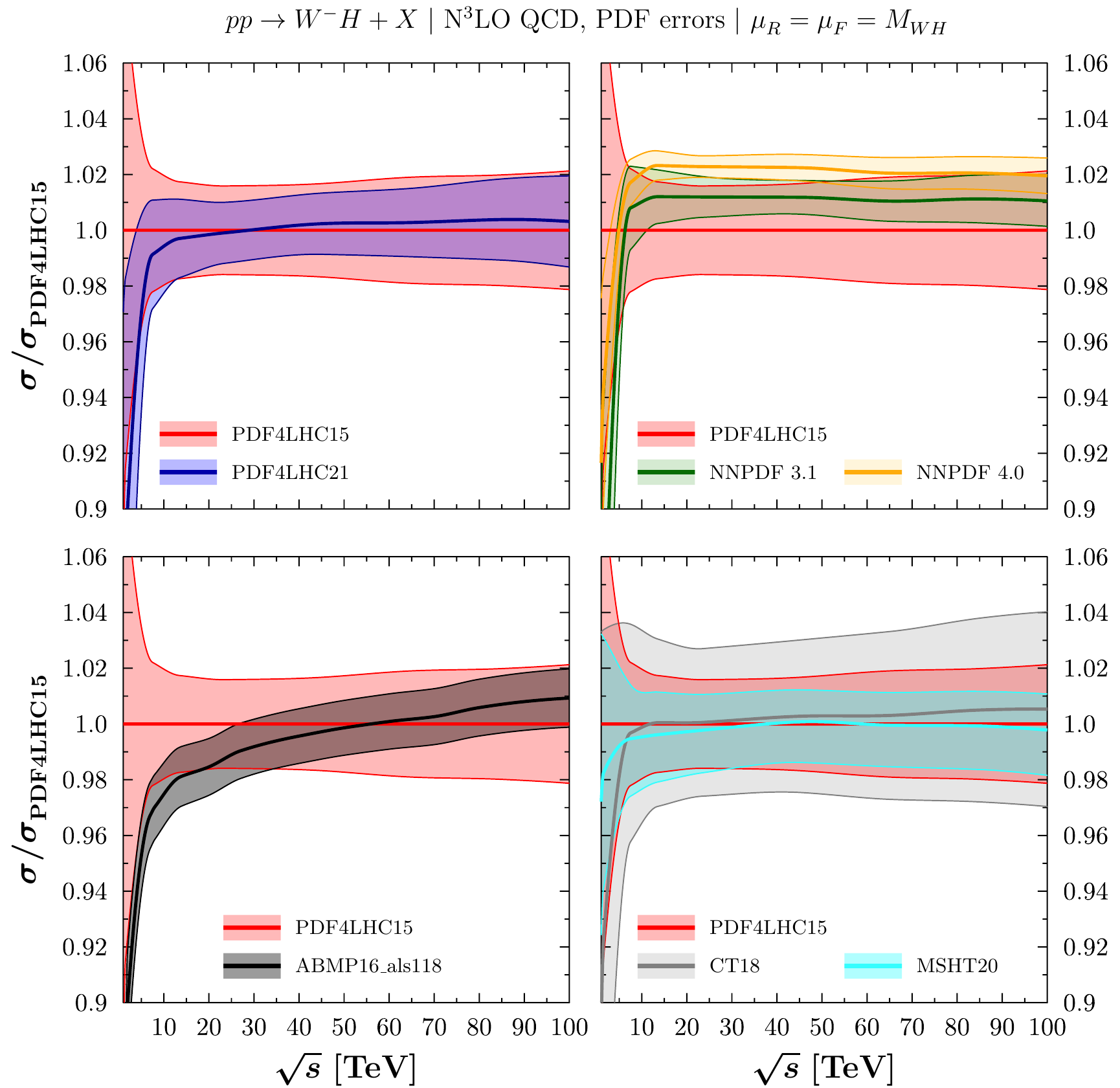}
\caption{\label{fig:PDF_WminusH} Same as in fig.~\ref{fig:PDF_WplusH}, but for $W^- H$ inclusive cross section.}
\end{figure}

\begin{figure}
\centering
\includegraphics[width=0.85\textwidth]{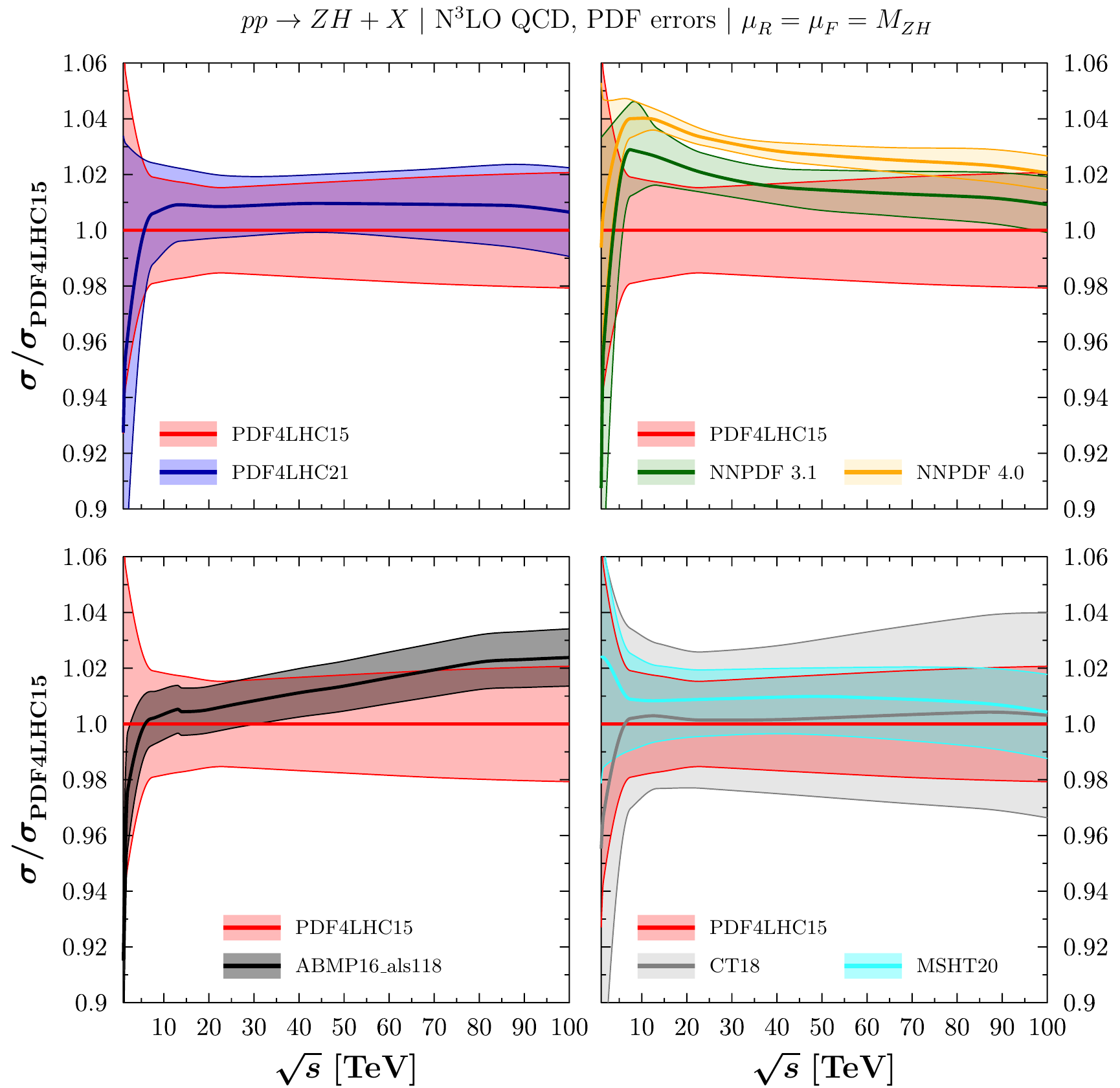}
\caption{\label{fig:PDF_ZH} Same as in fig.~\ref{fig:PDF_WplusH}, but for $Z H$ inclusive cross section.}
\end{figure}

We have also performed an extensive study of the various predictions using different PDF sets, including the 68\% confidence level (CL) PDF uncertainties calculated using the prescription associated with each PDF set, see ref.~\cite{Butterworth:2015oua} for more details. We use the same PDF sets as in section~\ref{sec:pheno}: our default set {\tt PDF4LHC15\textunderscore nnlo\textunderscore mc} (PDF4LHC15)~\cite{Butterworth:2015oua}, against which we compare its latest update  {\tt PDF4LHC21\textunderscore mc} (PDF4LHC21)~\cite{Ball:2022hsh}, the last MSHT update {\tt MSHT20nnlo\textunderscore as118} (MSHT20)~\cite{Bailey:2020ooq}, two NNPDF sets {\tt NNPDF31\textunderscore nnlo\textunderscore as\textunderscore 0118 } (NNPDF~3.1)~\cite{NNPDF:2017mvq} and the latest {\tt NNPDF40\textunderscore nnlo\textunderscore as\textunderscore 01180} (NNPDF~4.0)~\cite{NNPDF:2021njg}, as well as {\tt CT18NNLO} (CT18)~\cite{Hou:2019efy} and {\tt ABMP16als118\textunderscore 5\textunderscore nnlo} (ABMP16\textunderscore als118)~\cite{Alekhin:2017kpj}. Note again that the ABMP set we have chosen is not the nominal set, but the one in which $\alpha_S(m_Z) = 0.118$, thus putting on equal foot all sets with respect to their input value for the strong coupling constant. Figures~\ref{fig:PDF_WplusH}, \ref{fig:PDF_WminusH}, and \ref{fig:PDF_ZH} display the PDF uncertainties for the associated Higgs production with a $W^+$, $W^-$, and $Z$ boson respectively, as a function of the c.m. energy in TeV at a proton-proton collider. All N$^3$LO QCD predictions are normalized to the N$^3$LO QCD result using the central PDF4LHC15 set. We note an overall good agreement between the different PDF sets, though the NNPDF sets show about $1\sigma$ deviation in the range of energies relevant for the LHC when compared to the PDF4LHC15 set. This is a reflection of what has been observed for Drell-Yan processes in section~\ref{sec:sec:pdfDYandHiggs}. Except for very low energies, the PDF uncertainties are of the order of a few percent, and significantly larger for CT18 set than for all the other sets.

We have also performed a dedicated PDF+$\alpha_S$ analysis at the 13 TeV LHC with our nominal PDF4LHC15 set. Following the prescription of ref.~\cite{Butterworth:2015oua} we can obtain the correlated PDF+$\alpha_S$ uncertainty on our predictions, using dedicated sets with $\alpha_S^{1\sigma-}(m_Z) = 0.1165$ and $\alpha_S^{1\sigma+}(m_Z) = 0.1195$ as lower and upper deviations at $1\sigma$ from the central fitted value of the strong coupling constant $\alpha_S(m_Z) = 0.118$. This gives (in percent) the uncertainty
\beq
\delta(\alpha_S) = \frac{\sigma^{\textrm{N}^3\textrm{LO}}(\alpha_S^{1\sigma+}(m_Z)) - \sigma^{\textrm{N}^3\textrm{LO}}(\alpha_S^{1\sigma-}(m_Z))}{2},
\eeq
which combined with $\delta(\textrm{PDF})$ in quadrature gives the uncertainty (in percent) $\delta(\textrm{PDF}+\alpha_S)$,
\beq
\label{eq:deltaPDFas}
\delta(\textrm{PDF}+\alpha_S) = \sqrt{\delta(\textrm{PDF})^2 + \delta(\alpha_S)^2}.
\eeq
On top of this uncertainty, we have to take into account the mismatch at N$^3$LO between the perturbative order of the hard coefficients (at N$^3$LO), and the perturbative order of the PDFs (at NNLO). As currently no PDF set is obtained using data extracted at N$^3$LO accuracy yet, we can use the prescription introduced in ref.~\cite{Anastasiou:2016cez} to account for this mismatch and the error induced by the missing accuracy in the PDF determination and calculate a PDF-TH uncertainty $\delta(\textrm{PDF-TH})$ (in percent). We compare at NNLO the predictions obtained using NNLO and NLO PDF sets, and divide by two the result to account for the fact that the mismatch at N$^3$LO is expected to be smaller than at NNLO,
\beq
\label{eq:deltaPDFTH}
\delta(\textrm{PDF-TH}) = \frac12 \frac{\left|\sigma^{\textrm{NNLO}}(\textrm{NNLO PDF}) - \sigma^{\textrm{NNLO}}(\textrm{NLO PDF})\right|}{\sigma^{\textrm{NNLO}}(\textrm{NNLO PDF})}.
\eeq
Table~\ref{tab:WZH_pdfas} displays the predictions obtained at the LHC at 13 TeV for inclusive Higgsstrahlung cross sections with a dynamical central scale. The PDF+$\alpha_S$ and PDF-TH uncertainties are of comparable size, around $\pm 1.5\%$ to $\pm 2\%$. They are much bigger than the scale uncertainty at N$^3$LO and are the dominant source of theoretical errors at this perturbative order in the QCD expansion. Analogous results for the fixed scale are shown in Table~\ref{tab:WZH_pdfas_fixed}. As expected, the choice of the scale has minimal impact on the PDF and $\alpha_s$ uncertainties.

\begin{table}
\begin{centering}
\begin{tabular}{|c|c|c|c|c|}
\hline 
Process & $\sigma^{\textrm{N$^3$LO}}$ [pb] & $\delta(\textrm{PDF})$ [\%] & $\delta(\textrm{PDF}+\alpha_S)$ [\%] & $\delta(\textrm{PDF-TH})$ [\%]\tabularnewline
\hline 
\hline 
$W^{+}H$ & $0.884$ & $\pm 1.59$ & $\pm 1.80$ & $\pm 1.45$\tabularnewline
\hline 
$W^{-}H$ & $0.559$ & $\pm 1.76$ & $\pm 1.92$ & $\pm 1.63$\tabularnewline
\hline 
$ZH$ & $0.786$ & $\pm 1.77$ & $\pm 1.95$ & $\pm 1.53$\tabularnewline
\hline 
\end{tabular}
\par\end{centering}
\caption{\label{tab:WZH_pdfas} Total cross section for associated Higgs production at N$^3$LO in QCD, at a 13 TeV $pp$ collider for a dynamical central scale choice $\mu_0=M_{HV}$. The PDF set {\tt PDF4LHC15\textunderscore nnlo\textunderscore mc} has been used. The symmetrical PDF, PDF+$\alpha_S$, and PDF-TH uncertainties (in percent) are also given.}
\end{table}

\begin{table}
\begin{centering}
\begin{tabular}{|c|c|c|c|c|}
\hline 
Process & $\sigma^{\textrm{N$^3$LO}}$ [pb] & $\delta(\textrm{PDF})$ [\%] & $\delta(\textrm{PDF}+\alpha_S)$ [\%] & $\delta(\textrm{PDF-TH})$ [\%]\tabularnewline
\hline 
\hline 
$W^{+}H$ & $0.883$ & $\pm 1.59$ & $\pm 1.80$ & $\pm 1.45$\tabularnewline
\hline 
$W^{-}H$ & $0.558$ & $\pm  1.76$ & $\pm 1.93$ & $\pm 1.64 $\tabularnewline
\hline 
$ZH$ & $0.785$ & $\pm 1.82$ & $\pm 1.99$ & $\pm 1.54$\tabularnewline
\hline 
\end{tabular}
\par\end{centering}
\caption{\label{tab:WZH_pdfas_fixed} Total cross section for associated Higgs production at N$^3$LO in QCD, at a 13 TeV $pp$ collider for a fixed central scale choice $\mu_0=M_V + M_H$. The PDF set {\tt PDF4LHC15\textunderscore nnlo\textunderscore mc} has been used. The symmetrical PDF, PDF+$\alpha_S$, and PDF-TH uncertainties (in percent) are also given.}
\end{table}


\section{Comparison of N$^3$LO predictions}
\label{sec:N3LOComp}

In this paper, we have collected phenomenological predictions for a range of diverse processes computed to N$^3$LO in perturbative QCD.
While quantum corrections affect each process in a unique way and their particular form is subject to the kinematic restrictions placed on the process, it is informative to use this range of processes to learn about the qualitative features of perturbative corrections at this order (see also ref.~\cite{Caola:2022ayt}).
The processes considered here fall in the category of production processes - two highly energetic initial-state partons undergo a scattering process to produce a specific final state. 
They are paratypical examples of a much larger range of scattering processes studied at the LHC. 
The universality of QCD radiation leads us to expect similar features of N$^3$LO corrections in other processes that have not yet  been calculated at the same perturbative order.
In particular, examples are multi-boson production processes, like the cross section to produce two $Z$ or $W$ bosons.

\begin{table}[!h]
\begin{center}
\begin{tabular}{|c |c | c | c | c | c | c |}
\hline
 &$Q$ [GeV]&$\delta \sigma^{\text{N$^3$LO}}$&$\delta \sigma^{\text{NNLO}}$ &$\delta(\text{scale}) $&$\delta(\text{PDF}+\alpha_S)$&$\delta(\text{PDF-TH})$ \\
 \hline
{$gg\to ${ Higgs}} & $m_H$ &$ 3.5\%$ &$ 30 \%$ & ${}^{+0.21\%}_{-2.37\%}$ & $\pm 3.2 \%$ & $\pm 1.2 \%$ \\
 \hline
{$b\bar b \to ${ Higgs}} &$m_H$& -2.3\%&$ 2.1\% $ & ${}^{+3.0\%}_{-4.8\%}$ & $\pm 8.4 \%$ & $\pm 2.5 \%$ \\
 \hline
 \multirow{2}{5em}{NCDY}& 30&  -4.8\%&$-0.34\%$ & ${}^{+1.53 \%}_{-2.54 \%}$ & ${}^{+3.7\%}_{-3.8\%}$ & $\pm 2.8 \%$  \\
& 100 &  -2.1\% &$ -2.3\% $ & ${}^{+0.66 \%}_{-0.79 \%}$ & ${}^{+1.8\%}_{-1.9\%}$ & $\pm 2.5 \%$  \\
\hline
 \multirow{2}{5em}{CCDY$(W^+)$}& 30& -4.7\% &$ -0.1\% $ & ${}^{+ 2.5 \%}_{-1.7\%}$ & $\pm 3.95 \%$  & $\pm 3.2 \%$ \\
& 150& -2.0\%&$ -0.1 \%$ &${}^{+0.5 \%}_{-0.5\%}$ & $\pm 1.9 \% $ & $\pm 2.1 \%$\\
\hline
 \multirow{2}{5em}{CCDY$(W^-)$}& 30& -5.0\% &$- 0.1\%$ & ${}^{+ 2.6 \%}_{-1.6\%}$ & $\pm 3.7 \%$ & $\pm 3.2 \%$ \\
& 150& -2.1\%& $-0.6\%$ & ${}^{+0.6 \%}_{-0.5\%}$ & $\pm 2 \%$ & $\pm 2.13 \%$\\
\hline
\end{tabular}
\caption{\label{tab:summary_N3LO} Results for production processes obtained with {\tt n3loxs}. For details, see the discussion in the main text.}
\end{center}
\end{table}
In Table~\ref{tab:summary_N3LO} we compare predictions for N$^3$LO predictions for Higgs production in gluon- and bottom-quark-fusion and charged- and neutral-current Drell-Yan production.
These processes represent examples of gluon- and quark-induced processes at the LHC.
Note that we do not include all available N$^3$LO results for inclusive processes, but we focus on two-to-one inclusive processes at the LHC, which we believe to represent a set of hadron-collider observables that can be compared consistently. In particular, we do not include the $VH$ processes of section~\ref{sec:vh}, because they are based on the same coefficient functions as the Drell-Yan processes. We also do not include the N$^3$LO results for VBF~\cite{Dreyer:2016oyx,Dreyer:2018qbw}, because they are based on DIS coefficient functions, and thus the pattern of QCD corrections is expected to follow the pattern for DIS processes rather than hadron collider processes.

First let us comment on the size of N$^3$LO corrections. 
We define
\beq
\delta \sigma^{\text{N$^n$LO}}_X=\frac{\Sigma_X^{\text{N$^n$LO}}(Q^2)}{\Sigma_X^{\text{N$^{n-1}$LO}}(Q^2)}-1.
\eeq
We observe that for the production processes listed here, $\delta \sigma_X^{\text{N$^3$LO}}$ is at the order of several percent. 
We may contrast this with intrinsic limitations of absolute measurements at the LHC, like the determination of the overall scattering luminosity~\cite{ATLAS:2019pzw,CMS:2021xjt}  or our ability to resolve the energy of final-state radiation~\cite{CMS:2016lmd,ATLAS:2017bje}, and compare this with the anticipated precision that will be reached by the LHC~\cite{Dainese:2019rgk}.
We conclude that phenomenological comparisons at the percent level for certain processes will be possible and that N$^3$LO QCD corrections modify predictions at the same level of precision.
Even higher precision will be achievable by considering ratios of processes that eliminate some of the shared uncertainties and developing predictions for targeted observables to extract physical information will play a crucial role.

Assessing the uncertainty of predictions due to the use of perturbative QFT is crucial. 
N$^3$LO corrections for production processes represent the currently highest order achieved, and the size of these corrections can therefore inform us on the progression of corrections obtained with QFT perturbation theory.
Regarding results for $\delta \sigma_X^{\text{NNLO}}$, we observe that much smaller corrections were obtained at N$^3$LO than NNLO for gluon fusion Higgs production, and that the corrections are of comparable size for quark-initiated processes.
Overall, our interpretation is that N$^3$LO corrections lead to an improved description of scattering processes.

In previous sections, we studied the impact of perturbative scale variations on N$^3$LO processes and the use of these scale variations as uncertainty estimates.
We would like to re-emphasize here the importance of reliable uncertainty estimates and the exploration of alternative methods to set perturbative uncertainties beyond scale variations.
Correlations of uncertainties between different bins of a distribution or between different perturbative observable of the same or different processes will have to be addressed in the future. 
First steps towards alternative methods beyond scale variation were investigated for example in refs.~\cite{Cacciari:2011ze,Bonvini:2020xeo,Duhr:2021mfd}.
Nevertheless, using seven-point scale variations ($\delta(\text{scale})$) to estimate perturbative uncertainties, we observe in Table~\ref{tab:summary_N3LO} that these are at the percent level for the processes we consider.
This observation yet again emphasizes the importance for N$^3$LO QCD computations to achieve percent-level precision in LHC phenomenology.

In the right-most two columns of Table~\ref{tab:summary_N3LO} we collected uncertainties due to missing N$^3$LO PDFs (see eq.~\eqref{eq:deltaPDFTH}) and our imprecise knowledge of PDFs and the strong coupling constant (see eq.~\eqref{eq:deltaPDFas}). 
We observe that the associated uncertainties are of comparable size for all considered processes here and present currently some of the largest theoretical uncertainties on our predictions.
We have studied in previous sections the impact of more modern PDF sets on cross section predictions and found a slight reduction of PDF uncertainties.
Recently, new approximate N$^3$LO PDFs were obtained in ref.~\cite{McGowan:2022nag}, methods to quantify other theoretical uncertainties are discussed in ref.~\cite{Kassabov:2022orn} and new methodologies to fit PDFs are explored for example in ref.~\cite{NNPDF:2021njg}.
We leave a more detailed study of these developments to future work but emphasize the importance of reliable PDF uncertainties due to their dominating role in theoretical predictions at N$^3$LO.

\section{Summary and conclusions}
 \label{sec:conclusion}

We performed a comprehensive phenomenological analysis of the fully-inclusive color-singlet production cross section at hadron colliders up to the third order in the strong coupling constant. Together with this article, we introduce the computer program {\tt n3loxs}, which allows us to numerically calculate inclusive N$^3$LO cross sections for Higgs production in gluon fusion and $b\bar{b}$ annihilation and charged- and neutral-current Drell-Yan processes. In addition, our code allows us to integrate over the virtuality of the final state, which enables us to calculate associated Higgs boson production at N$^3$LO using the known corrections for Drell-Yan processes.
 
We investigated scale uncertainties as a method of assessing theoretical uncertainties due to missing higher orders at different values of the center-of-mass energy and for $pp$ and $p\bar{p}$ collisions. In most of the analyzed cases, the traditional 7-point scale variation at NNLO does not capture the N$^3$LO central value. 
As a consequence, uncertainty estimates for missing higher orders based on scale variation alone should be taken with a grain of salt, already at NNLO. We note that Bayesian methods may offer in the future a viable alternative as they typically provide more significant uncertainties than the scale variation at NNLO, cf., e.g., refs.~\cite{Cacciari:2011ze,Bonvini:2020xeo,Duhr:2021mfd}.

 We compared the N$^3$LO predictions for different PDF sets. We note significant overall progress in the past years in determining the PDF and their uncertainties. While minor observed discrepancies between various sets require dedicated studies within the PDF community, in the meantime a conservative approach would require taking the envelope of the predictions for all the modern sets to account for possible systematic bias. As the third-order QCD corrections become available and the LHC gathers more precise data, the missing N$^3$LO PDF sets are emerging as the biggest obstacle for further improvement of precision phenomenology at the hadron colliders.

 In addition to the Higgs and Drell-Yan cross sections, we presented for the first time associated $VH$ production at N$^3$LO. We find that the third-order correction amounts to about $\sim 1\%$ and is similar in size to the NNLO correction, similar to the case of Drell-Yan production. This is not surprising, given that the $VH$ process involves the same partonic coefficient functions as for the Drell-Yan process. We conclude that the inclusion of the third-order correction is vital for precise phenomenological predictions, and the current theoretical uncertainty is dominated by the parametric and PDF uncertainties, which are typically at the order of 2\%.
 
 Our results discussed in this article, together with the {\tt n3loxs} code which is available
 as a repository at \href{https://github.com/jubaglio/n3loxs}{https://github.com/jubaglio/n3loxs}, will help benchmark precision QCD calculations and phenomenological studies at the LHC. 

\subsection*{Acknowledgments}
RS is supported by the United States Department of Energy under Grant Contract DE-SC0012704.
BM is supported by the United States Department of Energy, Contract DE-AC02-76SF00515.


\appendix
\section{\texttt{n3loxs} }
\label{app:code}
 
{\tt n3loxs} is the cross section calculator developed
for this paper. This appendix presents the code, how to install it, and how to use
it. The default values of the physical parameters are presented in
appendix~\ref{app:parameters}.

{\tt n3loxs} is a public code written mainly in {C++} and
{Python} and allows the user to calculate up to N$^3_{}$LO in QCD
the following observables:
\begin{enumerate}
\item
  Neutral Drell-Yan differential production cross section $Q^2_{}
  \textrm{d}\sigma/\textrm{d}Q^2_{}$ as a function of the virtuality $Q$ of the gauge
  boson, in the leptonic channel\linebreak $p p (p\bar{p})\to \gamma^*_{}/Z^*_{}\to
  \ell^+_{}\ell^-_{} + X$;
\item
  Charged-current Drell-Yan differential production cross section $Q^2_{}
  \textrm{d}\sigma/\textrm{d}Q^2_{}$ as a function of the virtuality $Q$ of the $W$
  boson, in the leptonic channel\linebreak $p p (p\bar{p})\to W^+_{}\to
  \ell^+_{}\nu_\ell^{} + X$ ($p p (p\bar{p})\to W^-_{}\to
  \ell^-_{}\bar{\nu}_\ell^{} + X$);
\item
  Higgs boson inclusive production cross section in the gluon fusion
  channel,\linebreak $g g\to H + X$, both for $pp$ and $p\bar{p}$ colliders;
\item
  Higgs boson inclusive production cross section in the bottom-quark fusion
  channel, in the five-flavor scheme, $b\bar{b}\to H + X$, both for $pp$ and $p\bar{p}$ colliders;
\item
  Higgsstrahlung inclusive production cross section in the charged and
  neutral channels, $p p (p\bar{p})\to W^\pm_{} H + X$ and $p p
  (p\bar{p})\to Z H + X$.
\end{enumerate}
The user can also calculate binned cross sections for Drell-Yan processes
in a given range for $Q$, that is the production cross section
$\sigma(Q_{\text{min}}^{}\leq Q\leq Q_{\text{max}}^{})$, see appendix~\ref{app:binned}.

\subsection{Installation}

The program is written in C++ and Python. It requires a C++11
compatible compiler, Python 3.5 or higher, as well as the {\sc LHAPDF}
library already installed on the user machine. GNU Scientific Library
(gsl) version 2.6 or higher is also required and is shipped with the
code for convenience.

First {\tt gunzip} and {\tt untar} the file {\tt n3loxs.tar.gz} with
{\tt tar -xzf n3loxs.tar.gz}. This will unpack the program into the
directory {\tt ./n3loxs}. Enter the main directory with {\tt cd
  ./n3loxs}, containing the following source files and subdirectories,
\begin{itemize}
\item
  {\tt gsl-2.6}: the subdirectory containing the source code for the
  gsl library;
\item
  {\tt include}: the subdirectory containing the  C++ header files, in
  particular the various physical constants hardcoded in {\tt
    ./include/constants.h};
\item
  {\tt LICENSE}: the GPL 3 license of the program;
\item
  {\tt Makefile}: the script used to compile the program;
\item
  {\tt makegsl.sh}: the script used to compile the gsl library;
\item
  {\tt n3loxs}: the Python main executable of the program;
\item
  {\tt n3loxs\_parameters.in}: the default input file containing
  customizable physical parameters as well as various flags;
\item
  {\tt README.md}: the ReadMe file explaining how to compile and use the
  program;
\item
  {\tt src}: source code subdirectory containing sub-subdirectories
  according to the various processes implemented in the program as
  well as the source code for the strong coupling constant evolution.
\end{itemize}
The user shall start compiling the shipped gsl library with {\tt
  ./makegsl.sh}. Once gsl is compiled, make sure that the {\sc LHAPDF}
config program ({\tt lhapdf-config}) is correctly assigned in the {\tt
  ./Makefile}. This is usually found by your computer when typing {\tt
  lhapdf-config}, but in case the {\sc LHAPDF} library has been installed in
a custom directory, simply change the line
\cpcsub{%
  {\tt LHAPDFCONFIG = lhapdf-config}
}
in the Makefile to
\cpcsub{%
  {\tt LHAPDFCONFIG = [absolute path to the LHAPDF
    installation]/bin/lhapdf-config}
}
After this check all the subprograms needed by the main executable
{\tt n3loxs} can be compiled by typing {\tt make all}. This will
create two subdirectories, {\tt   ./build} containing the {\tt .o}
files, and {\tt ./subprogs} containing the C++ subprograms for each
process.

The main executable, {\tt n3loxs}, accesses the C++ subprograms
located in {\tt ./subprogs}. If the user changes the location of the
main executable {\tt n3loxs}, they have to update the 52nd line of the
script so that it accesses correctly the subprograms. Instead of the
line
\cpcsub{%
  {\tt maindir=maindirpid.stdout.strip('\textbackslash{}n')},
}
it has to be replaced by the absolute path to the directory {\tt
  n3loxs}
\cpcsub{%
  {\tt maindir='[absolute path of the directory n3loxs on the
    computer]'}
}

\subsection{Usage}

The program accepts up to three arguments on the command line, as well as
an optional flag:
\begin{itemize}
\item
  {\tt -lattice lattice}: The lattice size is an integer used for the
  integration. If this argument is omitted the default value is {\tt
    lattice=1000}.
\item
  {\tt -seed seed}: The seed is an integer that is used to initialize
  the pseudo-random-number generator. If the argument is omitted the
  default value is {\tt seed=1}.
\item
  {\tt  --filename filename}: The name of the input file if the user
  does not want to use the default file, {\tt
    n3loxs\_parameters.in}. Note that the user input file needs to have
  the same structure as the default input file, otherwise the program
  will not run correctly and produce ill-defined results.
\item
  {\tt  --scale} or {\tt --7point}: An optional flag to calculate in a
  single call to the program 16 different predictions for the
  renormalization scale being varied between 0.5 and 2 times the
  chosen central scales of the process, while the factorization scale
  is fixed by the user at a given value ({\tt --scale} flag); 
  or an optional flag to calculate the seven-point scale
  variation around the central scales of the process ({\tt --7point}
  flag). The flags are mutually exclusive.
\end{itemize}
Note that the default lattice may be too coarse for high precision
studies requiring a precision at the sub-percent level. This happens for example for scale variation studies. It is advised to use at least {\tt
  lattice=10000} in such cases.

The user can modify a handful of parameters. The default values for
the physical parameters are given in
appendix~\ref{app:parameters}. The other parameters give the user
control over the calculational framework:
\begin{itemize}
\item
  {\tt process}
    Selector of the desired process to be
  calculated. Following options are available:
  \begin{center}
  \begin{longtable}{p{0.14 \textwidth} p{0.75 \textwidth}}
  {\tt process=1}: & neutral Drell-Yan differential
  production cross section at a given $Q$ value; \\
  {\tt process=2}: & charged Drell-Yan $W^+_{}$ differential production
  cross section at a given $Q$ value; \\

  {\tt process=3}: & charged Drell-Yan $W^-_{}$ differential production
  cross section at a given $Q$ value; \\
  {\tt process=4}: & Higgsstrahlung $W^+_{} H$ inclusive production
  cross section, fixed scale;\\
  {\tt process=5}: & Higgsstrahlung $W^-_{} H$ inclusive production
  cross section, fixed scale;\\
  {\tt process=6}:  & Higgsstrahlung $Z H$ inclusive production
  cross section, fixed scale;\\
  {\tt process=7}: & Five-Flavor scheme $b\bar{b}\to H$ inclusive
  production cross section;\\
  {\tt process=8}: & Gluon fusion $g g\to H$ inclusive
  production cross section, either in the heavy-top limit (HTL) or in
  the Born-improved HTL;\\
  {\tt process=9}: &neutral Drell-Yan production cross section in the
  bin $Q_{\text{min}}^{}\leq Q\leq Q_{\text{max}}^{}$; \\
    {\tt process=10}: & charged Drell-Yan $W^+_{}$ production cross
  section in the bin $Q_{\text{min}}^{}\leq Q\leq Q_{\text{max}}^{}$;\\
  {\tt process=11}: & charged Drell-Yan $W^-_{}$ production cross
  section in the bin $Q_{\text{min}}^{}\leq Q\leq Q_{\text{max}}^{}$; \\
  {\tt process=12}: & Higgsstrahlung $W^+_{} H$ inclusive production
  cross section, dynamical scale;\\
  {\tt process=13}: & Higgsstrahlung $W^-_{} H$ inclusive production
  cross section, dynamical scale;\\
  {\tt process=14}: & Higgsstrahlung $Z H$ inclusive production
  cross section, dynamical scale.
      \end{longtable}
  \end{center}
\item
  {\tt order}: Perturbative order in QCD up to which the calculation
  will be performed. {\tt order=0} for LO calculation, up
  to {\tt order=3} for N$^3_{}$LO calculation. The default choice in
  the input file is {\tt order=3}.
\item
  {\tt collider}: Choice of the collider type. {\tt collider=0} for a
  $pp$ collider (LHC), {\tt collider=1} for a $p\bar{p}$
  (Tevatron). The default choice in the input file is {\tt
    collider=0}.
\item
  {\tt muf0}: User-defined central factorization scale $\mu_F^0$, in
  GeV. If {\tt muf0=-1} (default choice in the input file), $\mu_F^0$
  is fixed internally to the default central scale of the process (see below).
\item
  {\tt xmuf}: Ratio of the factorization scale $\mu_F^{}$ used in the calculation and the previously defined central factorization scale $\mu_F^0$, such that $x_{\mu_F} = \mu_F/\mu_F^0$. The default choice in the input file is {\tt xmuf=1.0}, which corresponds to the factorization scale used in the calculation equal to the central scale $\mu_F=\mu_F^0$. This parameter is useful for calculating the factorization scale variation. Note that this is equivalent to a direct change of the central scale. 
\item
  {\tt mur0}: User-defined central renormalization scale $\mu_R^0$, in
  GeV. If {\tt mur0=-1} (default choice in the input file), $\mu_R^0$
  is fixed internally to the default central scale of the process (see below).
\item
  {\tt xmur}: Ratio of the renormalization scale $\mu_R^{}$ used in the calculation and the previously defined central renormalization scale $\mu_R^0$, such that $x_{\mu_R} = \mu_R/\mu_R^0$. The default choice in the input file is {\tt xmur=1.0}, which corresponds to the renormalization scale used in the calculation equal to the central scale $\mu_R=\mu_R^0$. This parameter is useful for calculating the renormalization scale variation. Note that this is equivalent to a direct change of the central scale. .

 The default central scales are defined as follows: 
 \begin{itemize}
\item
  Neutral- and charged-current Drell-Yan processes: $\mu_F^0=\mu_R^0=Q$.
\item
  Charged Higgsstrahlung processes (fixed scale): $\mu_F^0=\mu_R^0=m_H^{}+m_W^{}$.
\item
  Charged Higgsstrahlung processes (dynamical scale): $\mu_F^0=\mu_R^0=M_{HW}^{}$.
\item
  Neutral Higgsstrahlung process (fixed scale): $\mu_F^0=\mu_R^0=m_H^{}+m_Z^{}$.
\item
  Neutral Higgsstrahlung process (dynamical scale): $\mu_F^0=\mu_R^0=M_{HZ}^{}$.
\item
  Bottom-quark fusion Higgs process: $\mu_F^0 = \frac14\left(m_H^{}+2
    m_b^{}\right)$, $\mu_R^0 = m_H^{}$.
\item
  Gluon fusion Higgs process: $\mu_F^0=\mu_R^0=\frac12 m_H^{}$.
\end{itemize}
 
\item
  {\tt mt\_scheme}: Flag to control the framework of the gluon fusion
  process. If {\tt mt\_scheme=0}, the calculation is performed in the
  on-shell scheme for the top-quark mass, if {\tt mt\_scheme=1}, the
  $\overline{\text{MS}}$ scheme is used. The default choice in the
  input file is {\tt mt\_scheme=1}.
\item
  {\tt htl\_flag}: Flag to control the framework of the gluon fusion
  process. If {\tt htl\_flag=0}, the calculation is performed in the
  pure heavy-top limit (HTL). If {\tt htl\_flag=1}, the calculation is
  performed in the Born-improved HTL, where the results are rescaled
  by the full LO Born matrix elements. The default choice in the input
  file is {\tt htl\_flag=1}.
\item
  {\tt ncdy\_flag}: Flag to control the framework of the neutral
  Drell-Yan process. If {\tt ncdy\_flag=0}, only the off-shell photon
  contribution is taken into account. If {\tt ncdy\_flag=1}, all
  contributions are taken into account ($Z$, $\gamma^*_{}$, the
  interference terms, as well as the non-decoupling top contributions due to the axial anomaly). The
  default choice in the input file is {\tt ncdy\_flag=1}.
\end{itemize}

\section{Parameters}
\label{app:parameters}
Below we summarize all parameters used throughout the paper. These
parameter values are used as default values in the \texttt{n3loxs}
program. The code uses some hardwired parameters located in the file
\texttt{include/constants.h}, whose modification requires
recompilation of the code. In particular, it uses $n_f^{}=5$ for the
number of massless quarks used in the perturbative coefficients as
well as in the $\beta$ functions and anomalous dimensions. The
Cabibbo–Kobayashi–Maskawa matrix elements are also stored in the same
file and read
\begin{align}
  |V_{ud}^{}| = 0.97446,\quad  |V_{us}^{}| = 0.22452,\quad  |V_{ub}^{}|
  = 0.00365,\nonumber\\
  |V_{cd}^{}| = 0.22438, \quad |V_{cs}^{}| = 0.97359,\quad |V_{cb}^{}|
  = 0.04214, \nonumber\\
  |V_{td}^{}| = 0.00896, \quad |V_{ts}^{}| = 0.04133,\quad |V_{tb}^{}|
  = 0.999105.
\end{align}

In addition, there are other physical parameters that can be changed
by the user in the file \texttt{n3loxs\_parameters.in} which is the
main input file of the code. Any parameter and flag in this file can
be modified at will without recompiling the code. The default values
of these parameters are as follow,
\bea
v&=&246.221~\text{GeV}\,,\nonumber\\
m_W&=&80.398~\text{GeV}\,,\nonumber\\
m_Z&=&91.1876~\text{GeV}\,,\nonumber\\
m_H&=&125.09~\text{GeV}\,,\nonumber\\
m_t&=&172.5~\text{GeV}\,,\nonumber\\
m_b&=&4.58~\text{GeV}\,,\nonumber\\
\overline{m}_t(\overline{m}_t)&=&162.7~\text{GeV}\,,\nonumber\\
\overline{m}_b(\overline{m}_b)&=&4.18~\text{GeV}\,,\nonumber\\
\Gamma_W&=&2.085~\text{GeV}\,,\nonumber\\
\Gamma_Z&=&2.4952~\text{GeV}\,,\nonumber\\
\alpha_S(m_Z)&=&0.118\,, \nonumber\\
1/\alpha(0)  &=&137.035999084\,, \nonumber\\
E_{\text{collider}}&=&13~\text{TeV} \,,\nonumber\\
Q&=&100~\text{GeV}\,, \quad Q_{\text{min}}=80~\text{GeV}\,, \quad
Q_{\text{max}}=90~\text{GeV} \,, \nonumber\\
\text{PDF}&=&\text{PDF4LHC15\_nnlo\_mc}\,, \quad \text{PDFset}=0\,.
\eea
Note that $\alpha_S(m_Z)$ is automatically chosen by the code in
accordance with the PDF set selected by the user: This
parameter cannot be changed independently and does not appear in the
file \texttt{n3loxs\_parameters.in}. For Drell-Yan processes, either
$Q$ or the set $Q_{\text{min}}^{}/Q_{\text{max}}^{}$ is used, but not
both at the same time, depending on the process chosen.
%
%
Finally, let us note following relations:
\bea
&&g_W=\sqrt{\frac{8 m_W^2 G_F}{\sqrt{2}}},\hspace{1cm}m_Z=\frac{m_W}{\cos\theta_W},\hspace{1cm}G_F=\frac{1}{\sqrt{2}v^2},\hspace{1cm}\nonumber\\
&&e=g_W\sin\theta_W,\hspace{1cm}\alpha=\frac{m_W^2\left(1-\frac{m_W^2}{m_Z^2}\right)}{\pi v^2}.
\eea
Note that the code uses either $1/\alpha(0)$ or the vacuum expectation
value $v$, depending on the process chosen, but not both parameters at
the same time.
\section{\boldmath Inclusive cross sections for $VH$ processes at
  various additional center-of-mass energies}
\label{app:vhadditional}

We present in this appendix inclusive cross sections at N$^3$LO in QCD
for Higgsstrahlung production, $p p\to V
H$, for c.m. energies $\sqrt{s} = 7, 8, 14$~TeV (LHC), 27~TeV (HE-HLC)
and 100~TeV (FCC-hh). Results for the $W^+H$ process can be found in
Table~\ref{tab:WplusH_additional}, for the $W^-H$ process in
Table~\ref{tab:WminusH_additional}, and for the $ZH$ process in
Table~\ref{tab:ZH_additional}.

\begin{table}[!ht]
\begin{centering}
\begin{tabular}{|c|c|c|c|c|c|}
\hline 
$\sqrt{s}$ [TeV] & $\sigma^{\textrm{N$^3$LO}}(W^+H)$ [pb] & $\delta(\textrm{scale})$ [\%]
  & $\delta(\textrm{PDF})$ [\%] & $\delta(\textrm{PDF}+\alpha_S)$ [\%] & $\delta(\textrm{PDF-TH})$ [\%]\tabularnewline
\hline 
\hline 
7 & $0.390$ & $^{+0.15}_{-0.23}$ & $\pm 1.92$ & $\pm 2.02$ & $\pm 0.91$\tabularnewline
\hline 
8 & $0.470$ & $^{+0.18}_{-0.25}$ & $\pm 1.86$ & $\pm 1.98$ & $\pm 1.05$\tabularnewline
\hline 
14 & $0.969$ & $^{+0.29}_{-0.31}$ & $\pm 1.51$ & $\pm 1.75$ & $\pm 1.49$\tabularnewline
\hline 
27 & $2.11$ & $^{+0.38}_{-0.37}$ & $\pm 1.45$ & $\pm 1.79$ & $\pm 1.70$\tabularnewline
\hline 
100 & $8.56$ & $^{+0.49}_{-0.52}$ & $\pm 2.11$ & $\pm 2.51$ & $\pm 1.36$\tabularnewline
\hline 
\end{tabular}
\par\end{centering}
\caption{\label{tab:WplusH_additional} Total cross section for
  associated Higgs production with a $W^+$ boson at N$^3$LO in QCD, at
  a proton-proton collider at various c.m. energies $\sqrt{s}$ in TeV
  for a dynamical central scale choice $\mu_0=M_{HW}$. The PDF set
  {\tt PDF4LHC15\textunderscore nnlo\textunderscore mc} has been
  used. The 7-point scale uncertainty as well as the symmetrical PDF,
  PDF+$\alpha_S$, and PDF-TH uncertainties (in percent) are also
  given.}
\end{table}

\begin{table}[!ht]
\begin{centering}
\begin{tabular}{|c|c|c|c|c|c|}
\hline 
$\sqrt{s}$ [TeV] & $\sigma^{\textrm{N$^3$LO}}(W^-H)$ [pb] & $\delta(\textrm{scale})$ [\%]
  & $\delta(\textrm{PDF})$ [\%] & $\delta(\textrm{PDF}+\alpha_S)$ [\%] & $\delta(\textrm{PDF-TH})$ [\%]\tabularnewline
\hline 
\hline 
7 & $0.218$ & $^{+0.22}_{-0.31}$ & $\pm 2.28$ & $\pm 2.35$ & $\pm 1.24$\tabularnewline
\hline 
8 & $0.271$ & $^{+0.24}_{-0.32}$ & $\pm 2.14$ & $\pm 2.23$ & $\pm 1.34$\tabularnewline
\hline 
14 & $0.620$ & $^{+0.32}_{-0.36}$ & $\pm 1.72$ & $\pm 1.90$ & $\pm 1.66$\tabularnewline
\hline 
27 & $1.48$ & $^{+0.39}_{-0.43}$ & $\pm 1.57$ & $\pm 1.85$ & $\pm 1.82$\tabularnewline
\hline 
100 & $6.79$ & $^{+0.46}_{-0.58}$ & $\pm 2.13$ & $\pm 2.50$ & $\pm 1.58$\tabularnewline
\hline 
\end{tabular}
\par\end{centering}
\caption{\label{tab:WminusH_additional} The same as in
  Table~\ref{tab:WplusH_additional}, but for the $W^-H$ process.}
\end{table}


\begin{table}[!ht]
\begin{centering}
\begin{tabular}{|c|c|c|c|c|c|}
\hline 
$\sqrt{s}$ [TeV] & $\sigma^{\textrm{N$^3$LO}}(ZH)$ [pb] & $\delta(\textrm{scale})$ [\%]
  & $\delta(\textrm{PDF})$ [\%] & $\delta(\textrm{PDF}+\alpha_S)$ [\%] & $\delta(\textrm{PDF-TH})$ [\%]\tabularnewline
\hline 
\hline 
7 & $0.322$ & $^{+0.15}_{-0.22}$ & $\pm 1.94$ & $\pm 2.02$ & $\pm 1.01$\tabularnewline
\hline 
8 & $0.395$ & $^{+0.18}_{-0.24}$ & $\pm 1.88$ & $\pm 1.98$ & $\pm 1.15$\tabularnewline
\hline 
14 & $0.869$ & $^{+0.28}_{-0.30}$ & $\pm 1.69$ & $\pm 1.88$ & $\pm 1.58$\tabularnewline
\hline 
27 & $2.01$ & $^{+0.36}_{-0.36}$ & $\pm 1.56$ & $\pm 1.87$ & $\pm 1.80$\tabularnewline
\hline 
100 & $8.93$ & $^{+0.43}_{-0.48}$ & $\pm 2.07$ & $\pm 2.47$ & $\pm 1.54$\tabularnewline
\hline 
\end{tabular}
\par\end{centering}
\caption{\label{tab:ZH_additional} The same as in
  Table~\ref{tab:WplusH_additional}, but for the $ZH$ process and with $\mu_0=M_{HZ}$.}
\end{table}


\begin{table}[!ht]
    \begin{centering}
    \begin{tabular}{|c|c|c|c|c|c|}
    \hline 
    $\sqrt{s}$ [TeV] & $\sigma^{\textrm{N$^3$LO}}(W^+H)$ [pb] & $\delta(\textrm{scale})$ [\%]& 
    $\delta(\textrm{PDF})$ [\%] & $\delta(\textrm{PDF}+\alpha_S)$ [\%] & $\delta(\textrm{PDF-TH})$ [\%]\tabularnewline
    \hline 
    \hline
7 & $ 0.390 $ & $^{+ 0.19 }_{ -0.28 }$ & $\pm  1.94 $ & $\pm 2.04 $ & $ \pm  0.92 $\tabularnewline 
\hline
8 & $ 0.470 $ & $^{+ 0.22 }_{ -0.30 }$ & $\pm  1.87 $ & $\pm 1.99 $ & $ \pm  1.06 $\tabularnewline 
\hline
14 & $ 0.968 $ & $^{+ 0.30 }_{ -0.35 }$ & $\pm  1.57 $ & $\pm 1.80 $ & $ \pm  1.50 $\tabularnewline 
\hline
27 & $ 2.11 $ & $^{+ 0.38 }_{ -0.41 }$ & $\pm  1.47 $ & $\pm 1.81 $ & $ \pm  1.71 $\tabularnewline 
\hline
100 & $ 8.56 $ & $^{+ 0.46 }_{ -0.57 }$ & $\pm  2.21 $ & $\pm 2.59 $ & $ \pm  1.34 $\tabularnewline 
\hline
\end{tabular}
    \par\end{centering}
    \caption{\label{tab:WplusH_additional_fixed} The same as in
  Table~\ref{tab:WplusH_additional}, but with $\mu_0=M_{H}+M_W$.}
\end{table}


\begin{table}[!ht]
    \begin{centering}
    \begin{tabular}{|c|c|c|c|c|c|}
    \hline 
    $\sqrt{s}$ [TeV] & $\sigma^{\textrm{N$^3$LO}}(W^-H)$ [pb] & $\delta(\textrm{scale})$ [\%]& 
    $\delta(\textrm{PDF})$ [\%] & $\delta(\textrm{PDF}+\alpha_S)$ [\%] & $\delta(\textrm{PDF-TH})$ [\%]\tabularnewline
    \hline 
    \hline
7 & $ 0.218 $ & $^{+ 0.22 }_{ -0.31 }$ & $\pm  2.28 $ & $\pm 2.35 $ & $ \pm  1.25 $\tabularnewline 
\hline
8 & $ 0.271 $ & $^{+ 0.24 }_{ -0.32 }$ & $\pm  2.15 $ & $\pm 2.23 $ & $ \pm  1.35 $\tabularnewline 
\hline
14 & $ 0.620 $ & $^{+ 0.32 }_{ -0.36 }$ & $\pm  1.74 $ & $\pm 1.92 $ & $ \pm  1.67 $\tabularnewline 
\hline
27 & $ 1.48 $ & $^{+ 0.39 }_{ -0.43 }$ & $\pm  1.70 $ & $\pm 1.96 $ & $ \pm  1.83 $\tabularnewline 
\hline
100 & $ 6.78 $ & $^{+ 0.45 }_{ -0.58 }$ & $\pm  2.18 $ & $\pm 2.54 $ & $ \pm  1.57 $\tabularnewline 
\hline
\end{tabular}
    \par\end{centering}
    \caption{\label{tab:WminusH_additional_fixed} The same as in Table~\ref{tab:WminusH_additional}, but with $\mu_0=M_H +M_W$.}
\end{table}


\begin{table}[!ht]
    \begin{centering}
    \begin{tabular}{|c|c|c|c|c|c|}
    \hline 
    $\sqrt{s}$ [TeV] & $\sigma^{\textrm{N$^3$LO}}(ZH)$ [pb] & $\delta(\textrm{scale})$ [\%]& 
    $\delta(\textrm{PDF})$ [\%] & $\delta(\textrm{PDF}+\alpha_S)$ [\%] & $\delta(\textrm{PDF-TH})$ [\%]\tabularnewline
    \hline 
    \hline
7 & $ 0.322 $ & $^{+ 0.18 }_{ -0.27 }$ & $\pm  1.98 $ & $\pm 2.06 $ & $ \pm  1.02 $\tabularnewline 
\hline
8 & $ 0.395 $ & $^{+ 0.21 }_{ -0.29 }$ & $\pm  1.88 $ & $\pm 1.98 $ & $ \pm  1.16 $\tabularnewline 
\hline
14 & $ 0.868 $ & $^{+ 0.29 }_{ -0.33 }$ & $\pm  1.71 $ & $\pm 1.90 $ & $ \pm  1.58 $\tabularnewline 
\hline
27 & $ 2.01 $ & $^{+ 0.35 }_{ -0.39 }$ & $\pm  1.57 $ & $\pm 1.87 $ & $ \pm  1.81 $\tabularnewline 
\hline
100 & $ 8.92 $ & $^{+ 0.40 }_{ -0.52 }$ & $\pm  2.11 $ & $\pm 2.50 $ & $ \pm  1.52 $\tabularnewline 
\hline
\end{tabular}
    \par\end{centering}
    \caption{\label{tab:ZH_additional_fixed} The same as in Table~\ref{tab:ZH_additional}, but for $\mu_0=M_H+M_Z$. }
\end{table}


\section{Binned cross sections for Drell-Yan processes}
\label{app:binned}
\begin{figure}[!h]
\begin{center}
\includegraphics[width=0.32 \textwidth]{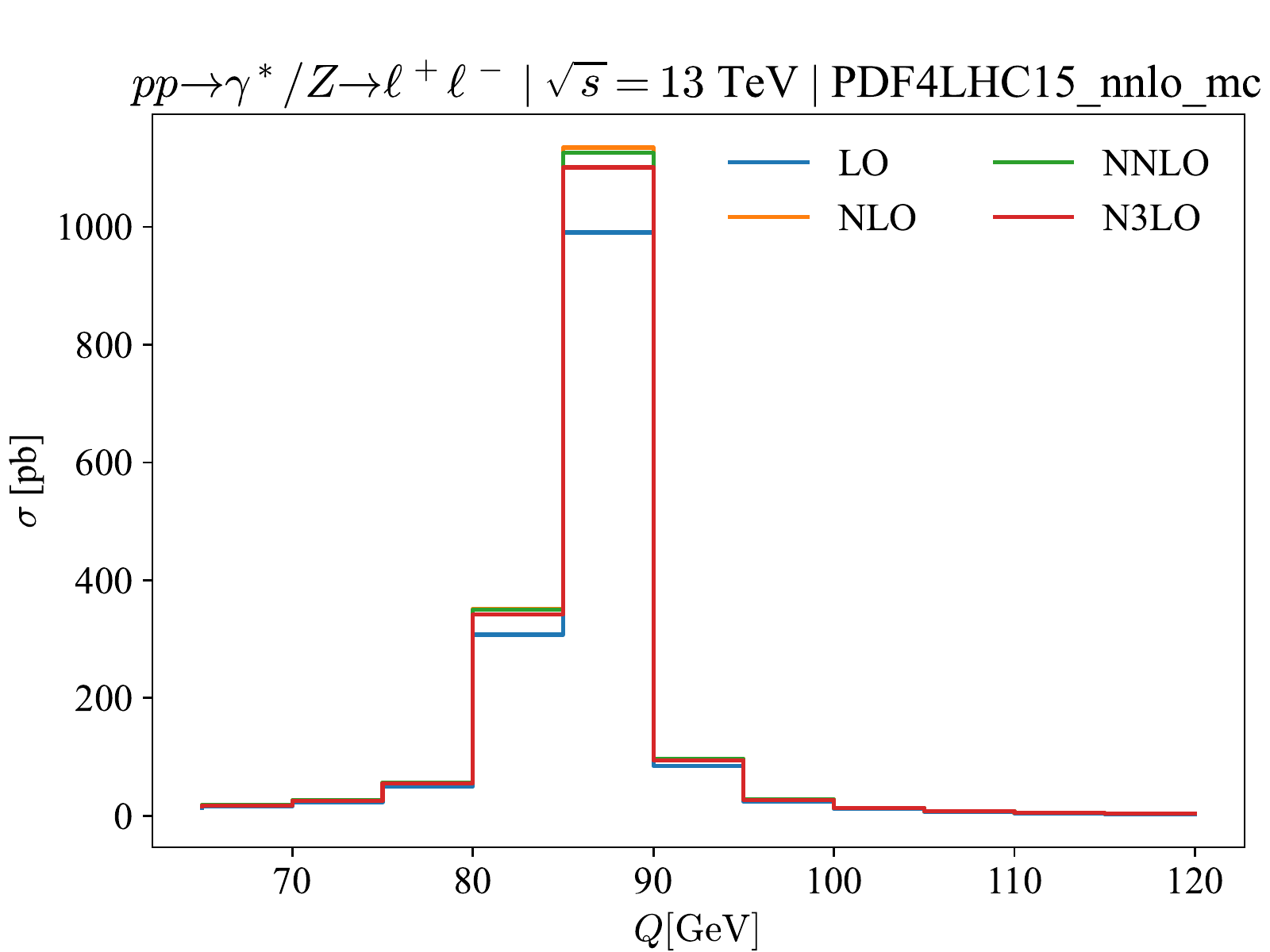} 
\includegraphics[width=0.32 \textwidth]{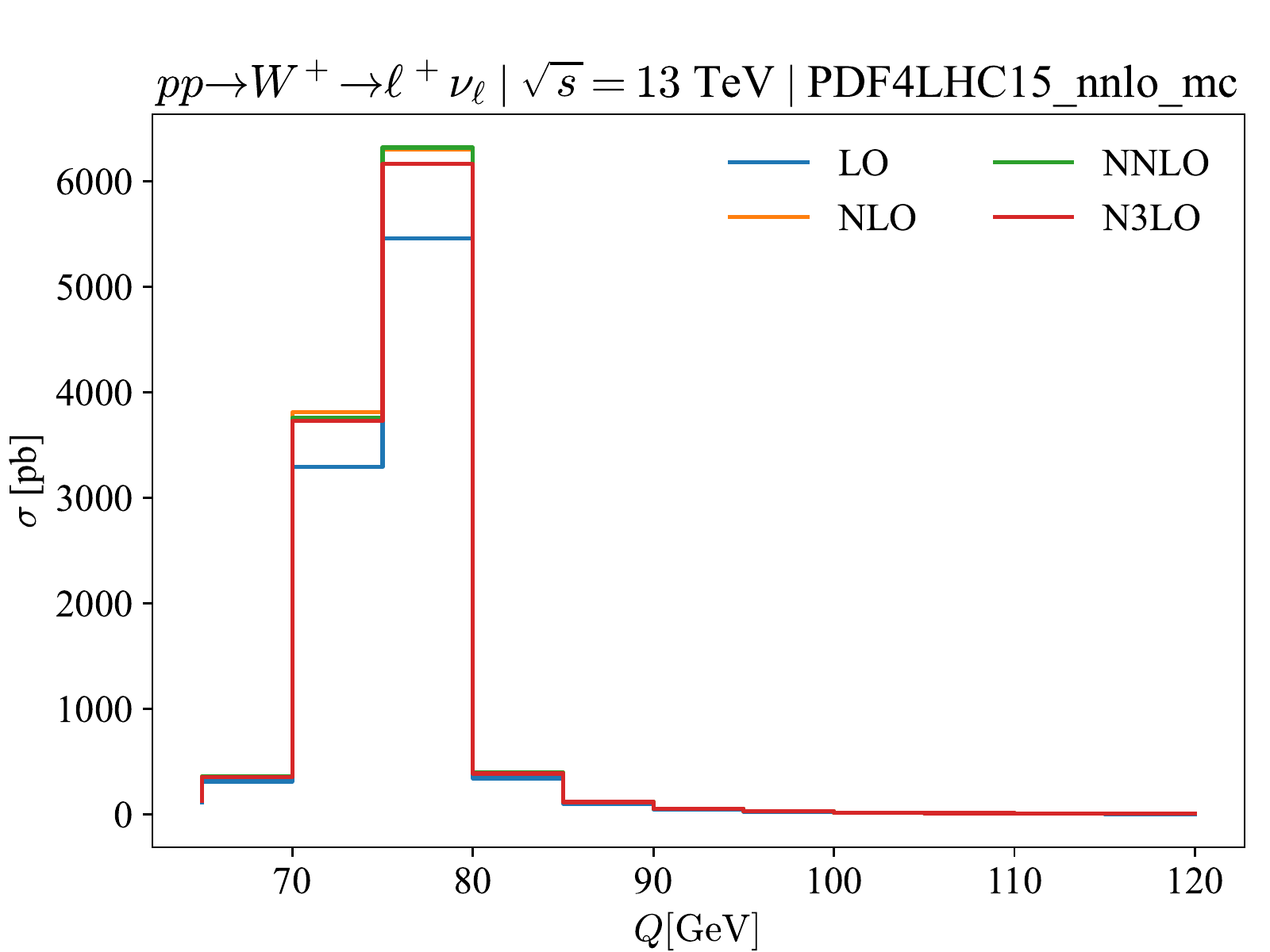}
\includegraphics[width=0.32 \textwidth]{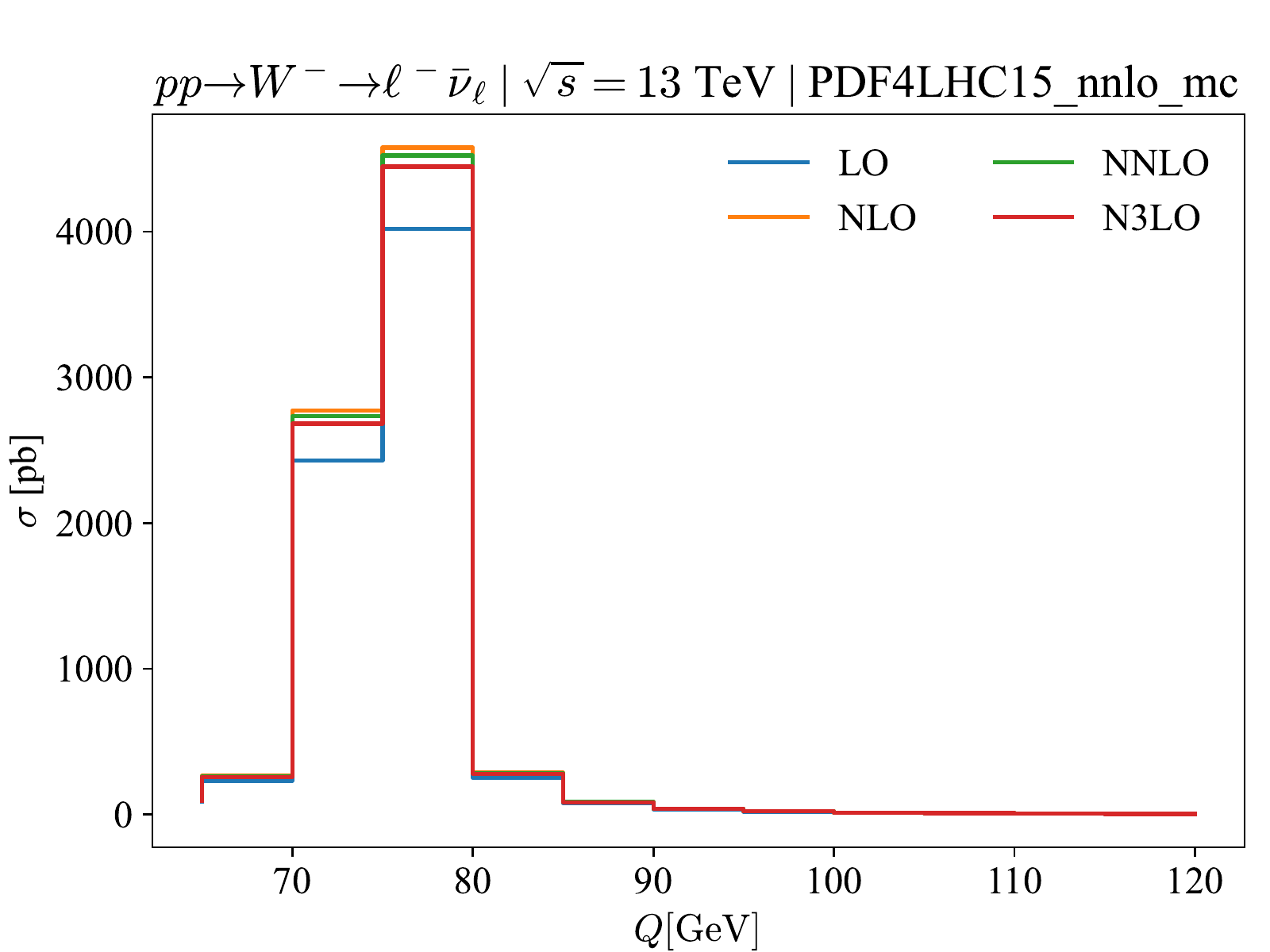}\\
\includegraphics[width=0.32 \textwidth]{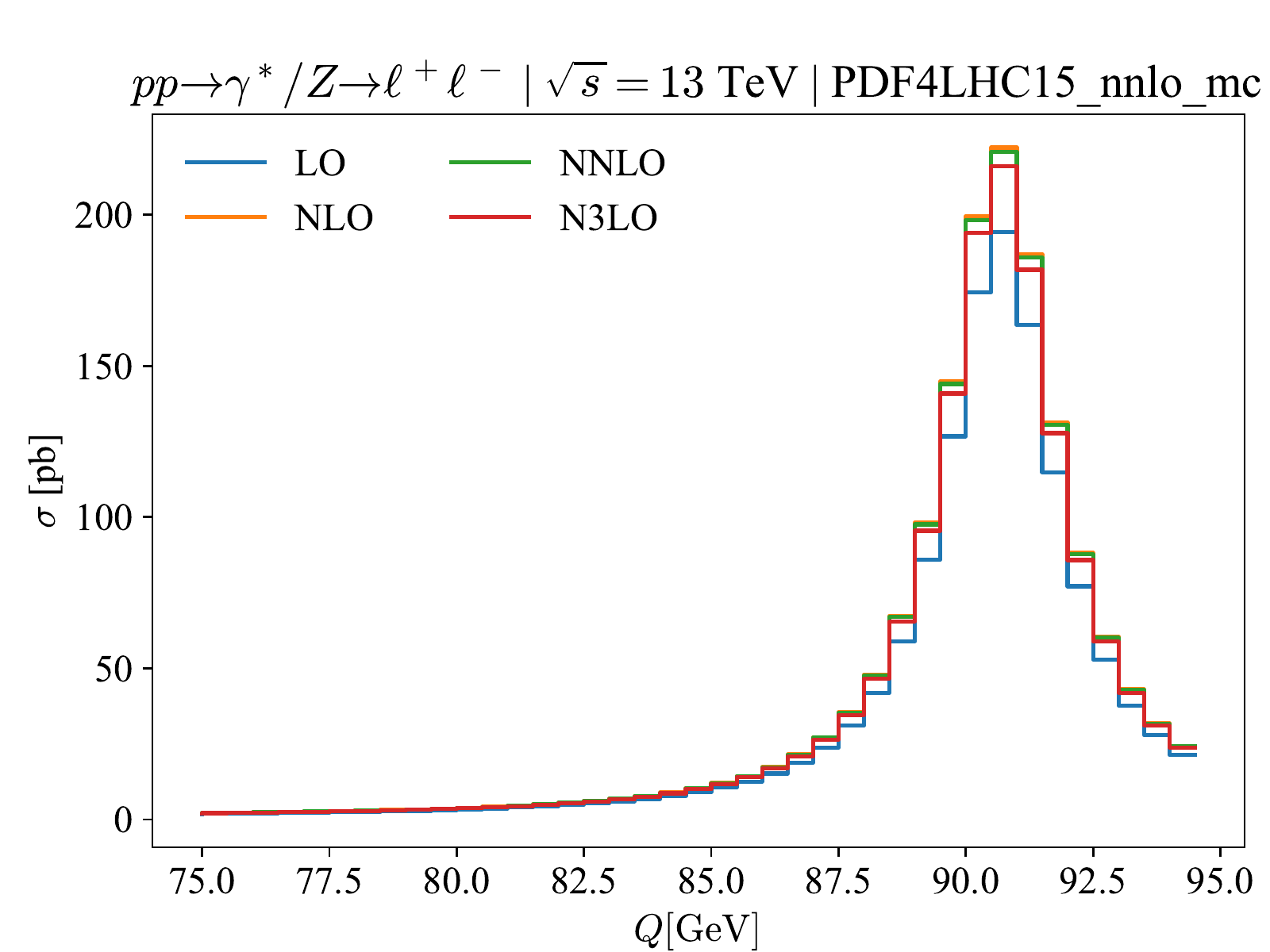} 
\includegraphics[width=0.32 \textwidth]{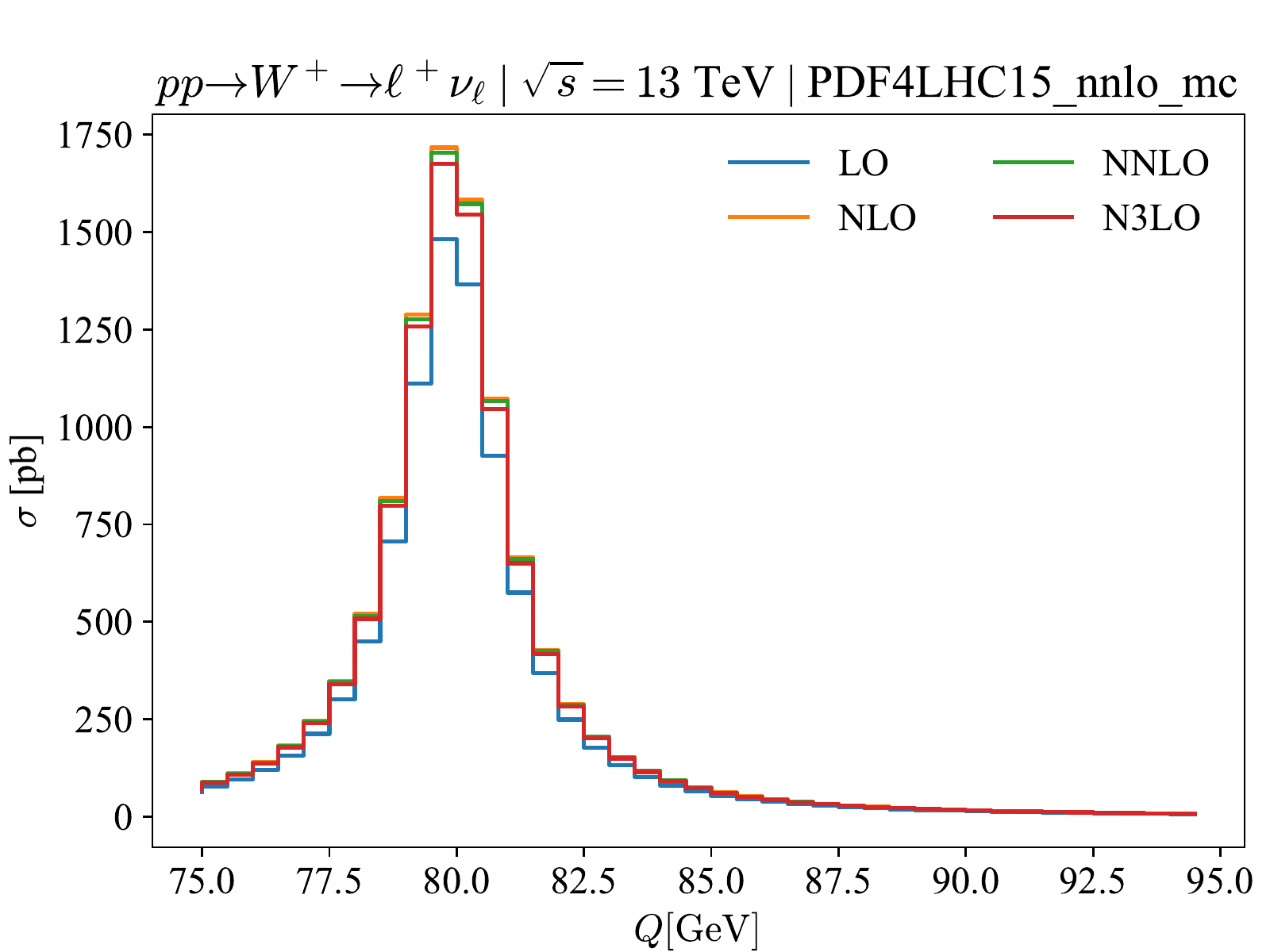}
\includegraphics[width=0.32 \textwidth]{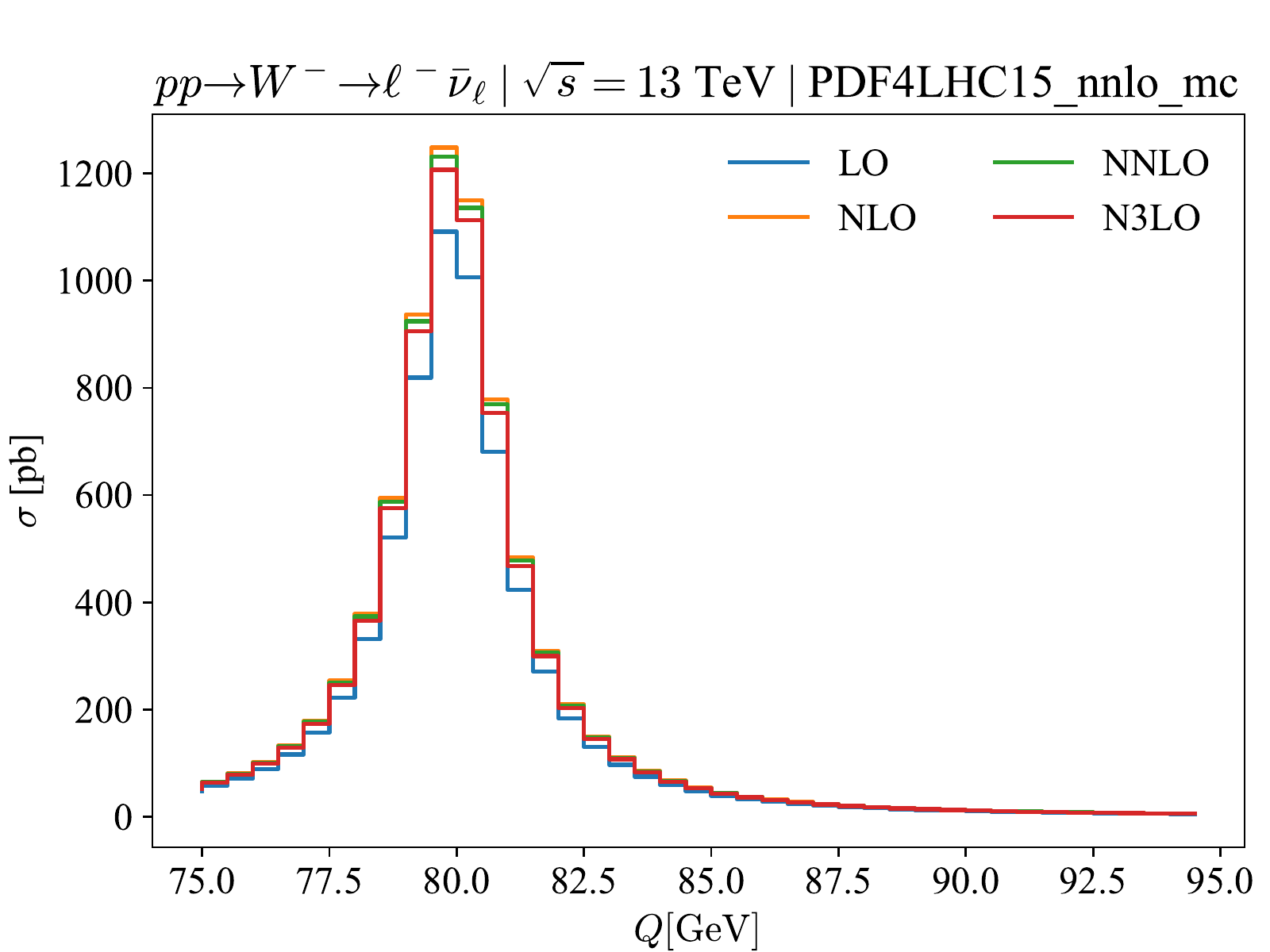}\\
\caption{\label{fig:bins_DY}The invariant-mass distribution around the $Z$ or $W$-peak for the neutral- and charged-current Drell-Yan processes at the 13 TeV LHC, in bins of size 5 GeV (upper panels) and 0.5 GeV (lower panels) respectively. Note that the $y$-axis on the upper panel is logarithmic.}
\end{center}
\end{figure}

{\tt n3loxs} can also calculate Drell-Yan cross sections in a given bin range of the invariant-mass distribution $Q_1 \leq Q \leq Q_2$:
\beq\label{eq:bins}
\int_{Q_1^2}^{Q_2^2}\df Q^2\,\frac{\df \sigma_{N_1 N_2\to f_1 f_2+X}}{\df Q^2}\,.
\eeq
Using eq.~\eqref{eq:bins} we can produce the invariant-mass distribution in a binned form, in windows around the $Z$- or $W$-peak. As an example, we display in fig.~\ref{fig:bins_DY} at the 13 TeV LHC for bin sizes of 5 GeV and 0.5 GeV respectively. By choosing the largest bin range possible, it is possible to obtain the total cross section for the processes $pp\to Z^*/\gamma^*\to \ell^+\ell^-$ and $pp\to W\to  \ell \nu_\ell$ up to N$^3$LO in QCD, using dynamical renormalization and factorization scales $\mu_R=\mu_F=Q$. 

\bibliography{inclusive}
\bibliographystyle{JHEP}

\end{document}